\documentclass[%
aps,
prx,
reprint,
superscriptaddress,
nofootinbib,
amsmath,
amssymb
]{revtex4-2}
\usepackage{
    physics,
    graphicx,
    multirow,
    tabularx,
    mathtools, 
    placeins, 
    nicefrac, 
    hyperref, 
    threeparttable, 
    siunitx,
    enumitem,
    makecell
}
\newcommand{\heading}[2]{\parbox[t][6ex][t]{#1}{\footnotesize #2}}

\usepackage[capitalize]{cleveref}
\Crefname{section}{Sec.}{Secs.}

\usepackage[dvipsnames]{xcolor}
\hypersetup{
    colorlinks,
    linkcolor={blue!50!black},
    citecolor={blue!50!black},
    urlcolor={blue!80!black}
}

\usepackage[caption=false]{subfig} 
\graphicspath{{Figures/}}

\newcommand{\tsh}{\tau_{\mathrm{sh}}}
\newcommand{\tmeas}{\tau_{\mathrm{meas}}}
\newcommand{\tmax}{\tau_{\mathrm{max}}}
\newcommand{\tgap}{\tau_{\mathrm{gap}}}
\newcommand{\tinit}{\tau_{\mathrm{init}}}
\newcommand{\tcz}{\tau_{\textsc{cz}}}
\newcommand{\tH}{\tau_{\textsc{h}}}
\newcommand{\Tcirc}{T_{\mathrm{circ}}}
\newcommand{\Tloop}{T_{\mathrm{loop}}}
\newcommand{\Tcycle}{T_{\mathrm{cycle}}}
\newcommand{\Kloop}{K_{\mathrm{loop}}}

\begin{document}


\title{Looped Pipelines Enabling Effective 3D Qubit Lattices in a Strictly 2D Device}

\newcommand{\qmaddress}{\affiliation{Quantum Motion, 9 Sterling Way, London N7 9HJ, United Kingdom}}
\newcommand{\oxddress}{\affiliation{Department of Materials, University of Oxford, Parks Road, Oxford OX1 3PH, United Kingdom}}

\author{Zhenyu Cai}
\email{cai.zhenyu.physics@gmail.com}
\qmaddress
\oxddress

\author{Adam Siegel}
\qmaddress
\oxddress

\author{Simon Benjamin}
\email{simon.benjamin@materials.ox.ac.uk}
\qmaddress
\oxddress

\date{\today}

\begin{abstract}
    Many quantum computing platforms are based on a two-dimensional physical layout. 
    Here we explore a concept called \emph{looped pipelines} which permits one to obtain many of the advantages of a 3D lattice while operating a strictly 2D device. The concept leverages qubit shuttling, a well-established feature in platforms like semiconductor spin qubits and trapped-ion qubits. The looped pipeline architecture has similar hardware requirements to other shuttling approaches, but can process \emph{a stack of qubit arrays} instead of just one.
    Even a stack of limited height is enabling for diverse schemes ranging from NISQ-era error mitigation through to fault-tolerant codes. For the former, protocols involving multiple states can be implemented with a space-time resource cost comparable to preparing \emph{one} noisy copy. For the latter, one can realise a far broader variety of code structures; as an example we consider layered 2D codes within which transversal CNOTs are available. 
    Under reasonable assumptions this approach can reduce the space-time cost of magic state distillation by two orders of magnitude.
    Numerical modelling using experimentally-motivated noise models verifies that the architecture provides this benefit without significant reduction to the code's threshold. 
\end{abstract}

\maketitle

\section{Introduction}
Many platforms that are being explored for quantum computing have the property that qubits can only interact with physically proximal partners. For such systems, one may ask about the importance of the dimensionality of the qubit array: Would 1D, 2D or 3D be required to achieve a given task efficiently? There are a number of studies related to this question. For example, within a certain framework
it is possible to achieve quantum advantage using a noisy circuit of constant depth on a 3D qubit lattice~\cite{bravyiQuantumAdvantageNoisy2020}, but it is not possible on a 2D qubit array if we are considering tasks like variational quantum algorithms~\cite{bravyiClassicalAlgorithmsQuantum2021}. When considering quantum error correction (QEC), moving from 2D qubit arrays to 3D qubit lattices enables 2D topological codes with transversal CNOT gates~\cite{dennisTopologicalQuantumMemory2002} or 3D topological codes with transversal non-Clifford gates~\cite{bombinTopologicalComputationBraiding2007,vasmerThreedimensionalSurfaceCodes2019}. In other applications like quantum error mitigation (QEM)~\cite{koczorExponentialErrorSuppression2021,hugginsVirtualDistillationQuantum2021} and quantum annealing~\cite{qiuProgrammableQuantumAnnealing2020}, 3D qubit lattices can lessen or remove the demand for long-range gates which are usually much noisier. However despite the advantages provided by the 3D qubit lattices mentioned above, most hardware platforms nowadays are still confined to a 2D layout due to technological challenges. Hence, it is interesting to explore methods to efficiently implement an effective 3D qubit lattice on a physical 2D hardware substrate.

In many state-of-the-art hardware platforms, besides the qubit connectivity limitations, another key challenge is the conflict between providing spaces for classical control and maintaining a high qubit density. In hardware platforms where native entangling operations are short-range, like silicon spin systems and trapped ions, we might expect to have an array of closely-packed qubits. However, addressing qubits in such a closely-packed array might lead to crosstalk. Moreover, the number of classical control elements required can scale with the area of the array (number of qubits) while the available space for these control elements may only scale with the perimeter of the array, leading to challenges in the wiring routing and heat dissipation~\cite{vandersypenInterfacingSpinQubits2017}. To tackle the dichotomy of short-range interactions versus the desire for well-spaced structures, a natural solution is expanding the qubit array using shuttling tracks: Qubits that are scheduled to interact are brought together through qubit shuttling. As a result, qubit shuttling forms the basis for many of the most promising scalable architectures for platforms like semiconductor spin qubits~\cite{buonacorsiNetworkArchitectureTopological2019,boterSpiderwebArraySparse2022,jnaneMulticoreQuantumComputing2022} and trapped-ion qubits~\cite{kielpinskiArchitectureLargescaleIontrap2002,lekitschBlueprintMicrowaveTrapped2017,kaushalShuttlingbasedTrappedionQuantum2020,ryan-andersonRealizationRealTimeFaultTolerant2021,hilderFaultTolerantParityReadout2022}. The related question of how to efficiently exploit the shuttling capabilities using suitable control sequences has also been considered in Refs~\cite{tomitaComparisonAncillaPreparation2013,bermudezAssessingProgressTrappedIon2017,gutierrezTransversalityLatticeSurgery2019,bermudezFaulttolerantProtectionNearterm2019}. However, as we space out the qubit array using shuttling tracks, it appears that the qubit density will decrease accordingly, reducing the number of qubits that we can fit onto a single integrated platform (e.g. a silicon chip).

In this article, we study a pipelining architecture which is obtained by replacing the commonly-considered linear shuttling tracks with shuttling loops. In this way, instead of storing and processing \emph{a single 2D qubit array} in the shuttling-based architecture, we can effectively store and process \emph{a stack of 2D qubit arrays} using a similar amount of physical resources, i.e. we aim to increase the qubit density for shuttling-based architecture without compromising any of the advantages brought by shuttling. Then, with the simple generalisation of enabling (even limited) interactions between qubits in the same shuttling loop, we can perform transversal interactions between different layers in the stack of qubit arrays. This transforms the stack of qubit arrays into an effective 3D qubit lattice and enables computing tasks that are impossible before in a strictly 2D lattice of static, locally interacting qubits. Inevitably there are practical constraints on the height of the effective 3D lattice one can achieve, depending on the hardware and the circuits we try to run. However as we will see later, despite such limitations, very significant advantages can still emerge in various applications.

Our interest is in performing circuit-based quantum computation on matter qubits like silicon spin and trapped-ion qubits. Howoever before we proceed, we also want to take note of the extensive literature for building 3D cluster states in the photonic platform, which can be used for performing fault-tolerant measurement-based quantum computation to overcome challenges like non-deterministic two-qubit gates and photon losses~\cite{raussendorfFaulttolerantOnewayQuantum2006,raussendorfTopologicalFaulttoleranceCluster2007,raussendorfFaultTolerantQuantumComputation2007}. Due to the ``flying'' nature of photonic qubits, in order to perform a given sequence of operations in a photonic processor, one must physically arrange a series of optical components along the optical path; similarly, in conventional computers the classical circuits we want to perform are physically printed onto the classical processors. Due to such similarities, classical concepts like pipelining can be naturally adopted into photonic processors, turning into techniques like time-domain multiplexing and optical loops, which are then used in a range of proposed schemes for building 3D photonic cluster states~\cite{liResourceCostsFaultTolerant2015,augerFaulttolerantQuantumComputation2018,herrLocalScalableLattice2018,fukuiTemporalmodeContinuousvariableThreedimensional2020,bourassaBlueprintScalablePhotonic2021,larsenFaultTolerantContinuousVariableMeasurementbased2021}. On the other hand, typical architectures for \emph{stationary} matter qubits operate on a completely different paradigm with the target circuit emerging from the time direction rather than physically appearing `baked into' the processor layout. Hence, the adoption of classical computing techniques for matter qubits is much less straightforward. In this article, we will construct a hybrid approach between flying and stationary qubits. This approach allows one to apply the classical pipelining concept to matter qubits and support features native to \emph{3D circuit-based} scalable quantum computation, whereas existing schemes for performing \emph{measurement-based} quantum computation using 3D photonic cluster states are effectively equivalent to \emph{2D circuit-based} scalable quantum computation.

We start by looking at the general construction of the pipelining architecture in \cref{sec:pipeline_array}. It is then used to implement a stack of 2D topological codes in \cref{sec:QEC}, and more specifically for the task of magic state distillation in \cref{sec:magic_state_distill}. We consider the specific case of semiconductor platforms in \cref{sec:silicon_implementation}, where we present numerical simulations to determine thresholds for tolerable shuttling errors. In \cref{sec:QEM}, we look at a near-term application of the looped pipelining architecture in QEM. We conclude with some general remarks and a discussion of possible future directions in \cref{sec:concl}.

\section{Pipelining 2D Qubit Arrays}\label{sec:pipeline_array}
\subsection{Classical Linear Pipeline}\label{sec:classical_pipeline}
\begin{figure*}[htbp]
    \centering
    \includegraphics[width = 0.7\textwidth]{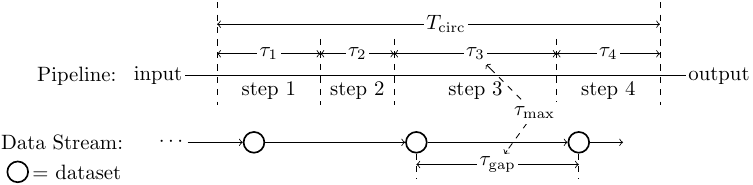}
    \caption{An example of a classical data pipeline consists of $4$ steps. Here we have a steady flow of datasets with a time gap $\tmax$ between consecutive datasets. }
    \label{fig:classical_pipeline}
\end{figure*}

Let us suppose we want to implement a classical data-processing circuit on many datasets. To convert the circuit into a \emph{data-processing pipeline}, we will divide it into multiple \emph{steps} (also called \emph{stages}) with each step only able to operate on \emph{one dataset at a time} as shown in \cref{fig:classical_pipeline}. Now instead of inputting the second dataset only after the first dataset completed the whole circuit, we can input the second dataset into the first data-processing step as soon as the first dataset leaves the first step, similarly for all the subsequent datasets and the subsequent steps. In this way, all data-processing steps can be working on different datasets in parallel, increasing the throughput of the system. This is the concept of \emph{pipelining}. In fact, it is often the case that each step can actually perform a range of different operations with the same efficiency when using different control parameters. Hence, we can implement different circuits on the different datasets using the same data-processing pipeline. 

The processing time needed for the $m^\text{th}$ step in the pipeline is denoted as $\tau_m$. Suppose there are $M$ steps in total, then the time taken for \emph{one} dataset to flow through the entire pipeline and execute the whole circuit is simply:
\begin{align}\label{eqn:circ_time}
    \Tcirc = \sum_{m = 1}^{M} \tau_m.
\end{align}
This is the processing time required for \emph{every} dataset without pipelining, and it is also the time required to process the \emph{first} dataset in the pipeline. The additional time required to process the second dataset is determined by the slowest (rate-limiting) step in the pipeline with a processing time of
\begin{align}\label{eqn:rate_limiting_time}
    \tmax = \max_{i} \tau_i,
\end{align}
which is called \emph{the rate-limiting step time}. This is because the second dataset can only enter the slowest step when the first dataset exits, similarly for any additional datasets.  Hence, the total time needed to process $k$ datasets using such a pipelining scheme is at least
\begin{equation}\label{eqn:steady_flow_pipeline_time}
    \begin{aligned}
        T_{\mathrm{pipe}}(k) &= \Tcirc + (k-1)\tmax.
    \end{aligned}
\end{equation}
We will call this the \emph{steady-flow pipelining scheme} since the data stream can flow through the entire pipeline without needing to modify the time gap between adjacent datasets or be put on hold anywhere along the pipeline. 

As we deviate from the steady-flow scheme, we will often need to temporarily put the datasets on hold within the pipeline. To achieve this, we would need to add \emph{buffer}s in between the steps, which can hold multiple datasets temporarily and require no processing time.

\subsection{Looped Qubit Pipeline}\label{sec:qubit_pipeline}
\begin{figure}[htbp]
    \centering
    \subfloat[Linear qubit pipeline \label{fig:linear_qubit_pipeline}]{\includegraphics[width = 0.4\textwidth]{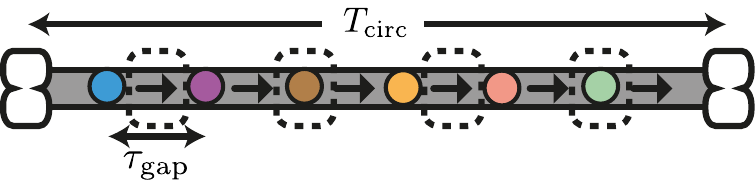}}\\
    \subfloat[Looped qubit pipeline \label{fig:qubit_pipeline}]{\includegraphics[width = 0.4\textwidth]{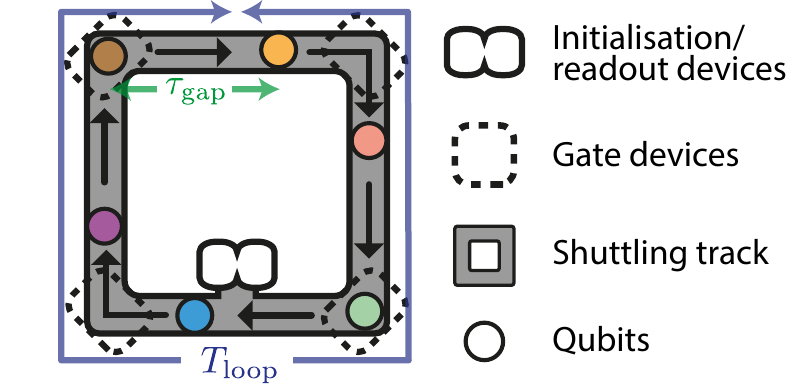}}
    \caption{Pipelines for applying single-qubit circuits on a stream of qubits.}
\end{figure}

Similar to the classical pipeline, we can define a linear qubit pipeline as shown in \cref{fig:linear_qubit_pipeline}~\cite{patomakiPipelineArchitectureSilicon},  with the datasets being qubits and the processing steps being shuttling steps, quantum gates devices, initialisation and measurement devices. Each shuttling step here is a small section of shuttling track that can hold one and only one qubit at a time. We can also use the steady-flow scheme for the qubit pipeline, for which the time required for processing $k$ qubits is simply given by \cref{eqn:steady_flow_pipeline_time}. Note that in practice, each shuttling step is usually a very short shuttling section that performs relatively quickly compared to the other processes (see, e.g. \cref{sec:silicon_implementation}), thus the rate-limiting step is often not shuttling. We will assume without loss of generality that shuttling tracks can hold the qubits without pushing them forward, thus they can also act as buffering regions for the pipeline when needed. 

Going beyond a linear pipeline, we can have a looped pipeline, which can be implemented using the structure shown in \cref{fig:qubit_pipeline}. This piece of hardware, which we will simply refer to as a ``\emph{loop}'', is made out of initialisation and measurement devices connecting to an outer shuttling loop with gate devices. Unlike the linear qubit pipeline, in which the circuit depth we can perform is limited by the number of gate devices in the hardware, by allowing the qubits to go around the loop and reuse the gate devices, we can now effectively perform circuits of \emph{any depth} even though we only have a small number of gate devices in the hardware. 

Let us focus on the first qubit in the qubit stream, the time needed for it to wrap around the loop will be called the \emph{cycling period} and denoted as $\Tloop$. Assume that all additional qubits in the looped pipeline follow behind with a constant time gap of $\tgap$ between the consecutive qubits. When the qubit stream wraps around the loop, the first qubit in the qubit stream may collide with the last qubit. To avoid such \emph{qubit collision}, we need to ensure the cycle period $\Tloop$ is larger than the time gap between the first and the last qubit, which is $(k-1)\tgap$ with $k$ being the number of qubits:
\begin{align*}
    \Tloop \geq  (k-1)\tgap.
\end{align*}

The cycling period $\Tloop$ may change from one round to another since different steps on the loop can be activated at different rounds. We might want to apply $3$ gates in this round, but only $2$ gates in the next round; moreover certain rounds may not involve measurement or initialisation. Let us denote the \emph{minimum cycling period} throughout the whole pipeline process as $\Tloop^{\mathrm{min}}$, then to avoid qubit collision throughout the pipeline, we need to have:
\begin{align}
    \Tloop^{\mathrm{min}} &\geq  (k-1)\tgap \nonumber\\
    k &\leq   \Tloop^{\mathrm{min}}/\tgap + 1 = \Kloop \label{eqn:max_qubit_on_loop}
\end{align}
where $\Kloop$ is the maximum number of qubits we can fit on the loop. 

In the steady-flow scheme, we have $\tgap = \tmax$. Therefore, the maximum number of qubits we can fit in the loop in this case is simply:
\begin{align}\label{eqn:max_qubit_steady_flow}
    \Kloop =  \Tloop^{\mathrm{min}}/\tmax + 1.
\end{align}

If $\Kloop$ is too small and we want to fit in more qubits into the pipeline, we can try to increase 
$\Tloop^{\mathrm{min}}$ by adding buffering times, which we will simply call \emph{collision buffering}. In this article, in most of the cases, we will try to keep the qubit number in the loop low enough to satisfy \cref{eqn:max_qubit_steady_flow} so that no collision buffering will be needed.

\begin{figure}[htbp]
    \centering
    \includegraphics[width = 0.35\textwidth]{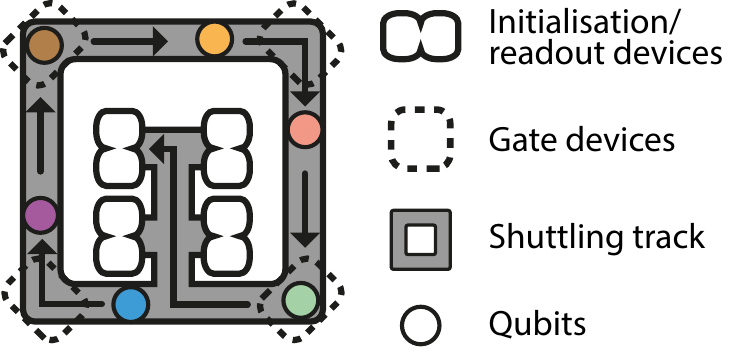}
    \caption{An example of a looped qubit pipeline with multiple initialisation/measurement devices.}
    \label{fig:multi_meas_qubit_pipeline}
\end{figure}

We can also try to increase the number of pipelined qubits by reducing $\tmax$. When the rate-limiting step is measurement or initialisation, we can reduce their effective processing time by adding more initialisation/measurement devices, for example as shown in \cref{fig:multi_meas_qubit_pipeline}, and operating them in parallel. With $m$ times more initialisation/measurement devices in pipeline, the effective measurement/initialisation time in the pipeline, and thus rate-limiting step time $\tmax$, will be reduced by a factor of $m$ as long as we are operating on more than $m$ qubits: $\tmax = \frac{\tau_{\mathrm{init/meas}}}{m}$. It is also possible to reduce the measurement/initialisation time to the point that they are not the rate-limiting step anymore. It is worth noting that \emph{without pipelining}, naively putting in more initialisation/measurement devices will not reduce the processing time. 

In the case that the rate-limiting step is the gate operation, we may be able to reduce its effective processing time by, for example, applying the gate operation while the qubit(s) involved travelling along a substantial portion of the loop. Then, the gate operation can be carried out alongside the shuttling process, instead of localising the gate operation at a corner. As will be further discussed in \cref{sec:edge_interaction}, this effectively decomposes the gate operation into many small gate steps to reduce the rate-limiting step time $\tmax$. 

One other possibility to fit more qubits in the loop is by adding bypasses to allow the qubits at the front of the qubit stream to lap the qubits at the back, so that the length of the whole qubit stream can wrap around the outer loop more than once. However, this could lead to more complex scheduling of the pipeline and possibly additional time cost. 

It is also possible to fit in more qubits \emph{without making any modification to the hardware} and instead by simply reducing the pipelining time gap between the qubits to below $\tmax$ as mentioned in \cref{sec:steady_flow_around_loop}. As we will see later in our discussion in \cref{sec:silicon_code_cycle}, this can be one of the most practical ways for solving the qubit collision problem. 

\subsection{Qubit-array Pipeline}\label{sec:array_pipeline}
\begin{figure}[htbp]
    \centering
    \subfloat[Processing a five-qubit array. \label{fig:array_shuttle}]{\includegraphics[width = 0.45\textwidth]{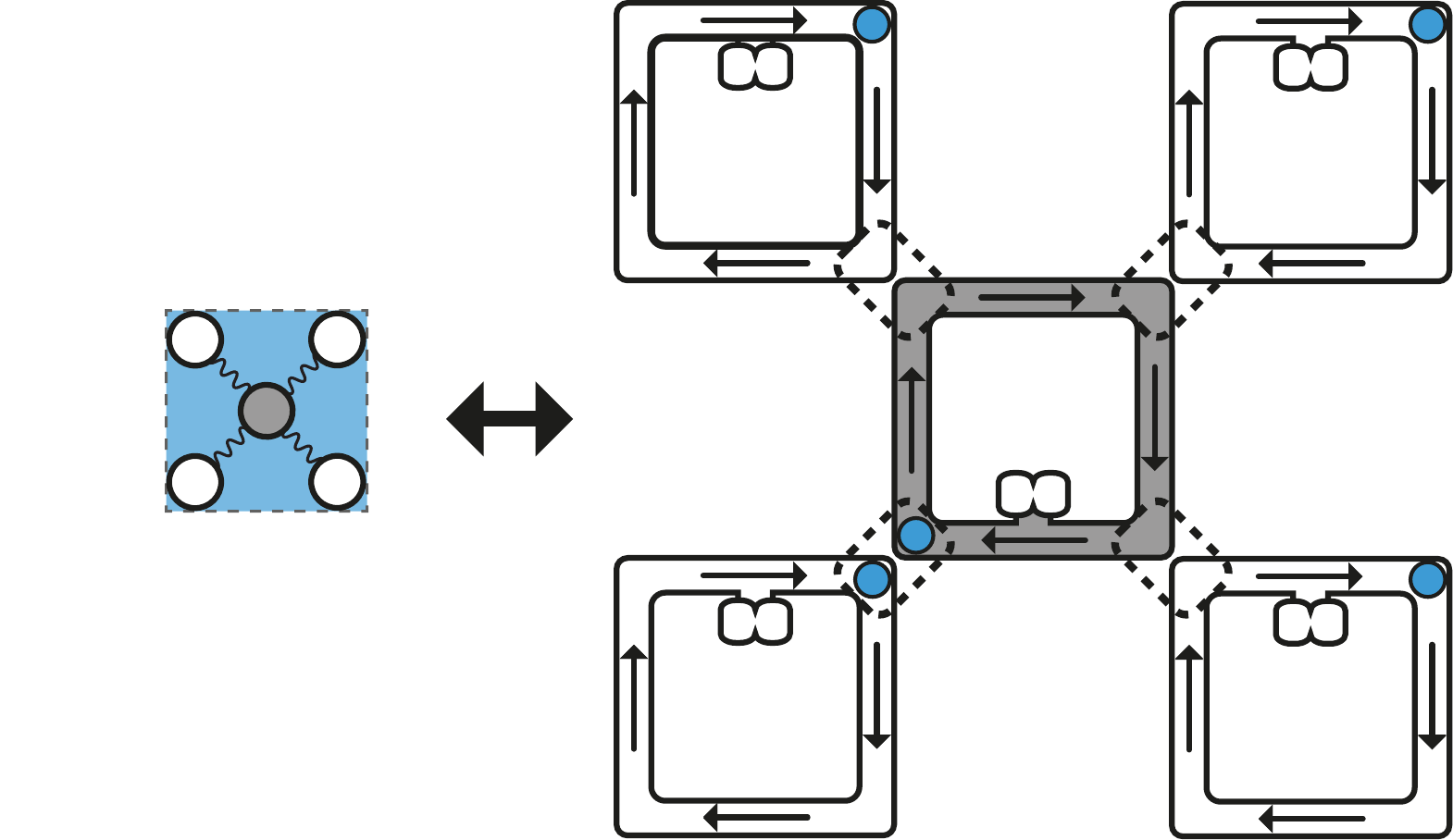}}\\
    \subfloat[Processing 4 five-qubit arrays using pipelining.\label{fig:array_pipeline}]{\includegraphics[width = 0.45\textwidth]{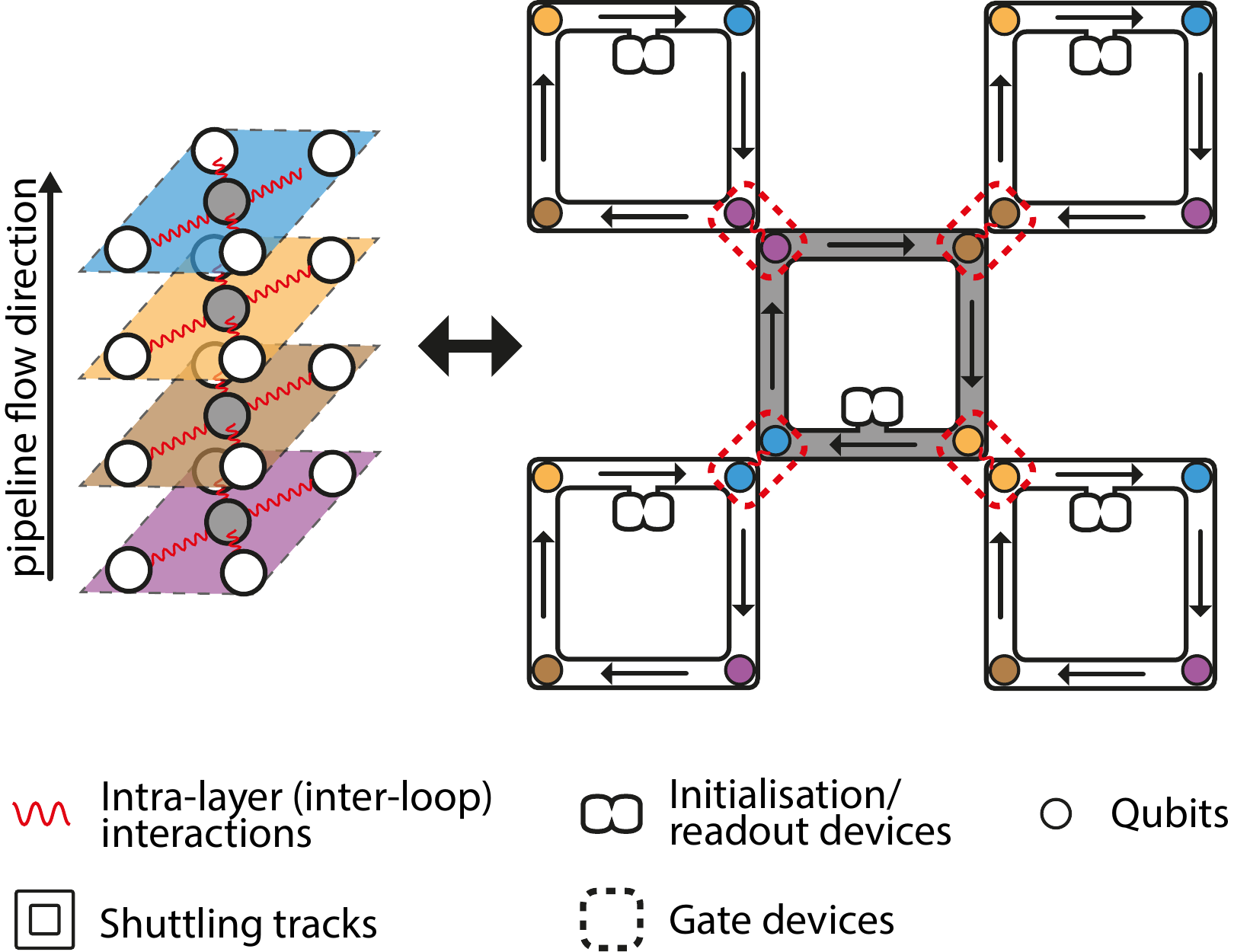}}
    \caption{Five-qubit arrays stored in a loop array. Multiple qubits can be stored on the same loop through pipelining, which corresponds to multiple layers of five-qubit arrays.}
    \label{fig:array_pipeline_both}
\end{figure}

Let us consider a five-qubit array with a central qubit interacting with the four neighbouring qubits. In order to avoid crosstalk among them and to provide additional spaces for the wiring of the classical controls, we can space out the qubits using shuttling loops as shown in \cref{fig:array_shuttle}, where the qubit array is now stored in a \emph{loop array} formed from the loop elements mentioned in the last section. By synchronising the movement of qubits in different loops, if the central qubit moves through one round of the loop, it will come into contact with all of the surrounding qubits along the way and interact with them, mimicking the connectivity of the five-qubit array as shown in \cref{fig:array_shuttle}. A method for synchronising more complex qubit movements is outlined in \cref{sec:shuttling_sync}. Such an architecture can be easily extended to a large qubit array. 

Now instead of one qubit per loop, we can fit multiple qubits in each loop as shown in \cref{fig:array_pipeline}, essentially loading multiple qubit arrays (denoted by different colours) into the loop array \emph{without adding any additional components}. As shown in \cref{fig:array_pipeline}, by synchronising the movement of qubits among different loops (as in the pure shuttling case in \cref{fig:array_shuttle}) and allowing the various devices on the loops to operate in parallel, we can process a stack of qubit arrays from the top layer to the bottom layer in a pipelining manner. Hence, this loop array can be used to implement a \emph{qubit-array pipeline} and thus we will also refer to it as a \emph{pipelining architecture}. Even in the simple case that the various layers of the virtual stack do not interact, this is already interesting for applications such as VQAs where one needs to repeatedly prepare and measure the same quantum state (to adequately learn a set of obervables).

Let $A$ denote the \emph{number of (entire) loop arrays} we need to store and process all qubit arrays, which is proportional to the \emph{space overhead} needed. When the number of qubit arrays is $n$, we can store them in $A = n$ different loop arrays ($n$ copies of arrays similar to \cref{fig:array_shuttle}) and process them all in \emph{parallel}. Alternatively, we can store them all in a single loop array ($A = 1$) and process them using \emph{pipelining}, leading to \emph{a factor of $n$ saving in the spatial overhead}, i.e. a factor of $n$ increase in the qubit density. As given by \cref{eqn:steady_flow_pipeline_time}, comparing the pipelining scheme to the parallel scheme, the time needed for the main computation is the same, which is simply $\Tcirc$, but processing each additional qubit using pipelining will increment the total time required by an amount $\tmax$ as in \cref{eqn:steady_flow_pipeline_time}.

In an application such as measuring observables by repeated sampling, it is likely that the number of qubit arrays $n$ we want to process exceeds the maximum number of qubits we can fit in the pipeline $\Kloop$ given by \cref{eqn:max_qubit_on_loop}. Then need to distribute the $n$ qubit arrays into at least $A = \lceil \nicefrac{n}{\Kloop}\rceil$ different loop arrays so that the number of qubit arrays stored in each loop array will not exceed $\Kloop$. Nevertheless, we compress the spatial overhead by a factor of $\Kloop$. This will be further explored in the context of surface code computation in \cref{sec:multi_stack}.

\begin{figure}[bhtp]
    \centering
    \includegraphics[width = 0.4\textwidth]{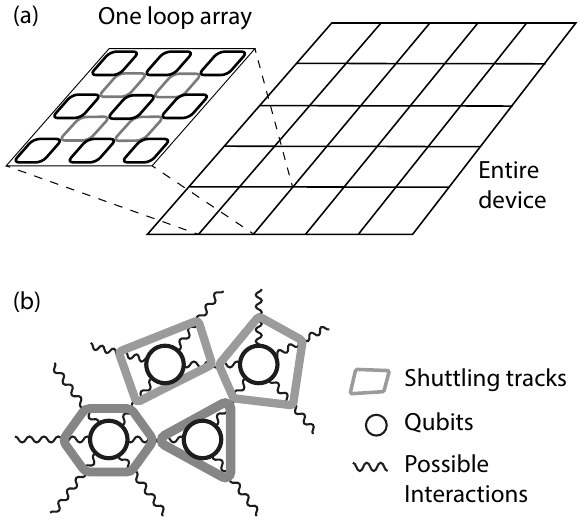}
    \caption{(a) An entire array of loops, sufficient for a given role (a NISQ calculation, or representing one logical qubit) may be only part of the entire device. (b) Transforming a given qubit connectivity graph to a pipelining architecture. The shuttling loop of a given qubit is simply constructed by connecting up the mid-points of all of its interaction edges.}
    \label{fig:graph_to_shuttling}
\end{figure}

The qubit-array pipeline can also be implemented on 2D qubit layouts beyond valence-4 connectivity as illustrated in \cref{fig:graph_to_shuttling}. We can see that since the inter-loop interactions are carried out at the corners of the loop, the qubits with $n$ neighbouring qubits will simply be transformed into $n$-gon loops, whose $n$ corners are connected to the corners of the $n$ neighbouring loops for interaction. There are of course other pipelining architectures possible as will be discussed in \cref{sec:edge_interaction}. 

\begin{figure}[htbp]
    \centering
    \includegraphics[width = 0.45\textwidth]{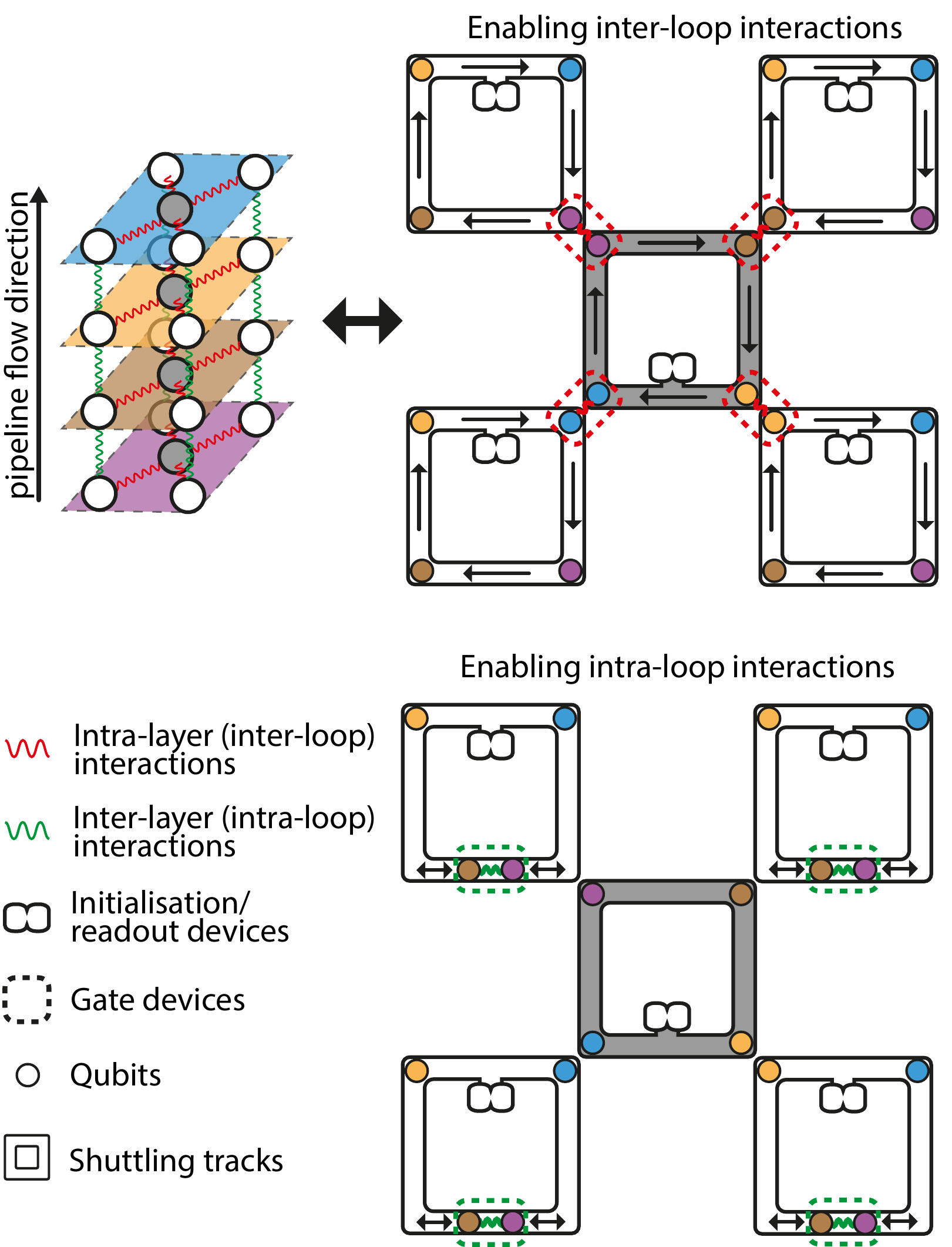}
    \caption{Interactions within layers are enabled by inter-loop (red) interactions while transversal interaction in between layers are enabled by intra-loop (green) interactions.
    }
    \label{fig:array_pipeline_intra}
\end{figure}

Generalising the architecture described above, we can add components to allow interactions between specific qubits within the same loop; in the stack picture this corresponds to interactions between qubits in different layers. If the same pattern of intra-loop interaction is repeated over a complete set of loops (as in the brown-purple pairing within the white loops in the lower panel of \cref{fig:array_pipeline_intra}) then we implement `transversal' interactions between layers. In this way, the qubit connectivity is effectively extended beyond a 2D array to a 3D lattice. The ``height'' of this 3D lattice cannot be increased indefinitely due to the limitation on the number of qubits in each loop as discussed in \cref{sec:qubit_pipeline}. Nonetheless, the increased connectivity can expand the computation power of the device by for example, enabling transversal CNOT gates in QEC codes and an efficient implementation of purification-based QEM as will be discussed later. 

The simplest way of scheduling these intra-loop interactions is just halting the pipelining flow, effectively halting all inter-loop interactions, and then bringing the corresponding qubits in the loop together for interaction. We can have a circuit structure of alternating layers of intra-layer (inter-loop) and inter-layer (intra-loop) interactions just like the standard gate scheduling for a 3D qubit lattice. In this article, we will be considering applications with mostly intra-layer (inter-loop) operations and rare inter-layer (intra-loop) operations. Hence, we will ignore the time cost for inter-layer (intra-loop) operations in the analysis for those applications.

\begin{figure}[htbp]
    \centering
    \includegraphics[width = 0.4\textwidth]{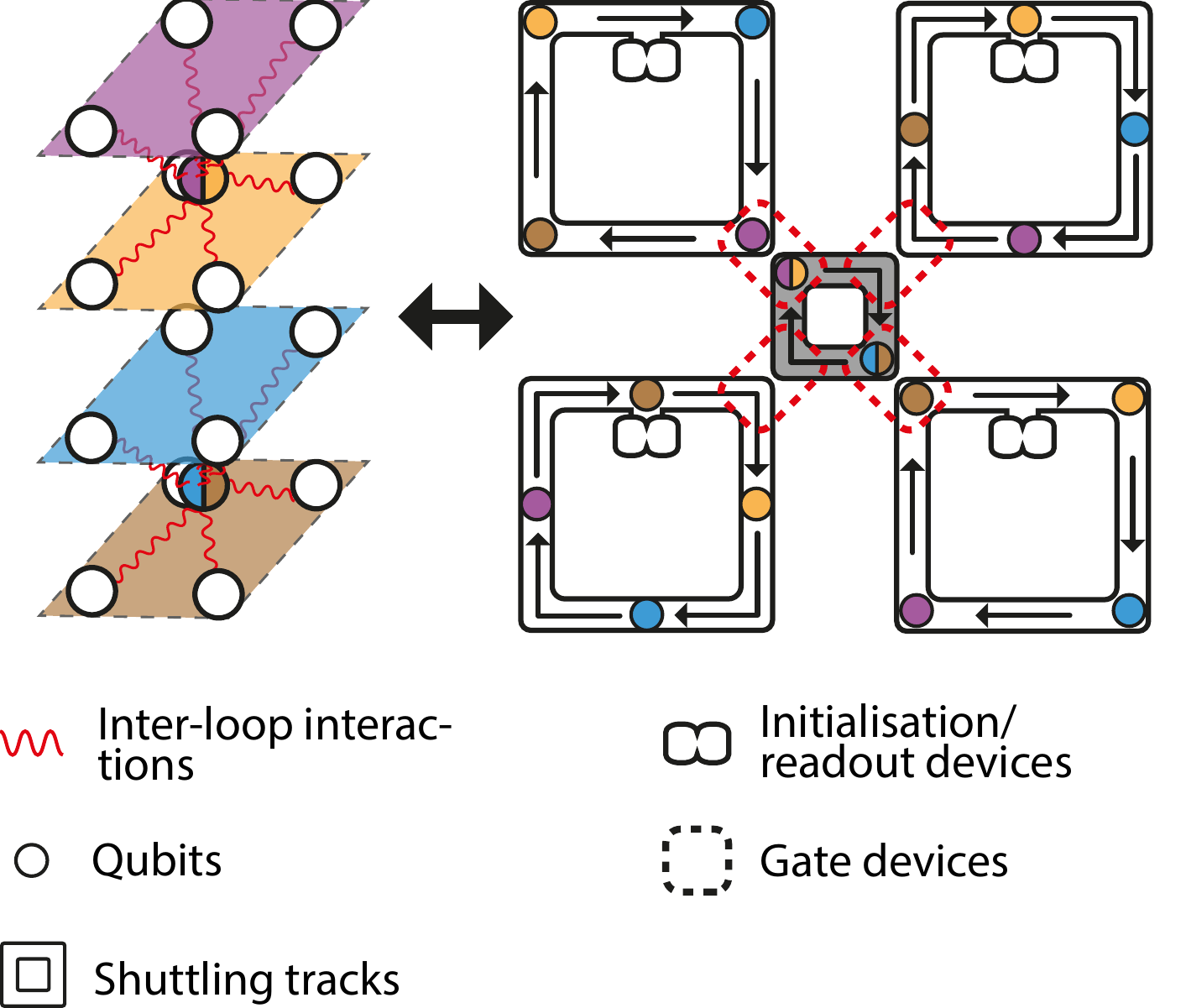}
    \caption{Constructing 3D qubit connectivity without using intra-loop interactions. Here the cycling frequency of the centre (grey) loop is twice as fast as that of the surrounding (white) loops. Interactions between the middle two layers can be enabled if we allow qubit swaps within the same loops as shown in \cref{fig:3D_qubit_connectivity}.}
    \label{fig:loop_different_size}
\end{figure}

We can also have different numbers of qubits and/or different cycling frequencies in different loops. In such a way, it is possible to construct 3D qubit structures beyond a stack of aligned layers, even without using intra-loop interaction. One such example is shown in  \cref{fig:loop_different_size}. If indeed we do also have intra-loop interactions available, then we can construct qubit structures that are yet more sophisticated (see e.g. \cref{fig:3D_qubit_connectivity} in Appendix).

\subsection{Alternative Qubit-array Pipeline Implementations}\label{sec:edge_interaction}

The shape of the loop we have constructed in \cref{fig:graph_to_shuttling} is merely one of the more simple implementation possibilities. Instead of a simple loop of the shape of an $n$-gon, we can have more complicated shapes like, e.g. a ``$8$''-shape with a crossing at the middle (i.e. instead of having the shape of a graph \emph{cycle}, we can have the shape of a graph \emph{circuit}).

\begin{figure}[htbp]
    \centering
    \includegraphics[width = 0.25\textwidth]{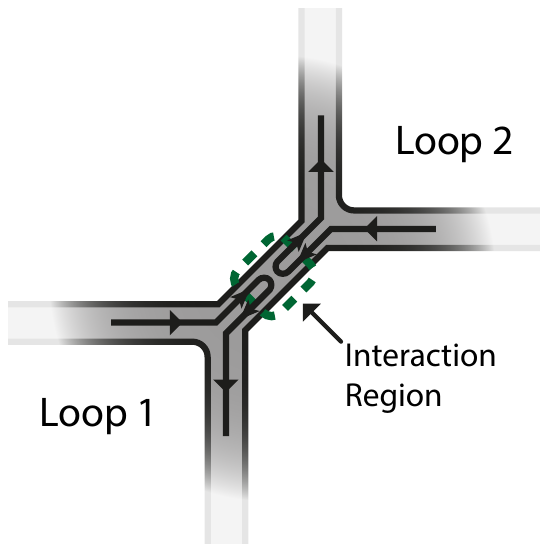}
    \caption{One possible way to implement the interactions between two loops using shuttling junctions.}
    \label{fig:corner_structure}
\end{figure}

\begin{figure}[htbp]
    \centering
    \subfloat[Partial edge overlaps.\label{fig:edge_interaction_partial}]{\includegraphics[width = 0.45\textwidth]{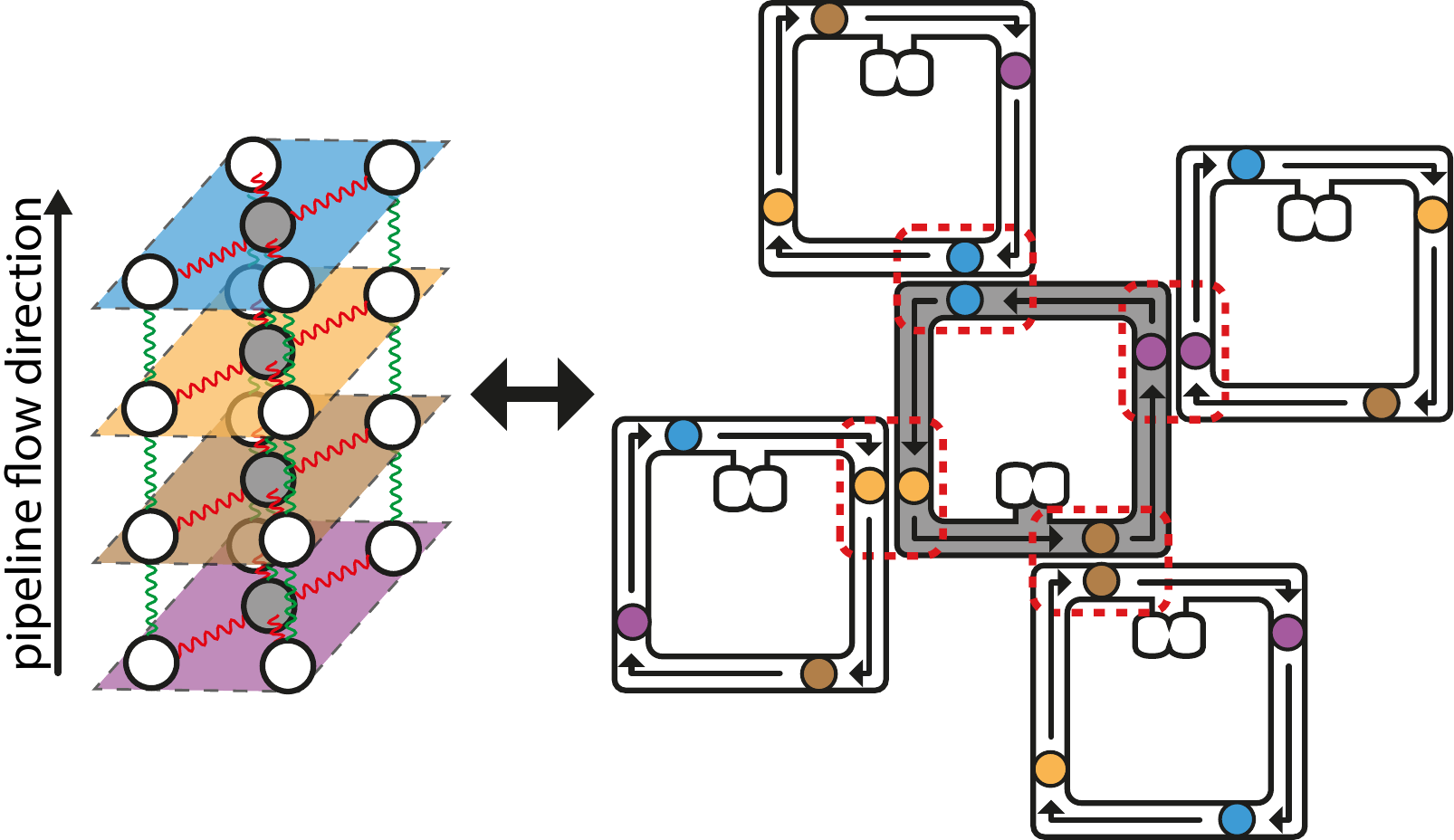}}\\
    \subfloat[Complete edge overlaps.\label{fig:edge_interaction}]{\includegraphics[width = 0.45\textwidth]{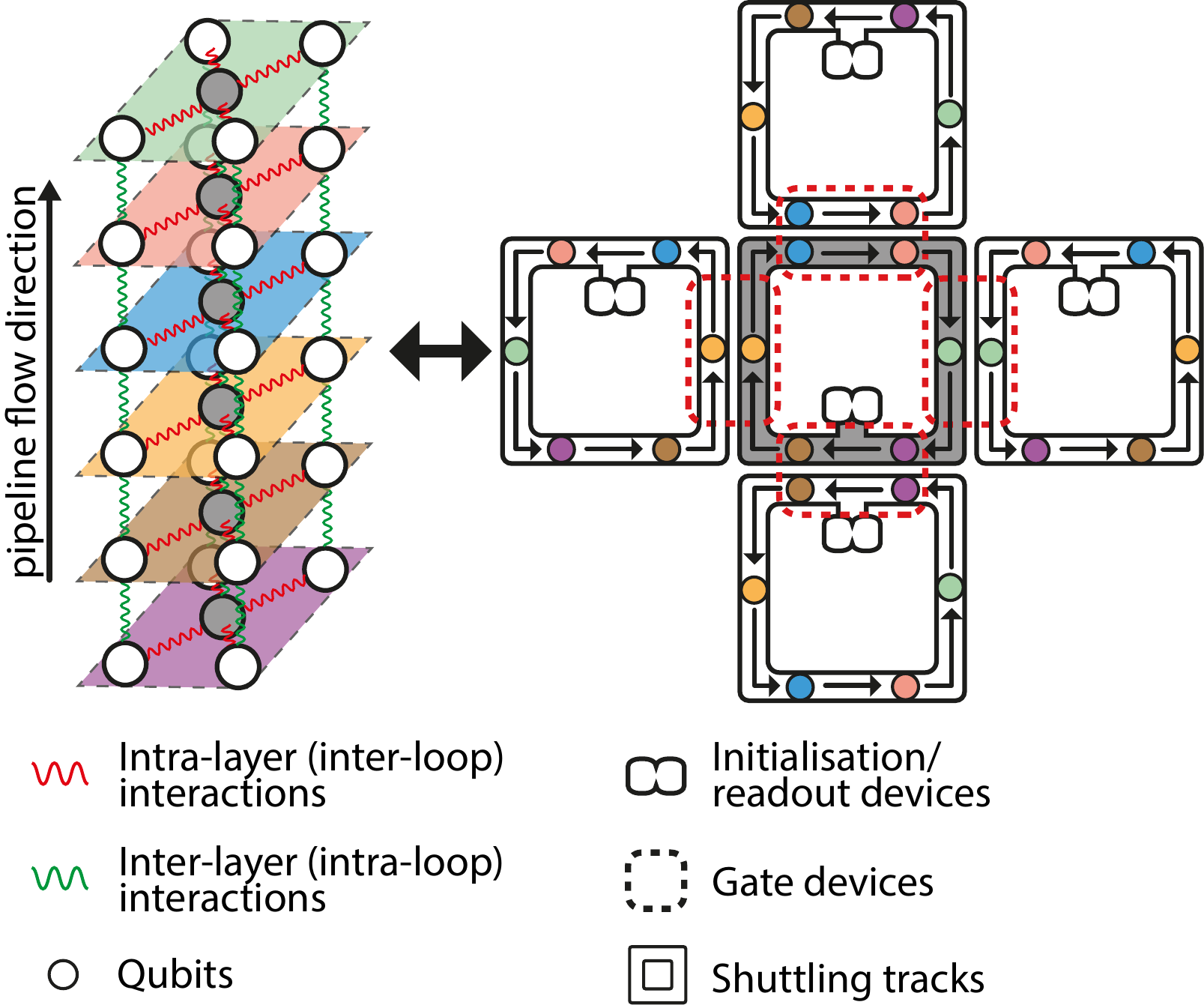}}
    \caption{Edge-interacting qubit-array pipelines. Intra-loop gate devices are not explicitly shown for visual clarity. We see that multiple qubit pairs can interact simultaneously along a given edge.}
\end{figure}

In \cref{fig:array_pipeline_both}, the corner interaction is illustrated as two separate tracks coming into proximity. In practice, such corner interaction can also be implemented using shutting junctions connecting the two loops as shown in \cref{fig:corner_structure}, utilising Y-junctions that have been extensively considered for trapped ion devices~\cite{bermudezAssessingProgressTrappedIon2017,gutierrezTransversalityLatticeSurgery2019,bermudezFaulttolerantProtectionNearterm2019}. It is also possible to extend the interaction regions between the loops beyond the corners by increasing the overlaps between the edges of different loops as shown in \cref{fig:edge_interaction_partial}. Pushing this to the extreme, we then have complete overlaps between the edges of different loops as shown in \cref{fig:edge_interaction}, which we will simply call the \emph{edge-interacting pipeline}. 

Here we see that the qubit interactions are now extending over the whole edge of the loop instead of simply at the corner. It is now possible to operate on multiple qubit pairs simultaneously along a given edge. Such an extended gate-operation node along the whole edge can be viewed as many consecutive small gate-operation steps, with each small gate-operation step only able to operate on one qubit at a time. If the gate step is the rate-limiting step in the corner-interacting pipeline, by decomposing each gate operation into the $m$ small gate steps in the edge-interacting pipeline, we can reduce the time required for the rate-limiting step by a factor of $m$. Of course, this assumes that the time needed for the gate operation for both the corner-interacting and edge-interacting pipelines are the same. In the edge-interacting pipeline, since we are carrying out shuttling and qubit interaction in parallel, there is a time saving on removing the shuttling time cost. The density of the loops in the loop array may also increase compared to the corner-interacting case. On the other hand, a qubit-qubit interaction realised while both qubits are transiting along the edge may not be possible in certain platforms, or may have lower fidelity than stationary qubit interaction at the corner. 
\section{Stacking 2D Topological Codes}\label{sec:QEC}

A class of QEC codes that are particularly relevant to physical implementation is \emph{2D (planar) topological codes}. These codes can naturally match to the geometric layouts of physical qubits to ensure \emph{local} stabiliser checks of \emph{constant weights}~\cite{bravyiQuantumCodesLattice1998,bombinTopologicalQuantumError2006}. When performing fault-tolerant quantum computation using 2D topological codes, we will have a set of 2D qubit arrays (storing different logical qubits) that need to be processed in the exact same way to perform stabiliser checks for quantum memory. This is a perfect setting for applying the qubit-array pipeline discussed in \cref{sec:array_pipeline}, which will \emph{increase the density of the logical qubits}. Furthermore, when we enable intra-loop operations (or equivalent extensions), then we will be able to perform transversal operations between the logical qubits for \emph{faster fault-tolerant quantum computation},  or indeed to realise 3D codes. 

\subsection{Pipelining Time Cost for Quantum Error Correction}\label{sec:QEC_pipe_time}
We will assume the entire pipeline of QEC consists of $D$ code cycles and ignore the initialisation of all qubits at the beginning and the measurement of all qubits at the end since the associated time costs are negligible compared to the main code cycles. The time required for \emph{one} logical qubit to run through \emph{one} code cycle is denoted as $\Tcycle$ and called the \emph{code cycle time}. Following \cref{eqn:steady_flow_pipeline_time}, the time needed to pipeline $k$ logical qubits through $D$ code cycles is simply:
\begin{equation}\label{eqn:time_surface_code}
    \begin{aligned}
        T_{\mathrm{pipe}}(D, k) &= D\Tcycle + (k-1) \tmax\\
        &= \left(D+(k-1)f\right)\Tcycle
    \end{aligned}
\end{equation}
with $f = \tmax/\Tcycle$ being the fraction of the code cycle time taken by the rate-limiting step. 

The depth of the logical circuit $D_{\mathrm{circ}}$ is the number of layers of logical operations we need to perform, which is usually much smaller than the number of code cycles $D$. For most of the interesting applications, we will usually find circuit depth scales more than linearly with the number of logical qubits needed $n$: $D_{\mathrm{circ}} \gtrsim n$. Now using the fact that the total number of logical qubits $n$ is larger than the number of logical qubits in each qubit-array pipeline $k$, and $f \leq 1$, we have $D \gg D_{\mathrm{circ}}  \gtrsim n > k > (k-1)f$. When applied to \cref{eqn:time_surface_code}, this implies that \emph{in many fault-tolerant applications, the pipelining time cost is negligible.}

\subsection{Surface Codes}\label{sec:surface_code}

\begin{figure*}[!htbp]
    \centering
    \parbox{\textwidth}{
        \parbox{0.3\textwidth}{
            \subfloat[The surface Code \label{fig:surface_code}]{\includegraphics[width = 0.3\textwidth]{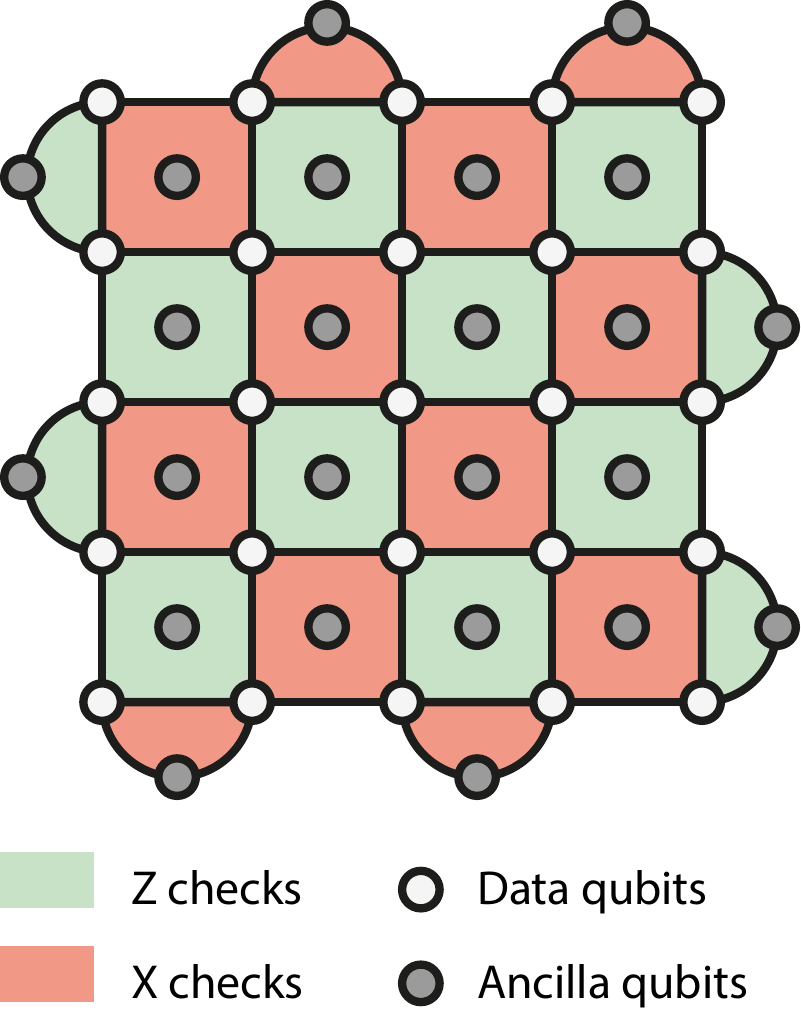}}
        }
        \hspace{4em}
        \parbox{0.27\textwidth}{
            \subfloat[X pairty check circuit]{\includegraphics[width = 0.27\textwidth]{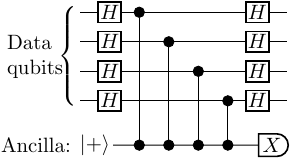}}
            \vspace{1em}
            \subfloat[Z pairty check circuit]{\includegraphics[width = 0.27\textwidth]{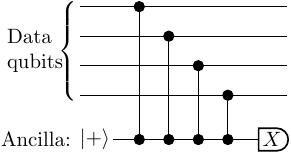}}
        }
    }
    \caption{The surface code and its parity check circuits.}
    \label{fig:parity_check}
\end{figure*}

Surface codes and colour codes are two of the most promising topological codes at the moment due to their 2D planar layout, high thresholds and the existence of efficient decoding algorithms~\cite{fowlerSurfaceCodesPractical2012,landahlFaulttolerantQuantumComputing2011}. Let us first use surface codes as an example. In surface codes we need to perform local $X$/$Z$ checks, which are represented using the red/green plaquettes in \cref{fig:surface_code}. For each $X$ check, we will use the circuit in \cref{fig:parity_check} to measure the $X$ parity of the data qubits located at the vertices of red plaquettes (semi-circles at the boundary), similarly for $Z$ checks.

\begin{figure*}
    \centering
    \includegraphics[width = 0.8\textwidth]{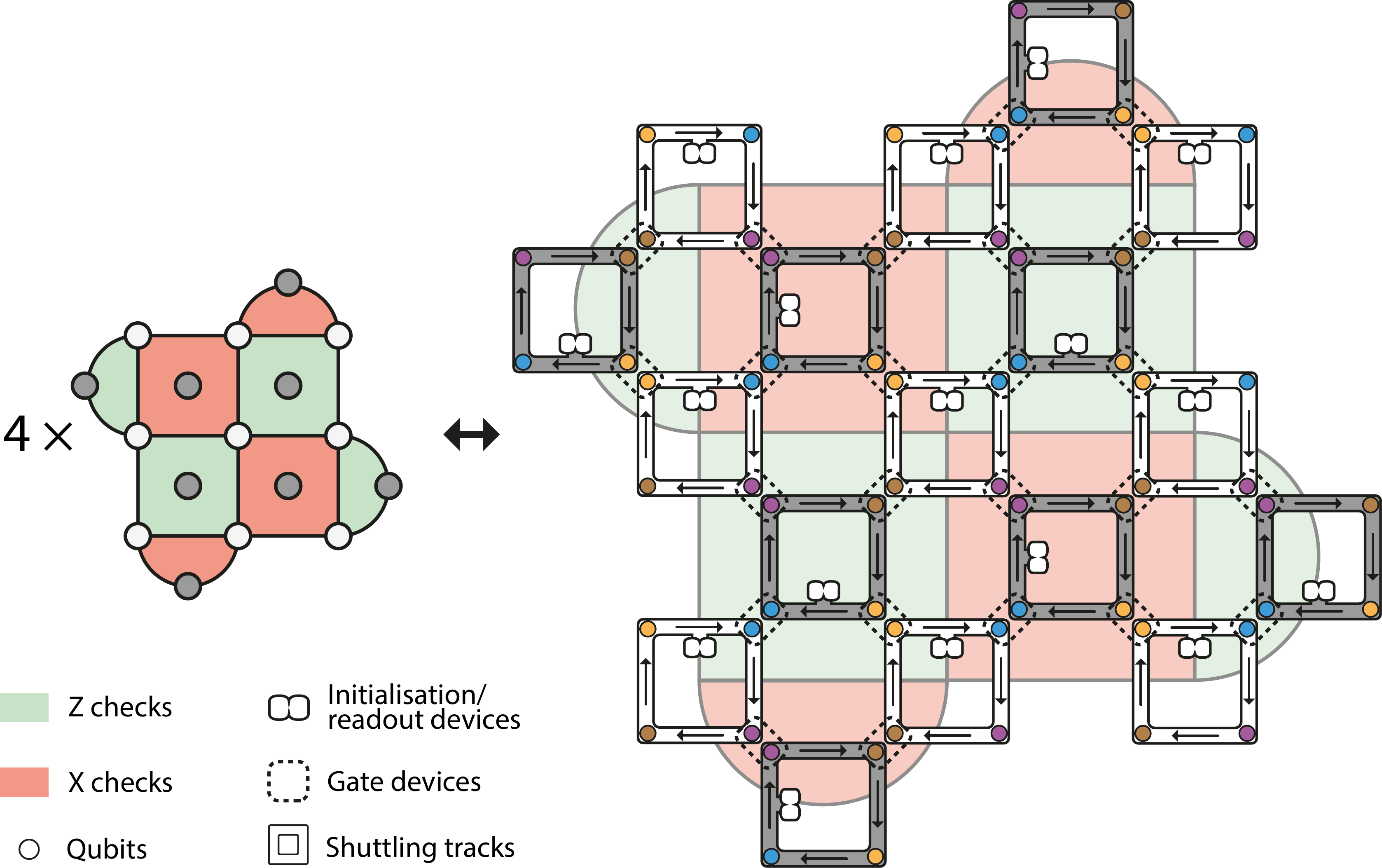}
    \caption{Pipelining architecture for surface codes. Intra-loop gate devices are not explicitly shown for visual clarity. The initialise/readout devices on the data qubit tracks would not be used during a normal stabiliser cycle. More specialised hardware design with heterogeneous data and ancilla loops is possible, but is not shown here for simplicity. Note that here all data and ancilla pipelines are flowing in the clockwise direction. It is also possible to run in a configuration that all data pipelines flow clockwise while all ancilla pipeline flow anti-clockwise.}
    \label{fig:surface_code_pipeline}
\end{figure*}

Each stabiliser check is a five-qubit process that can be implemented in the pipelining architecture in \cref{fig:array_pipeline}. There the data qubits are transformed into the surrounding (white) \emph{data loops} and the ancilla qubit is transformed into the central (grey) \emph{ancilla loop}. After a given ancilla qubit moves around the loop, it will be able to interact with every surrounding data qubits and carry out the parity check circuit in \cref{fig:parity_check}. 

The pipelining structure can be easily extended beyond the five-qubit check array to the whole surface code patch as shown in \cref{fig:surface_code_pipeline}. It is shown in \textcite{fowlerSurfaceCodesPractical2012} that $X$ and $Z$ checks in the surface code can be carried out in parallel if we perform each check in a zigzag pattern across the four data qubits. The way we perform each check in a circular manner in a loop in \cref{fig:array_pipeline} means that we can only perform $X$ and $Z$ checks in a staggered manner instead of in parallel. Of course, this is a feature of the round-shaped shuttling track rather than of pipelining, and it can be prevented by having a zigzag shuttling track instead. It is also worth noting that even without pipelining, we might want to stagger $X$ and $Z$ checks anyway since it can improve code performance by preventing error propagations and/or help with combating special kinds of errors~\cite{caiSiliconSurfaceCode2019,xuHighThresholdCodeModular2019,higgottSubsystemCodesHigh2021,strikisQuantumComputingScalable2021}. 

In a given stabiliser check, we need to be careful about ``hook'' errors~\cite{dennisTopologicalQuantumMemory2002} in which single-qubit errors on the ancilla qubits propagate and become weight-two errors on data qubits. To mitigate the damage due to hook errors, when performing checks around a loop, we need to start the $X$ checks at the position of the purple or orange dot in \cref{fig:surface_code_pipeline}, while the $Z$ checks need to start at the position of the blue or brown dot, such that the resultant weight-two errors do not align with the logical operators. This is reflected through the different orientation of the initialisation/measurement devices in $X$ and $Z$ ancilla in \cref{fig:surface_code_pipeline}.

\begin{figure*}[!htbp]
    \centering
    \includegraphics[width = 0.8\textwidth]{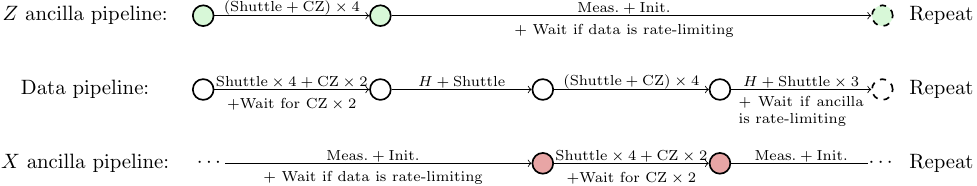}
    \caption{Pipeline workflow for a code cycle in surface codes. ``Shuttle'' here means shuttling along \emph{one edge} of the square shuttling loop. ``Wait'' here means buffering and it needs to be added into the pipeline for synchronisation depending on which pipeline is the rate-limiting pipeline. We have not included the initialisations of qubits before all code cycles and the measurement of all qubits after all code cycles.
    }
    \label{fig:code_cycle_pipeline}
\end{figure*}

We have constructed a pipeline for one surface code cycle taking into account the restrictions above as shown in \cref{fig:code_cycle_pipeline}, which is written as a combination of the data pipelines, $X$ ancilla pipelines and $Z$ ancilla pipelines. We will denote the time required for qubit measurements, initialisation, CZ gates, $H$ gates and shuttling through \emph{one full round} of the loop as $\tmeas$, $\tinit$, $\tcz$, $\tH$ and $\tsh$, respectively. The rate-limiting step time is given by
\begin{align*}
    \tmax = \max(\tmeas, \tinit, \tcz, \tH).
\end{align*}
Note that $\tsh$ contains many small shuttling steps and we have assumed that shuttling is not the rate-limiting step as mentioned before (this is merely to simplify our analysis and is not essential).

The code cycle time $\Tcycle$ will be obtained by looking at the data pipeline and the ancilla pipeline in one code cycle. In each code cycle, the data qubits will flow through three rounds of the loop with $8$ CZs and 2 $H$ applied along the way. Thus the circuit time needed for the data qubits for one code cycle is:
\begin{align}\label{eqn:data_cycle_1}
    \Tcirc^{\mathrm{data}} = 3\tsh + 8 \tcz + 2\tH.
\end{align}

On the other hand , the ancilla qubits (both $X$ and $Z$) will flow through one round of the loop with initialisation, $4$ CZs and measurement applied along the way. Thus the circuit time needed for the ancilla qubits for one code cycle is:
\begin{align}\label{eqn:anc_cycle_1}
    \Tcirc^{\mathrm{anc}} = \tsh + 4 \tcz + \tinit + \tmeas.
\end{align}

The code cycle time is simply determined by the rate-limiting pipeline
\begin{align}\label{eqn:cycle_time_surface}
    \Tcycle = \max(\Tcirc^{\mathrm{data}}, \Tcirc^{\mathrm{anc}})
\end{align}
with buffering added to the faster pipeline to achieve synchronisation. We can then use \cref{eqn:time_surface_code,eqn:cycle_time_surface} to obtain the time needed for pipelining $k$ surface code patches through $D$ cycles.  

For the data pipeline, the data qubits go around the loop three rounds in each code cycle, and the minimum cycling period is given by in last round in \cref{fig:code_cycle_pipeline} in which only one CZ gate and one $H$ gate are applied: $\Tloop^{\mathrm{min}} = \tsh + \tcz + \tH$. Hence, the maximum number of qubits we can fit into the pipeline is given by \cref{eqn:max_qubit_steady_flow}:
\begin{align}\label{eqn:max_qubit_surface}
    \Kloop = \frac{\Tloop^{\mathrm{min}}}{\tmax} + 1 = \frac{\tsh + \tcz + \tH}{\tmax} + 1.
\end{align}
For the ancilla pipeline, the ancilla qubit goes through one round of the loop without wrapping around, thus there is no danger of the front of the qubit stream colliding with the rear. 

When the rate-limiting step is measurement or initialisation, we can reduce the effective rate-limiting step time by adding more initialisation/measurement devices as discussed in \cref{sec:qubit_pipeline}. Note that in many platforms, one can perform non-destructive projective measurements (e.g. Pauli spin blockade measurement for semiconductor spin qubits in \cref{sec:silicon_code_cycle}), which will act as the initialisation step for the next code cycle and thus we have $\tinit = 0$. When the CZ or $H$ gate is the rate-limiting step, we might switch to a loop array with edge interactions to reduce the effective rate-limiting step time as discussed in \cref{sec:edge_interaction}. 

The pipeline we proposed in \cref{fig:code_cycle_pipeline} is simply one of the possible workflows. A simpler (but not necessarily efficient) stabiliser check scheme can be mimicking the conventional zigzag checking pattern~\cite{fowlerSurfaceCodesPractical2012} by flowing the qubits through two rounds of the loops and activating the gates at the appropriate moments. There are also other possibilities depending on the exact processing steps and the structure of the loops, which would be an interesting future direction to investigate. 

\subsection{Colour Codes}\label{sec:colour_code}
\begin{figure}[htbp]
    \centering
    \includegraphics[width = 0.35\textwidth]{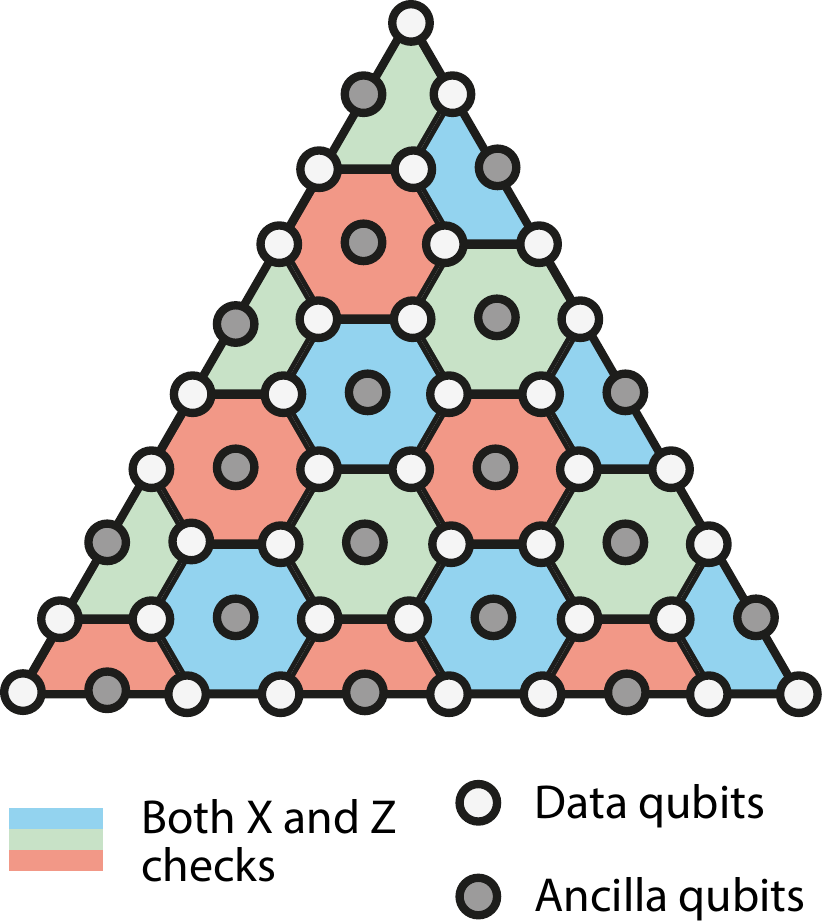}
    \caption{A logical qubit in the colour code.}
    \label{fig:colour_code}
\end{figure}

The structure of the colour code~\cite{bombinTopologicalQuantumDistillation2006} is shown in \cref{fig:colour_code}. In colour codes, instead of performing $X$ and $Z$ checks separately on different plaquettes like in surface codes, we need to perform both $X$ and $Z$ checks for all plaquettes. Such a structure is why we can implement transversal $H$ and $S$ logical gates in colour codes. The local checks in colour codes are weight-$6$ (weight-$4$ at the boundary), and thus can be carried out using a parity check circuit similar to \cref{fig:parity_check} but with $6$ data qubits. As the parity check circuit involves more qubits and gates, the error threshold of colour codes is usually lower than that of the surface codes. 

\begin{figure}[htbp]
    \centering
    \includegraphics[width = 0.49\textwidth]{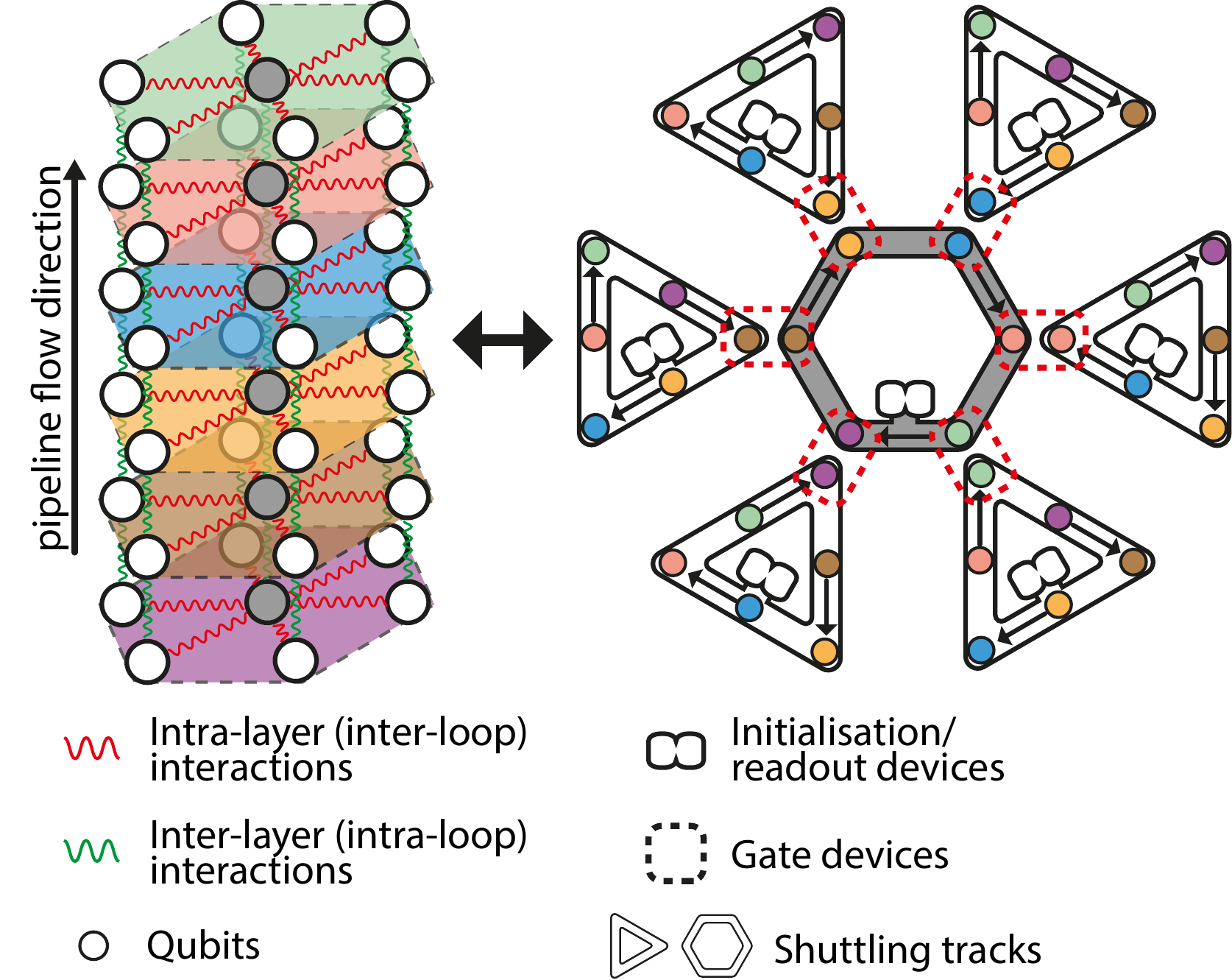}
    \caption{Pipelining architecture for colour code parity checks with $6$ qubits in the pipeline. Intra-loop gate devices are not explicitly shown for visual clarity. Here we effectively have $6$ layers of parity check units stacking on top of each other. }
    \label{fig:colour_code_pipeline}
\end{figure}

Colour codes can be transformed into a pipelining architecture following \cref{sec:pipeline_array}. Each stabiliser check will become the structure in \cref{fig:colour_code_pipeline} in which the data qubits are transformed into the surrounding (white) data loops and the ancilla qubit is transformed into the central (grey) ancilla loop. Since we have weight-$6$ checks (weight-$4$ at the boundary) for colour codes, we see that the ancilla loops are hexagons. The data loops are triangles since each data qubit is connected to $3$ ancilla qubits. Note that performing checks using a single ancilla in triangular colour codes is prone to hook errors just as in surface codes. This can be avoided by adding extra ancilla qubits called \emph{flag qubits} to detect these hook errors~\cite{chamberlandTriangularColorCodes2020}, and the new qubit layouts with the additional flag qubits can still be pipelined using the method discussed in \cref{sec:pipeline_array}. 

In surface codes, each ancilla loop is either responsible for the $X$ checks or the $Z$ checks, thus the operations in the $X$ checks and the $Z$ checks can be carried out in parallel. On the other hand, in colour codes, every ancilla loop is responsible for both the $X$ and $Z$ checks, thus in the simplest case, we can only carry out the $X$ and $Z$ checks in a strictly sequential manner. A possible code cycle pipeline for the colour codes is discussed in \cref{sec:colour_code_pipe}.

\subsection{Transversal Logical CNOT}\label{sec:transversal_cnot}
Most of the state-of-the-art topological codes like surface codes and colour codes are Calderbank-Shor-Steane (CSS) codes for which the stabiliser checks are either $X$ parity checks or $Z$ parity checks, i.e. there are no checks consisting of a mixture of $X$ and $Z$ operators. For CSS codes, the logical CNOT can be implemented transversally, i.e. it can be implemented by performing physical CNOT between the corresponding physical qubits in the two logical qubits~\cite{gottesmanStabilizerCodesQuantum1997}. Such a transversal logical CNOT is usually not considered in practical implementations since it is challenging to link the corresponding physical qubits within two topological code patches in a 2D architecture. Therefore instead, people turn to defect-based methods~\cite{raussendorfFaultTolerantQuantumComputation2007,raussendorfTopologicalFaulttoleranceCluster2007} or lattice surgeries~\cite{horsmanSurfaceCodeQuantum2012,landahlQuantumComputingColorcode2014}. However, these methods are based on \emph{code deformation}, which will require $\order{d}$ code cycles to be performed before the CNOT can be successfully completed, where $d$ is the \emph{distance of the code}. In contrast, transversal CNOTs and more generally any transversal gates can be directly implemented in the looped pipeline paradigm; one means is simply to modify a single code cycle as discussed in Sec.~\ref{sec:array_pipeline}. If this modified cycle introduces only a typical level of noise to the device, then no additional code cycles are required -- this is the assumption we make presently in our resource estimations. However, even if higher-than-normal noise is introduced by the modified cycle, we would only require a small fixed ($d$-independent) number of additional cycles~\cite{beverlandCostUniversalityComparative2021}, which would not significantly alter our conclusions.

\begin{table*}[tbhp!]
    \centering
    \begin{threeparttable}
        \begingroup
        \renewcommand{\arraystretch}{2}
        \setlength{\tabcolsep}{1.3em}
        \begin{tabular}{c | cccccc}\toprule
            &$H$&$S$&CNOT&$\ket{T}$ Init.&$\ket{0}$/$\ket{+}$ Init.&$X$/$Z$ Meas.\\ \hline
            Surface Code&$3d$~\cite{fowlerLowOverheadQuantum2019,litinskiGameSurfaceCodes2019}\tnote{a}&$d$~\cite{brownPokingHolesCutting2017,fowlerLowOverheadQuantum2019}\tnote{b}&\multirow{2}{*}{\parbox{25ex}{$2d$~\cite{fowlerLowOverheadQuantum2019} (lattice surgery)\\or $0$ (transversal)}} &\multirow{2}{*}{6~\cite{landahlQuantumComputingColorcode2014}} &\multirow{2}{*}{$0$~\cite{litinskiGameSurfaceCodes2019}\tnote{c}}&\multirow{2}{*}{0}\\
            Colour Code&0 &0 &&&&\\
            \botrule
        \end{tabular}
        \endgroup
        \begin{tablenotes}\footnotesize
            \item[a] $3d$ code cycles are needed for rotating a surface code in place (with help of ancilla patches). 
            \item[b] Code distance is halved during the gate.
            \item[c] We need $\order{d}$ code cycles for projecting the incoming state into the code space. However, this can be done simultaneously with all the other operations after initialisation. Hence, as long as there are $\order{d}$ code cycles between the qubit initialisation and measurement,  the initialisation time cost can be taken to be $0$, otherwise we need to take it to be $\order{d}$.
        \end{tablenotes}
        \caption{Number of code cycles needed for different logical operations. The correspondence is not exact due to reasons including the following: 1. we have assumed that $d$ code cycles would be needed for fault-tolerant syndrome extraction, which might change with e.g. the measurement error rate; 2. we have assumed transversal operations have low enough noise such that the usual frequency of code cycles for memories is enough for correcting their errors. However, this table will give us a good order-of-magnitude estimate for the operational speed of a given scheme.}
        \label{tab:op_time_cost}
    \end{threeparttable}
\end{table*}

The numbers of code cycles required for the different logical operations in the universal set are summarised in \cref{tab:op_time_cost}. We see that CNOTs, if implemented via lattice surgeries, can be a big part of the time cost for both surface codes and colour codes. For colour codes, CNOT via lattice surgeries would be the \emph{only} operation requiring $\order{d}$ code cycles and thus optimising it via pipelining can potentially speed up the computation by a large factor. For the surface code, CNOT is one of the slowest steps, but there are other steps requiring $\order{d}$ code cycles like the Hadamard gate $H$ and the phase gate $S$. One must also note that lattice surgeries will also require a larger qubit overhead (more ancilla patches) than transversal operations.

\subsection{Multiple code stacks}\label{sec:multi_stack}

Due to the qubit collision constraint outlined in \cref{sec:qubit_pipeline}, there is a limit to the number of logical qubits (2D code layers) that we can fit into the same pipeline (stack). When the number of logical qubits exceeds the capacity of a single stack we will employ multiple stacks forming an array, as we now discuss.

For a conventional 2D architecture, we might partition the device into multiple zones, or \emph{ patches}, each of which can store a separate logical qubit. In our looped pipeline architecture, we may also employ patches whose lateral dimensions are sufficient to represent one logical qubit; however those dimensions are measured now in terms of a certain number of complete loops. Therefore in each patch we can in fact store \emph{a stack} of $k$ logical qubits (layers), where $k$ is also the number of physical qubits within each loop. Denoting by $A$ the number of distinct patches, each now seen as supporting a stack, we have $kA$ logical qubits in total. Assuming that $k$ is fixed for a given hardware technology and application, then we scale the device by increasing $A$, i.e. adding to the number of hardware patches -- this is the spatial overhead of a given architecture.

Within the same stack, fast transversal CNOTs between logical qubits are available, while in between different stacks, logical CNOTs need to be carried out through e.g. lattice surgery. 
Within the same stack, swapping logical qubits can be carried out by transversally swapping the positions of the corresponding qubits in each loop, and thus can be implemented using only shuttling operations which are likely to be much faster than other steps in the pipeline. Such fast and reliable swap operations within the stack, along with the constant height of the stack, mean that we effectively have all-to-all connectivity within the stack. On the other hand, the connectivity between logical qubits in different stacks is dependent on the layout of the data and ancilla patches. If we have a chequerboard pattern of data and ancilla patches, then we will just have 2D nearest neighbour connectivity between data stacks via lattice surgery. 

\begin{figure}[htbp]
    \centering
    \includegraphics[width = 0.49\textwidth]{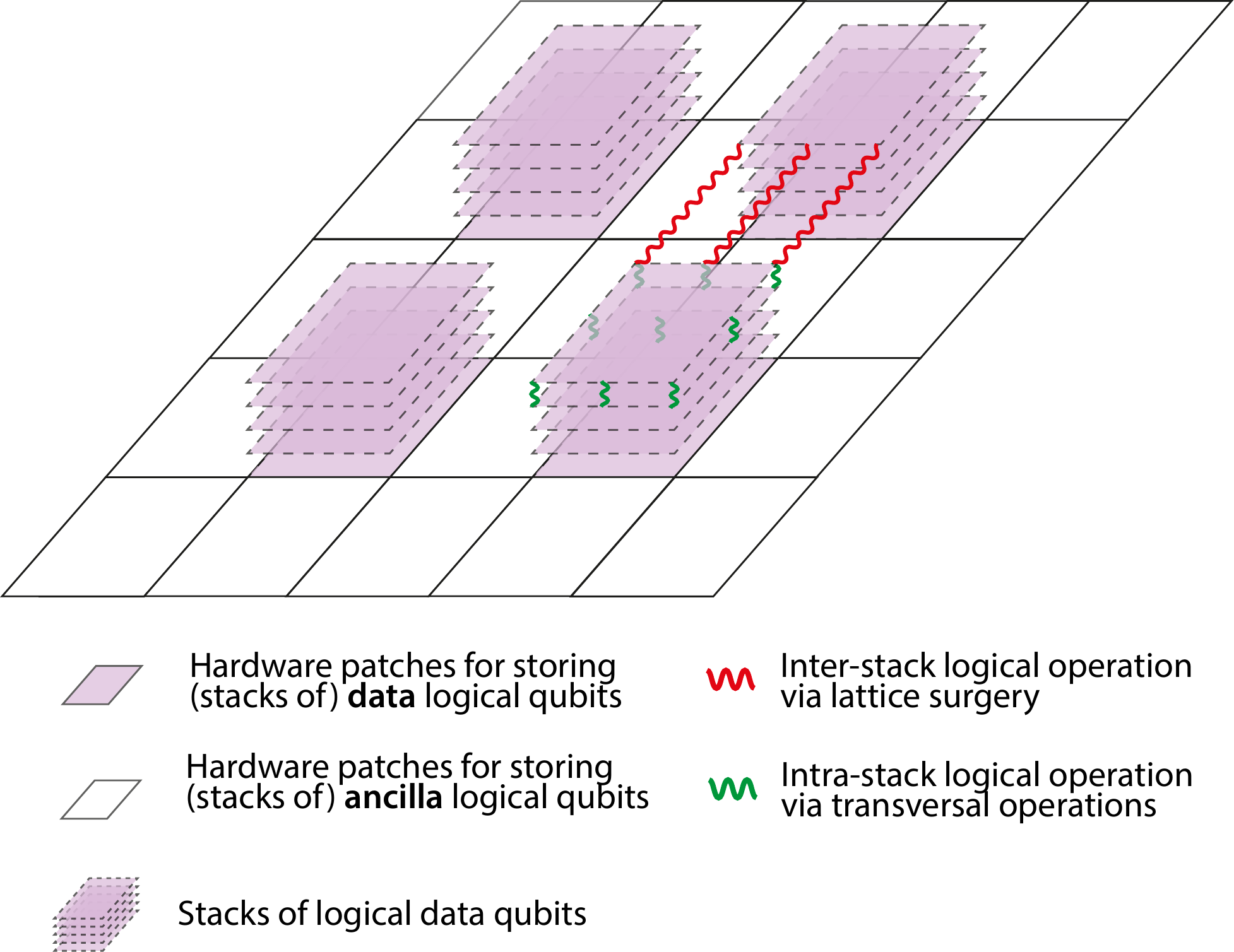}
    \caption{A possible layout for storing logical information in multiple stacks. Here each square is an array of loops that can store a stack of logical qubits. The ancilla logical qubits (stacks) do not need to be always present and thus are not explicitly shown here.}
    \label{fig:code_stacks}
\end{figure}

\begin{figure}[thbp]
    \centering
    \includegraphics[width = 0.49\textwidth]{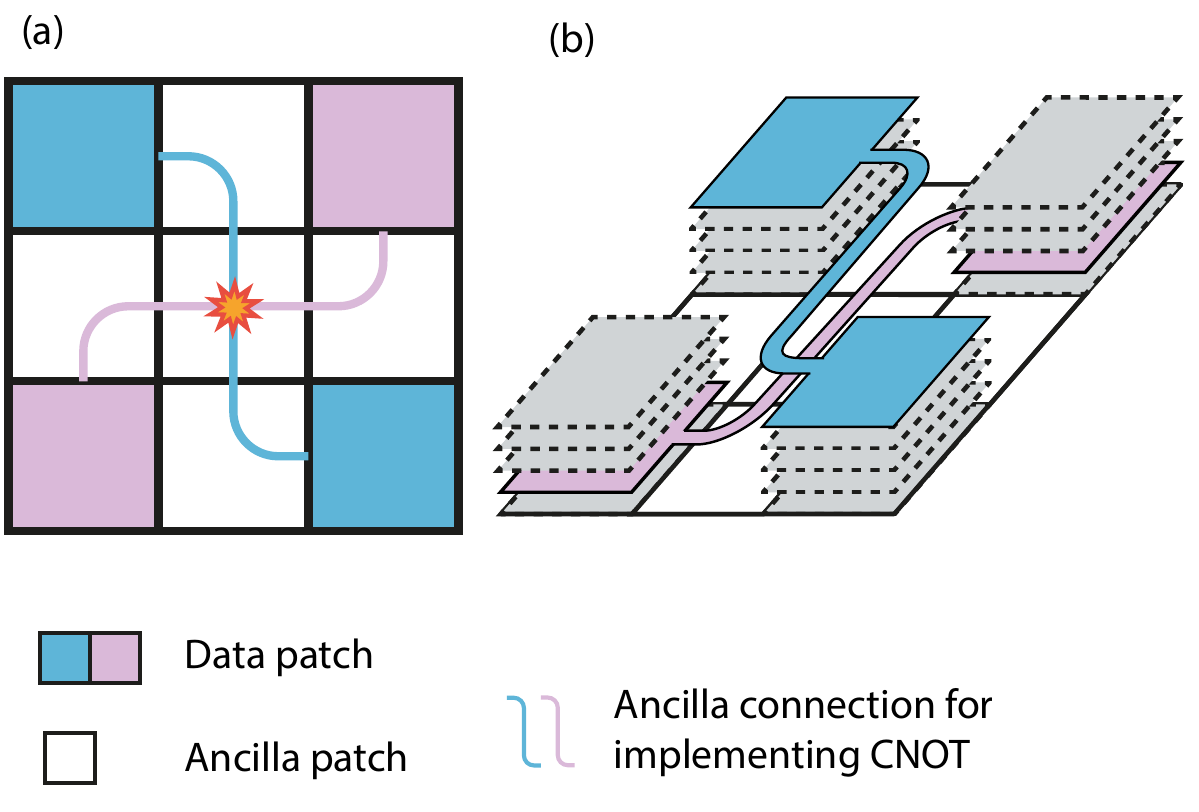}
    \caption{An illustration of how CNOTs can fail to be parallelised when using lattice surgery (panel (a)), when using a simple 2D architecture. Here we are trying to perform CNOTs using lattice surgery between the two blue qubits and the two purple qubits. They cannot be carried out in parallel because they both need to make use of the centre ancilla patch. In panel (b), we see this is possible with the virtual stack.}
    \label{fig:cnot_crossing}
\end{figure}

The simple solution mentioned above can be improved at the cost of using some of our resource as ancillas. A suitable layout is shown in \cref{fig:code_stacks} where we have $3$ ancilla patches per data patch; we then ensure direct connectivity between any two data stacks via lattice surgery. This can be achieved by using the ancilla space (white regions of  \cref{fig:code_stacks}) around the two data stacks to create a Bell pair through which the long-range CNOT can be performed~\cite{litinskiBraidingMajoranaTracking2017,beverlandSurfaceCodeCompilation2022}. Such a layout can of course be used without pipelining -- i.e. using single-layer data patches instead of stacks -- but then there is a limitation to the parallelisability of the lattice-surgery CNOT gates since the connecting ancilla patches cannot cross each other~\cite{beverlandSurfaceCodeCompilation2022}, see \cref{fig:cnot_crossing}(a). By adopting pipelining, lattice surgeries can be carried out in parallel in different layers of the stacks, and thus having $k$ layers in each stack will increase the parallelisability of the CNOT gates by $k$-fold (on top of enabling transversal CNOTs within each stack). A loose analogy is the use of flyovers to permit highways to cross one-another without intersecting.

While the advantages of introducing additional ancilla space are attractive, we stress that this is not essential for the pipelining paradigm: Pipelining $k$ layers in each stack immediately leads to a $k$-fold reduction in the spatial overhead compared to the other shuttling-based architectures. Indeed, the improved logical qubit connectivity in the pipelining architecture also means that a chequerboard arrangement of the data and ancilla stacks could be sufficient which can lead to a further $2$ times reduction in the spatial overhead compared to the arrangement in \cref{fig:code_stacks}. 

The impact of the pipelining paradigm on execution speed depends on the configuration details, but two limits are trivial: In the extreme case in which the circuit consists of only intra-stack CNOTs, we can achieve $\order{d}$ reduction in the time overhead by using transversal CNOTs instead of lattice surgery. On the other hand, for circuits consisting of only inter-stack CNOTs, pipelining enables the parallelisation of CNOTs in $k$ different layers, which can potentially reduce the CNOT depth by $k$ times, achieving a $k$ times reduction in the time overhead. 

Beyond these simple observations, it is apparent that savings are possible by tailoring the specific circuit we want to implement towards the pipelining architecture. In the following section, we look at one of the most important subroutines for fault-tolerant computation: magic state distillation. For a given well-studied approach, we explore the exact space-time overhead saving achievable through pipelining.

\section{Application to magic state distillation}\label{sec:magic_state_distill}
\begin{figure}[htbp]
    \centering
    \includegraphics[width = 0.3\textwidth]{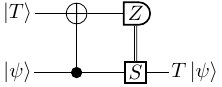}
    \caption{Gate teleportation circuit for implementing $T$ gate. The $S$ gate ($S = e^{-i\frac{\pi}{4}Z}$) is only applied when we obtain the $1$ outcome from the $Z$ measurement. If we change the correction to applying an $S^\dagger$ gate when we obtain the $0$ outcome from the $Z$ measurement, we would be implementing $T^\dagger \ket{\psi}$ instead.}
    \label{fig:T_tele}
\end{figure}

For full fault-tolerant quantum computation we need to perform logical gates beyond Clifford gates. One such non-Clifford gate, suitable for completing the universal set of operations, is the $T$ gate which is a $\frac{\pi}{4}$ rotation around the $Z$ axis in the Bloch sphere: $T = e^{-i\frac{\pi}{8}Z}$. It can be implemented using Clifford gates and a supply of $T$ states: $\ket{T}= T\ket{+}$ through gate teleportation shown in \cref{fig:T_tele}.

The circuit in \cref{fig:T_tele} is a logical circuit consisting of only logical Clifford gates. Hence, provided we can implement all the Clifford gates fault-tolerantly, the problem of implementing fault-tolerant $T$ gates becomes the problem of fault-tolerant preparation of logical $\ket{T}$. This is usually done through \emph{magic state distillation}, which is a process that consumes multiple noisy logical $\ket{T}$ and outputs fewer logical $\ket{T}$ of higher fidelity using logical Clifford operations and post-selection. The standard distillation protocol consists of the concatenation of two QEC codes. One is the \emph{base code} that we use to ensure the fault-tolerant implementation of Clifford gates, e.g. surface codes and colour codes as we described above. The other one is the \emph{distillation code} which has transversal $T$ gates.

\begin{figure}[htbp]
    \centering
    \includegraphics[width = 0.45\textwidth]{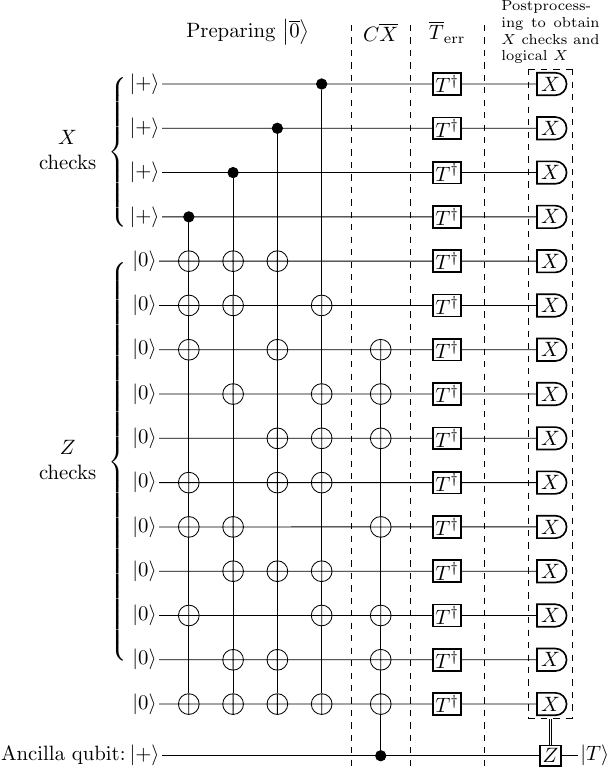}
    \caption{Circuit to perform $15$-to-$1$ magic state distillation from Ref.~\cite{fowlerSurfaceCodesPractical2012}. It starts with preparing the logical state of the distillation code $\ket{\overline{0}}$, then we prepare a Bell state between this logical qubit and the ancilla qubit $\propto \ket{\overline{0}0} + \ket{\overline{1}1} = \ket{\bar{+}+} + \ket{\bar{-}-}$. After that we implement the transversal $T$ gates by consuming noisy $\ket{T}$ using the circuit in \cref{fig:T_tele}. Note this requires an additional qubit at each location marked $T^\dagger$ in the figure, 15 in total, implying 31  qubits to execute the circuit. All qubits above are encoded in the base code (none are simply `physical' qubits). At the end, we perform local $X$ measurements to obtain the $X$ stabiliser checks and logical $X$ measurement of the distillation code. Any circuit runs with failed $X$ stabiliser checks are discarded. }
    \label{fig:fowler_circ}
\end{figure}

\begin{figure*}[htbp]
    \centering
    \includegraphics[width = 0.95\textwidth]{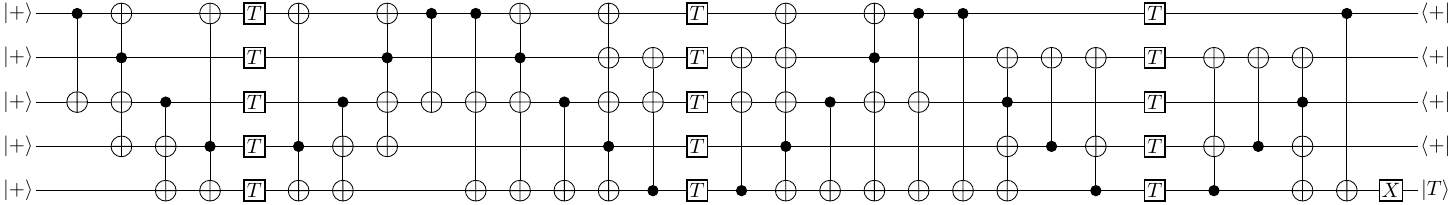}
    \caption{Circuit to perform $15$-to-$1$ magic state distillation with fewer qubits but a deeper circuit from Ref.~\cite{haahCodesProtocolsDistilling2018}. In this case, the $15$ noisy $\ket{T}$ are not consumed in parallel, but in 3 rounds. i.e. we only need to consume $5$ $\ket{T}$ at one time. Thus we can use $5$ qubits for the distillation code and $5$ qubits for generating the noisy $\ket{T}$. For the measurement at the end, we are only post-selecting the circuit runs that measure $\ket{+}$ for the first four qubits. The circuit here can be further optimised by removing some of the CNOT gates at the end that acts trivially on the post-selected states of the first four qubits $\ket{+}^{\otimes 4}$, and by recompiling the CNOT gates in between different rounds of $T$ gates. Note that all qubits above are encoded in the base code.}
    \label{fig:haah_circ}
\end{figure*}

One of the most widely studied distillation codes is the $15$-qubit Reed-Muller Code~\cite{bravyiUniversalQuantumComputation2005}, which is the smallest 3D colour code. Each execution of the distillation protocol using this code will consume $15$ noisy $\ket{T}$ and output a single $\ket{T}$ with higher fidelity if all checks are passed, thus it is called a $15$-to-$1$ distillation scheme. There are two different implementations of the distillation scheme, one requires more qubits and a shallower circuit as shown in \cref{fig:fowler_circ}~\cite{fowlerSurfaceCodesPractical2012}, while the other requires fewer qubits and a deeper circuit as shown in \cref{fig:haah_circ}~\cite{haahCodesProtocolsDistilling2018}. 

\subsection{The Space-time Overhead for Different Pipelining Schemes}
To implement the $15$-to-$1$ distillation scheme for colour codes and surface codes, there are many different possible schemes which include instances on the different extremes of the pipelining spectrum. As mentioned before, pipelining 2D codes can be essentially viewed as putting the 2D codes into stacks. Different pipelining schemes simply means different number of logical qubits (layers) in each stack. Here we will focus on three different pipelining schemes tailored to the $15$-to-$1$ distillation scheme.

\paragraph{One layer (no pipelining)} This means only $k = 1$ logical qubit in each stack (and thus $1$ qubit in each loop). We will carry out the distillation process using lattice surgeries, and thus the CNOTs will require $\order{d}$ code cycles (so constituting one of the time bottlenecks). In this case, it is natural to use the circuit in \cref{fig:fowler_circ} which requires fewer rounds of CNOTs than \cref{fig:haah_circ}. To implement the distillation circuit in \cref{fig:fowler_circ}, we need to store $31$ logical qubits in $31$ separate regions; for each of which we will use the term `stack' although this is the trivial $k=1$ layer limit. The resources needed for the ancilla logical qubits for implementing lattice surgeries will be of a similar order. Hence, in total we will need $A \sim 50$ `stacks'. 

\paragraph{Ten layers} We will put all the logical data qubits into one single stack ($A = 1$). In this way, all the CNOTs can be implemented transversally, and thus the circuit depth due to CNOT will not contribute much to the time cost. On the other hand, there is also a limit on the number of qubits we can fit in each stack following \cref{eqn:max_qubit_on_loop}. Hence, it is natural to use the circuit in \cref{fig:haah_circ}  which requires fewer qubits than \cref{fig:fowler_circ}. We need $5$ qubits for the distillation code and $5$ qubits for storing the noisy $T$ states, thus accounting for a total of $k =10$ logical qubits (layers) in the stack.

\paragraph{Five layers} We put the qubits for the noisy $T$ states in one stack and the qubits for the distillation code in another. Using the circuit in \cref{fig:haah_circ}, we will have $k = 5$ logical qubits in each stack. We also need another ancilla stack for the lattice surgeries between the two data stacks. Hence, in total we need $3$ stacks of logical qubits ($A = 3$).

By stepping through the distillation circuits in \cref{fig:fowler_circ,fig:haah_circ} and following the time costs of the different logical gates in \cref{tab:op_time_cost}, we can derive the number of code cycles $D$ required for the magic state distillation circuit for the different pipelining schemes. We specify the details in \cref{sec:time_overhead}, and summarise the results here in \cref{tab:msd_saving}. There we can see a decrease of $D$ as we increase $k$ due to the increased availability of transversal CNOTs. The code cycle time of the $k$-layer pipelining scheme is denoted as $\Tcycle^{(k)}$, which can be multiplied with the corresponding $D$ to obtain the full time overhead of the given pipelining scheme. Combining with the space overheads ($A$) discussed above, we can obtain the total space-time overhead ($AD\Tcycle^{(k)}$) for different pipelining schemes for implementing magic state distillation as summarised in \cref{tab:msd_saving}. There we see that if all code cycle times are similar $\Tcycle^{(1)} \simeq \Tcycle^{(5)} \simeq \Tcycle^{(10)}$, which is achievable for silicon platforms as we will discuss in \cref{sec:silicon_implementation}, then we can achieve substantial reductions in space-time overhead: by one order of magnitude for the $5$-layer stack and by two orders for the $10$-layer approach.

\begin{table}[tbhp!]
    \centering
    \renewcommand{\arraystretch}{2}
    \setlength{\tabcolsep}{0.2ex}
    \subfloat[Colour Codes]{
        \begin{tabular}{cccc}
            \toprule
            \heading{17ex}{$k$: $\#$ layers \\($\#$ qubits$/$loop)}& \heading{15ex}{$A$: $\#$ stacks\\ ($\#$ loop arrays)}&\heading{12ex}{$D$: $\#$ code cycles}& \heading{14ex}{Space-time overhead}\\
            \hline
            $1$&$50$&$12d$&$600d\Tcycle^{(1)}$\\
            $5$&$3$&$6d$&$ 18d\Tcycle^{(5)}$\\
            $10$&$1$&$d$&$d\Tcycle^{(10)}$\\
            \botrule
        \end{tabular}
    }\\
    \subfloat[Surface Codes]{
        \begin{tabular}{cccc}
            \toprule
            \heading{17ex}{$k$: $\#$ layers \\($\#$ qubits$/$loop)}& \heading{15ex}{$A$: $\#$ stacks\\ ($\#$ loop arrays)}&\heading{12ex}{$D$: $\#$ code cycles}& \heading{14ex}{Space-time overhead}\\
            \hline
            $1$&$50$&$13d$&$650d\Tcycle^{(1)}$\\
            $5$&$3$&$9d$&$27d\Tcycle^{(5)}$\\
            $10$&$1$&$3d$&$3d\Tcycle^{(10)}$\\
            \botrule
        \end{tabular}
    }
    \caption{A summary of the space-time overhead of the various pipelining schemes for the magic state distillation circuit. Here $\Tcycle^{(k)}$ is the code cycle time for the $k$-layer scheme for the given code. }
    \label{tab:msd_saving}
\end{table}

\subsection{Implication for surface code fault-tolerant computation}
In \cref{sec:full_FT}, we provide an analysis of the space-time saving achievable beyond just the magic state distillation stage. As discussed there, since magic state distillation is often the bottleneck of fault-tolerant computation~\cite{babbushEncodingElectronicSpectra2018,fowlerLowOverheadQuantum2019,litinskiGameSurfaceCodes2019}, the factor of time savings achieved by the magic state distillation circuit in \cref{tab:msd_saving} can be largely translated into the same time-saving factor for the overall fault-tolerant computation (barring some subtleties related to multi-round magic state distillation). The space saving factor is at least $k$ when we fit $k$ qubits into each loop ($k$ layers in each stack), and can be more if we optimise the circuit implemented to take advantage of the increased connectivity and faster CNOTs in the pipelining architecture, as we have done for the magic state distillation circuit. The final space-time saving achieved is of course highly dependent on the circuit we try to run and the implementation details. Nevertheless, our estimates in \cref{sec:full_FT} suggest that we should expect to achieve one-to-two orders of magnitude space-time saving for the {\em entire} fault-tolerant computation process using pipelining, just as we have achieved for magic state distillation specifically.

\section{Pipelining Surface Codes using Semiconductor Spin Qubits}\label{sec:silicon_implementation}
\subsection{Implementation of a Code cycle}\label{sec:silicon_code_cycle}
As discussed in \cref{sec:surface_code}, in a surface code cycle we want to perform one round of $X$ checks and $Z$ checks using the parity check circuits in \cref{fig:parity_check}. There are a number of specific implementation choices that are possible in a platform where qubits are embodied by semiconductor spins. In this Section we summarise one natural set of choices, so that threshold calculations can be presented in the next Section. Needless to say, multiple other options exist and we refer the reader to reviews such as Ref.~\cite{burkardSemiconductorSpinQubits2021}

\paragraph{Ancilla initialisation and measurements in $X$ basis.} One of the spins in a singlet/triplet pair is used as the effective ancilla qubit to flow around the loop and interact with the data qubits, while the other spin will be stored at the initialisation/readout node to act as the reference qubit for readout~\cite{jonesLogicalQubitLinear2018,caiSiliconSurfaceCode2019}. The singlet/triplet pair can be read out using Pauli spin blockade, which can achieve a measurement time of $\tmeas \approx 1\ \si{\us}$~\cite{schaalFastGateBasedReadout2020,jonesSpinBlockadeSpectroscopyMathrmSi2019,connorsRapidHighFidelitySpinState2020}. This is a projective non-destructive measurement that allows us to use the measured ancilla qubits directly in the next code cycle without initialisation, thus we have $\tinit = 0$. We note that the field is evolving rapidly and faster measurement time may well be possible without compromising fidelity. However it is instructive to take the time $1\mu$s and see the consequences.

\paragraph{Single-qubit gate.} We can implement single-qubit gates using electric dipole spin resonance (EDSR), which can achieve a Hadamard ($\pi/2$ $Y$ rotations) gate time of $\tH \approx 25\  \si{\ns}$~\cite{yonedaQuantumdotSpinQubit2018} and similarly for $\pi/2$ $Z$ rotations. We presently discuss a variant employing ESR.

\paragraph{Two-qubit gates.} A number of high-fidelity two-qubit gates have now been experimentally demonstrated through controlling the exchange interaction~\cite{millsTwoqubitSiliconQuantum2022,xueQuantumLogicSpin2022,noiriFastUniversalQuantum2022}. The specific physical operation may be  $\sqrt{\textsc{swap}}$ or a controlled-rotation, but generally we can assume that it is possible to implement (directly or as a composite) a CZ gate~\cite{lossQuantumComputationQuantum1998} with a gate time of $\tcz\sim 100\ \si{\ns}$.

\paragraph{Shuttling.}  In addition to the standard gate set above for the parity check circuit, we also need to include the shuttling operation in order to implement the pipelining architecture. In semiconductor spin qubits, each step of shuttling can be carried out by tipping the electrical potential of an occupied quantum dot to transfer the spin into an adjacent quantum dot adiabatically. Various modes such as `bucket brigade'~\cite{millsShuttlingSingleCharge2019} and `conveyor'~\cite{seidlerConveyormodeSingleelectronShuttling2021} have been explored, and high fidelity shuttling steps on the timescale of nanoseconds is expected to be possible~\cite{buonacorsiSimulatedCoherentElectron2020,krzywdaAdiabaticElectronCharge2020,krzywdaInterplayChargeNoise2021,jnaneMulticoreQuantumComputing2022,langrockBlueprintScalableSpin2022} with the corresponding shuttling speed being tens of $\si{\m\per\s}$. Hence, the shuttling step will not be the rate-limiting step in the pipeline. In Ref.~\cite{boterSpiderwebArraySparse2022}, it is estimated that a qubit separation of $\sim 10\ \si{\um}$ is needed for the local integration of various classical control electronics for the implementation of surface code. Such a qubit separation would correspond to a loop with an edge length of $\sim 7\ \si{\um}$ in the pipelined surface code architecture in~\cref{fig:surface_code_pipeline}. Assuming a shuttling speed of $\sim 25\ \si{\m\per\s}$, the time required for shuttling through one round of the loop will then be $\tsh \sim 1\ \si{\us}$.

In this case, the rate-limiting step is the measurement with a processing time of:
\begin{align*}
    \tmax = \tmeas \sim 1\ \si{\us}.
\end{align*}
Using \cref{eqn:data_cycle_1,eqn:anc_cycle_1}, the time needed to implement the parity check circuits for one code cycle for the data and ancilla pipelines are:
\begin{align}
    \Tcirc^{\mathrm{data}} &= 3 \tsh + 8 \tcz + 2\tH = 3.85\ \si{\us}\label{eqn:silicon_data_circ}\\
    \Tcirc^{\mathrm{anc}} &= \tsh + 4 \tcz + \tmeas = 2.4\ \si{\us}\label{eqn:silicon_anc_circ}.
\end{align}
In this scenario the data pipeline is the rate-limiting pipeline, which determines the code cycle time:
\begin{align*}
    \Tcycle = \max(\Tcirc^{\mathrm{data}}, \Tcirc^{\mathrm{anc}}) = 3.85\ \si{\us}. 
\end{align*}

We also need to know the maximum number of qubits we can fit into the pipeline without qubit collision. Following \cref{eqn:max_qubit_surface}, we have $\Kloop \approx 2$, i.e. we can only fit two qubits in each loop, which is not enough for implementing any of the pipelining magic state distillation schemes discussed in \cref{sec:magic_state_distill}. The most straightforward way to solve this is by adding more measurement devices as discussed in \cref{sec:qubit_pipeline}. If we equip each loop with $4$ measurement devices such that the rate-limiting step time is now $\tmax = \tmeas/4 = 0.25\ \si{\us}$, we will be able to fit in $5$ qubits in each loop following \cref{eqn:max_qubit_surface}, which is enough for carrying out the $5$-layer magic state distillation scheme in \cref{sec:magic_state_distill}. Note that the code cycle time $\Tcycle$ remains unchanged this way. 

However, there is a much more efficient scheme to fit in more qubits as discussed in \cref{sec:steady_flow_around_loop}. There we will deviate from the steady-flow scheme and reduce the pipelining qubit time gap below $\tmax = \tmeas$. No measurement steps occur in the data pipeline, thus this has no effect on the data pipeline. On the other hand, we need to add buffering to the ancilla pipeline since it contains measurement steps. If we add enough measurement devices such that the data pipeline remains the rate-limiting pipeline, the code cycle time \emph{$\Tcycle$ will remain unchanged}. In this way, we can fit $5$ qubits using only $2$ measurement devices in each loop. Hence, we only need $2$ measurement devices per loop for carrying out the $5$-layer magic state distillation scheme, instead of $5$ measurement devices mentioned above. With $3$ measurement devices in each loop, we can fit up to $10$ qubits, which will enable us to carry out the superior $10$-layer distillation scheme. 

Using the reduced time gap, if we \emph{do not add any measurement devices}, then the more qubits we fit in, the more buffering we need to add to the ancilla pipeline, which may then become the rate-limiting pipeline and lead to an increase in $\Tcycle$. As discussed in  \cref{sec:steady_flow_around_loop}, for our parameter regime, it is more efficient to add buffering into the data pipeline instead, which will increase the minimum cycling period $\Tloop^{\mathrm{min}}$ and thus increase the number of qubits we can fit in using \cref{eqn:max_qubit_on_loop}. As shown in \cref{sec:steady_flow_around_loop}, to implement the $5$-layer ($k=5$) and $10$-layer ($k=10$) distillation schemes without adding measurement devices, we need a code cycle time of $\Tcycle = 6\ \si{\us}$ and $\Tcycle = 10.5\ \si{\us}$, respectively. 
\begin{table*}[tbhp!]
    \centering
    \renewcommand{\arraystretch}{2}
    \setlength{\tabcolsep}{0.5ex}
    
    \begin{tabular}{cccccccc}
        \toprule
       \heading{14ex}{No. of qubits per loop: $k$}& \heading{12ex}{No. of loop arrays:$A$}&\heading{12ex}{No. of code cycles: $D$}&\heading{19ex}{No. of meas. devices per loop: $m$}&\heading{16ex}{Pipelined time gap: $\tgap$ $(\si{\us})$}& \heading{16ex}{Code cycle time: $\Tcycle$ $(\si{\us})$}&\heading{14ex}{Time saving factor}&\heading{14ex}{Space-time saving factor}\\
        \hline
        $1$ &$50$&$13d$ & $1$ & $1$& $3.85$&$1$ & $1$\\\hline
        \multirow{2}{*}{$5$}&\multirow{2}{*}{$3$}&\multirow{2}{*}{$9d$}&1&$0.4$&$4.8$&$1.2$ &$19$\\
        &&&2&$0.28$&$3.85$&$1.4$&$24$\\\hline
        \multirow{2}{*}{$10$}&\multirow{2}{*}{$1$}&\multirow{2}{*}{$3d$}&1&$0.32$&$8.55$&$2$&$100$\\
        &&&3&$0.125$&$3.85$&$4.3$&$200$\\
        \botrule
    \end{tabular}
    
    \caption{A summary of the space-time saving brought by pipelining for the surface code magic state distillation circuit implemented in the semiconductor spin qubit platform. The time overhead is simply $D\Tcycle$ and the factor of time saving is calculated using the no-pipeline ($k=1$) scheme as the baseline. Similarly, the space-time overhead is $AD\Tcycle$ and the factor of space-time saving is calculated using the no-pipeline ($k=1$) scheme as the baseline.}
    \label{tab:msd_saving_2}
\end{table*}

As mentioned before, the space-time overhead is simply the product of the number of loop arrays required ($A$), the number of code cycles required ($D$) and the code cycle time ($\Tcycle$). The space-time overhead savings achieved by applying pipelining to the magic state distillation circuit in a semiconductor spin qubit platform is summarised in \cref{tab:msd_saving_2}. There we see that without adding any measurement devices, the best we can achieve is an $100$ times reduction in the space-time overhead using the $10$-layer distillation scheme. If we add two more measurement devices per loop, we can achieve a $200$ times reduction in the space-time overhead using the $10$-layer distillation scheme. The $5$-layer scheme does not offer any advantages over the $10$-layer scheme in the cases that we have considered. The detailed space-time savings expected for the whole computation are discussed in \cref{sec:full_FT} for the cases in which additional measurement devices are added, as necessary, such that the code cycle time stays the same (the implications of using a single measuring device are also noted).

As mentioned above, we considered the case that the single-qubit gates are carried out via EDSR; this can necessitate the use of gate devices involving micromagnets placed close to the quantum dots. If such devices are placed on the shuttling loop, then our qubits may undergo unwanted rotations every time they pass these structures. To the extent that such rotations are deterministic, it may be possible to account for them within the algorithm we are implementing. We might also tackle such an issue at the architectural level by having these single-qubit gate devices branching off the shuttling loop (similar to the readout devices in earlier figures), so that the qubits will not go near these gate devices unless we want to apply gates on them. Since we expect the speed of the shuttling step to be relatively fast compared to the other operations in the pipeline, taking this detour need not have much effect on the arguments we made above.

As an alternative to the EDSR route, a platform might employ electron spin resonance (ESR) to implement the single-qubit gates instead. As shown in \cref{sec:ESR_spacetime_saving}, using ESR means a slower code cycle without pipelining $\Tcycle^{(1)} = 5.9\ \si{\us}$. However, an advantage to the slower code cycle is that we can actually fit more qubits into the pipeline without adding buffering and measurement devices. With the exact same hardware (without adding any measurement devices), we can carry out the $10$-layer distillation scheme with a code cycle of $9.15\ \si{\us}$, achieving a $140$ times reduction in the space-time overhead compared to the unpipelined scheme.

\subsection{Threshold Calculations}
In order to carry out parity checks for quantum error correction codes like the surface code, we need to use circuits like those in \cref{fig:parity_check} with the understanding that all components are at some level faulty. There exists a threshold for the error rate of these faulty physical components, below which the logical error rate of the code can be suppressed to any desired level by suitably scaling up the code. Hence, this error threshold implies a target component fidelity that experimentalists aim to surpass in order to scale up their system through error correction. The error threshold is different for different codes and is highly dependent on the exact implementation of the stabiliser check process. For the surface code with components suffering depolarising noise, various stabiliser check implementations have been studied and typical yield an error threshold of $0.5\%$ to $1\%$~\cite{stephensFaulttolerantThresholdsQuantum2014}. 

In this section, we will perform an error threshold simulation for the implementation of surface code pipelines using the semiconductor spin qubits as outlined in \cref{sec:silicon_code_cycle}. The noise model we use for the standard gate components are:
\begin{itemize}
    \item Measurement, initialisation and CZ gates experience fully depolarising noise with probability $p$. 
    \item Single-qubit gates experience completely depolarising noise with probability $p/10$.
\end{itemize}
This is consistent with some of the most widely used noise models for the standard gate set, so that our result can be compared to threshold results in other studies. On top of these standard gate noise sources, we will also consider the noise due to the shuttling process itself, and the noise due to placing multiple qubits in the same loop. 

\paragraph{Shuttling dephasing.} A leading source of errors for shuttling semiconductor spin qubits is phase rotation due to inhomogeneity of effective $g$-factors across different quantum dots in the loop~\cite{buonacorsiNetworkArchitectureTopological2019,buonacorsiSimulatedCoherentElectron2020,langrockBlueprintScalableSpin2022}. We can think of it as some deterministic phase rotation on the qubit after one round around the shuttling loop, which can be in principle be corrected by applying a calibrated inverse rotation at the end of each round. Phase rotations commute with CZ, thus as long as we correct it after each round so that the phase rotation on the data qubits does not go through the Hadamard gate and become bit rotation, ideally the shuttling noise can be perfectly removed. However, as time goes on, some shuttling loops may go out of calibration and there will be some remnant phase rotation after each round. If we \emph{twirl} this remnant noise by conjugating the circuit with random Pauli gates~\cite{wallmanNoiseTailoringScalable2016}, the noise will effectively become pure \emph{dephasing noise} on the qubits with the error probability $p_{\mathrm{sh}}$ in each code cycle~\cite{buonacorsiNetworkArchitectureTopological2019}. One must note, however, that the additional single-qubit Pauli gates used for twirling will also introduce single-qubit depolarising noise with the probability $p/10$. 

\paragraph{Shuttling leakage.} 
In silicon spin qubits, the mixture of two \emph{valley orbitals} (equivalent minima) in the bulk silicon conduction band gives rise to the ground orbital that our spin qubit lives in and an excited orbital. The amount of mixing between the two valley orbitals and the resultant energy separation (\emph{valley splitting}) between the ground orbital and the excited orbital is influenced by the heterointerface at the quantum dot. Due to variations of these heterointerfaces from one quantum dot to another, we will see variations in the ground orbital mixture from one dot to another. Such a variation means that as we shuttle the spin qubit from one dot to another, the spin qubit may interact with the excited orbitals in the two dots, resulting in a state that has qubit information `leaked' into the excited orbital. Such a leaked component is analogous to a spin qubit with a miscalibrated detuning, thus its interaction with the other qubits will also be erroneous~\cite{buterakosSpinValleyQubitDynamics2021}.
The leakage rate $p_{\mathrm{leak}}$ is highly dependent on the system's structure, and this is the free parameter we sweep in our modelling. A second key parameter is the inter-valley relaxation time, i.e. the typical duration for which a leaked state persists before relaxing back; we would wish this to be fast. This relaxation time can go below $100\ \si{\ns}$~\cite{langrockBlueprintScalableSpin2022} and even reach $10\ \si{\ns}$~\cite{yangSpinvalleyLifetimesSilicon2013,tahanRelaxationExcitedSpin2014}, which is much smaller than the code cycle time we considered ($\si{\us}$ scale). Hence, we expect the leaked qubits can be restored back into the computational subspace within each code cycle. In Ref.~\cite{langrockBlueprintScalableSpin2022} the authors argue that valley excitation and relaxation during shuttling will effectively lead to dephasing noise. Here we will make the pessimistic assumption of an even \emph{more damaging} model for the shuttling leakage errors: all qubits will leak with the probability $p_{\mathrm{leak}}$ at the start of each code cycle, and if leakage happens, the leaked qubit \emph{completely depolarises all qubits that it interacts with} and will become completely depolarised itself at the end of the code cycle. 

\paragraph{Unwanted intra-loop interaction.} Another possible noise source is the undesired long-range dipole-dipole interaction among qubits in the same loop. The strength of the dipole-dipole interaction is given by $J = \frac{\mu_0g_e^2\mu_B^2}{4\pi r^3}$ with $\mu_0$, $g_e$, $\mu_B$ and $r$ being the vacuum permeability, electron $g$-factor, Bohr magneton and the distance between the spins, respectively. As discussed in the last section, a reasonable assumption for the length of one edge of the loop is $10\ \si{\um}$. Even if we have $\sim 100$ qubits in the loop, it would mean a qubit spacing of $r \sim 0.1\ \si{\um}$, translating into $J \sim 100 \ \si{\Hz}$. Such a noise strength is negligible compared to the other energy scales (on the order of $\si{\MHz}$ to $\si{\GHz}$) in the scenario. 

\begin{figure}[htbp]
    \centering
    \includegraphics[width = 0.49\textwidth]{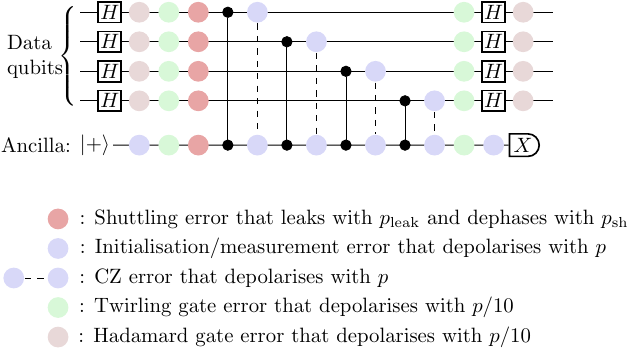}
    \caption{A diagram showing the different error locations in the X check circuits in our error model.}
    \label{fig:X_check_err_loc}
\end{figure}

The error locations in the $X$ parity check circuit is summarised in \cref{fig:X_check_err_loc}, similarly for the $Z$ checks.

\begin{figure*}[htbp]
    \centering
    \subfloat[Threshold for $p_{\mathrm{sh}}$ at $p = 0.5\%$, $p_{\mathrm{leak}} = p_{\mathrm{rash}} = 0$ \label{fig:thre_sh_005}]{\includegraphics[width = 0.3\textwidth]{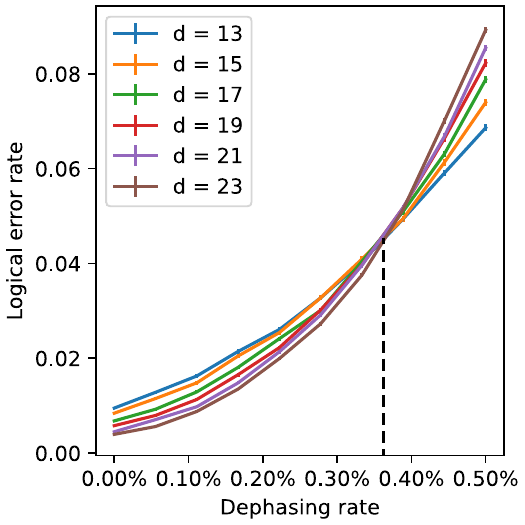}}
    \subfloat[Threshold for $p_{\mathrm{leak}}$ at $p = 0.5\%$, $p_{\mathrm{sh}} = p_{\mathrm{rash}} = 0$\label{fig:thre_lk_005}]{\includegraphics[width = 0.3\textwidth]{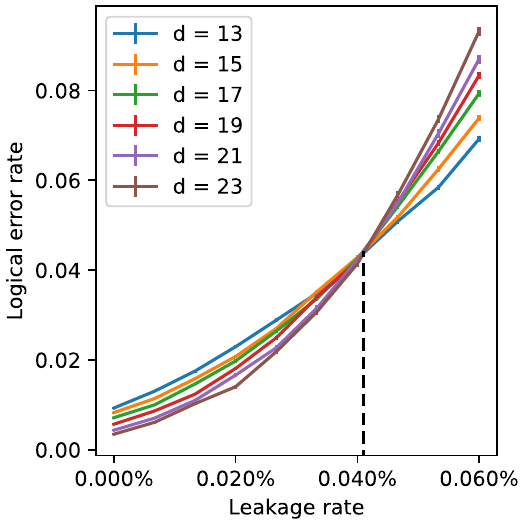}}
    \subfloat[Threshold for $p_{\mathrm{rash}}$ at $p = 0.5\%$, $p_{\mathrm{sh}} =  p_{\mathrm{leak}} = 0$ \label{fig:thre_rash_005}]{\includegraphics[width = 0.3\textwidth]{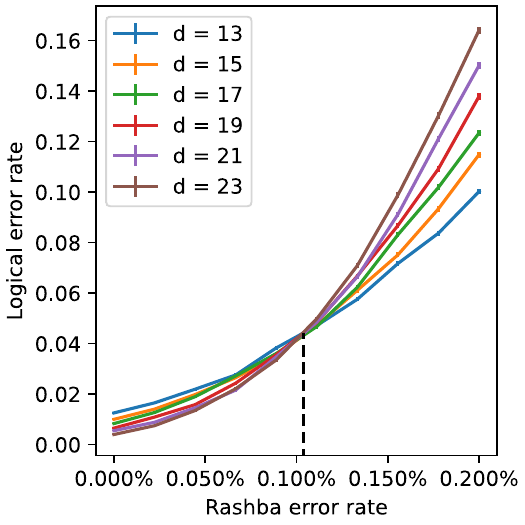}}\\
    \subfloat[Threshold for $p_{\mathrm{sh}}$ at $p = p_{\mathrm{leak}} = p_{\mathrm{rash}} = 0$\label{fig:thre_sh_0}]{\includegraphics[width = 0.3\textwidth]{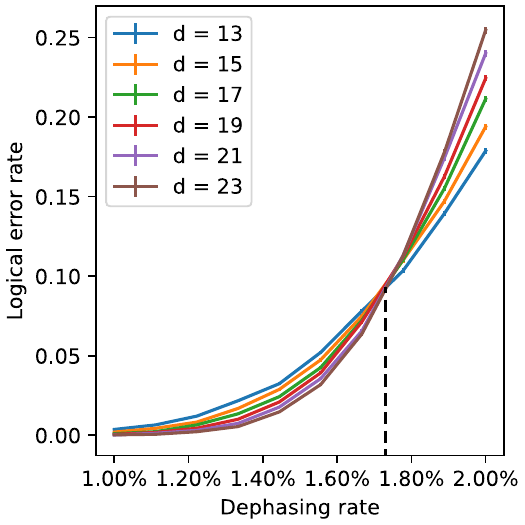}}
    \subfloat[Threshold for $p_{\mathrm{leak}}$ at \ $p = p_{\mathrm{sh}} = p_{\mathrm{rash}} = 0$\label{fig:thre_lk_0}]{\includegraphics[width = 0.3\textwidth]{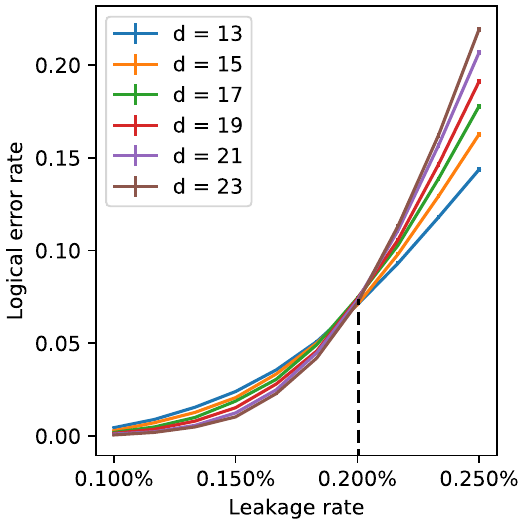}}
    \subfloat[Threshold for $p_{\mathrm{rash}}$ at \ $p = p_{\mathrm{sh}} =  p_{\mathrm{leak}} = 0$\label{fig:thre_rash_0}]{\includegraphics[width = 0.3\textwidth]{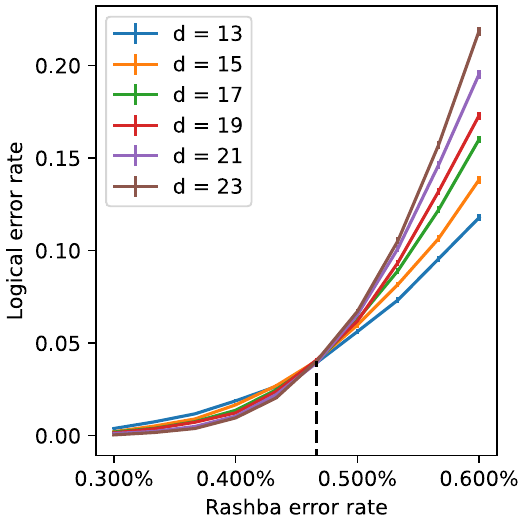}}
    
    \caption{Threshold plots for the dephasing errors ($p_{\mathrm{sh}}$), leakage errors ($p_{\mathrm{leak}}$) and the Rashba error ($p_{\mathrm{rash}}$) due to shuttling when implementing surface codes in semiconductor spin qubit using shuttling-based architectures. Note that in order to investigate the tolerance of the surface code against a given specific type of shuttling noise, the other types shuttling noise are turned off in our simulation. Each noise type is evaluated both for a gate error rate $p$ of zero, and for $p=0.5\%$.}
    \label{fig:threshold}
\end{figure*}

The error threshold we obtain for purely standard gate noise (without any of the shuttling noise) is 0.74\%, which agrees with previous well-known results~\cite{raussendorfTopologicalFaulttoleranceCluster2007}. Now suppose we manage to achieve a gate error rate of $p = 0.5\%$, which is below threshold and has been demonstrated in state-of-the-art experiments~\cite{xueQuantumLogicSpin2022,noiriFastUniversalQuantum2022}. Then the level of shuttling dephasing noise we can tolerate is found from the threshold plot in \cref{fig:thre_sh_005}, namely $p_{\mathrm{sh}} = 0.36 \%$. As mentioned, the shuttling dephasing noise is due to equipment drifting out of calibration and thus should be at a much lower level than the gate noise. Hence, the shuttling dephasing noise should not pose a problem for our implementation. This is consistent with the result in Ref.~\cite{buonacorsiNetworkArchitectureTopological2019}. 

Holding the gate error rate at $p = 0.5\%$, now let us shift our focus to the leakage noise. In the threshold plot in \cref{fig:thre_lk_005}, we obtain a leakage threshold of $p_{\mathrm{leak}} = 0.04 \%$ in the presence of gate noise. Ref.~\cite{langrockBlueprintScalableSpin2022} has argued that such a shuttling error rate is achievable when shuttling across tens of $\si{\um}$ at a speed of tens of $\si{\m\per\s}$, which is exactly the physical setting that we are considering in \cref{sec:silicon_code_cycle}. Note that the achievable error rate given in Ref.~\cite{langrockBlueprintScalableSpin2022} includes shuttling dephasing error, thus our achievable leakage error rate will be even lower. The leakage error model in Ref.~\cite{langrockBlueprintScalableSpin2022} is also less damaging than the error model we have assumed, hence our figures provide a pessimistic lower bound of the threshold for the model in Ref.~\cite{langrockBlueprintScalableSpin2022}.

If we can further suppress our gate noise below $p = 0.5\%$, we can further improve our tolerance against the dephasing noise and leakage noise due to shuttling. In the extreme of zero gate noise ($p = 0$), we can tolerate a level of dephasing noise $p_{\mathrm{sh}} = 1.75\%$ and a level of leakage noise $p_{\mathrm{leak}} = 0.20\%$. Overall, we see that the dephasing noise in shuttling is less damaging than standard gate noise, while the leakage noise, even assuming a very damaging noise model, is just one order of magnitude more damaging than the standard gate noise. Hence, they should not impose any fundamental limitation on implementing shuttling-based architectures in semiconductor spin qubits. It would be interesting to perform a similar analysis for trapped-ion shuttling, for example using the error model presented in Refs.~\cite{bermudezAssessingProgressTrappedIon2017,bermudezFaulttolerantProtectionNearterm2019}.

\begin{figure}[htbp]
    \centering
    \includegraphics[width = 0.49\textwidth]{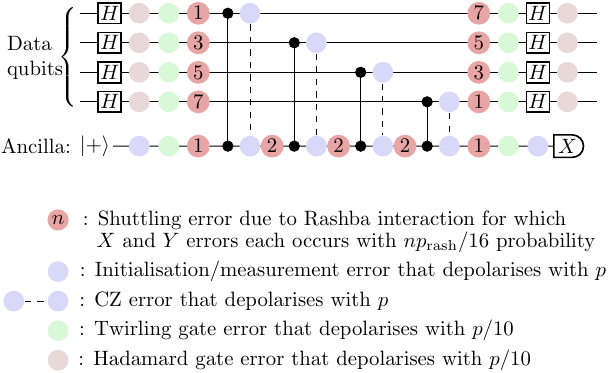}
    \caption{A diagram showing the different error locations in the X check circuits with the shuttling errors coming from Rashba interaction.}
    \label{fig:X_check_err_loc_rash}
\end{figure}

A further noise mechanism worth considering is the Rashba spin-orbit interaction due to the shuttling of the spin qubits. This will lead to coherent rotations of the qubits as they go around the loop, which can be characterised and corrected analogously to the shuttling dephasing~\cite{tanttuControllingSpinOrbitInteractions2019a}. We can again use twirling to decohere any uncompensated residual noise; in this case however it will give rise to stochastic Pauli $X$ and $Y$ noise. For simplicity, we will assume $X$ and $Y$ errors occur with the same probability $p_{\mathrm{rash}}/2$ each time we shuttle the qubit around the whole loop. In the context of performing $X$ stabiliser check, we have labelled the error locations of the Rashba errors in \cref{fig:X_check_err_loc_rash}, with the error strength in each location adjusted according to the corresponding shuttling length in that section of the circuit. Note that there we only show the shuttling errors due to Rashba interaction. The corresponding thresholds for the shuttling error due to Rashba interaction are shown in \cref{fig:threshold} where the threshold is $p_{\mathrm{rash}} = 0.1\%$ with gate noise being $p = 0.5\%$, and $p_{\mathrm{rash}} = 0.46\%$ in the absence of gate noise $p = 0\%$. These thresholds are lower than those for shuttling dephasing noise because we are using CZ gates in our parity check circuit as shown in \cref{fig:X_check_err_loc_rash} along which $X$ and $Y$ errors can propagate but $Z$ noise cannot. A higher threshold can be obtained if we are using CNOT gates in the parity check circuits instead. The thresholds given above effectively determine the required accuracy for the coherent correction we need to perform. They are of the similar order as the shuttling dephasing noise we studied above and should not present any fundamental roadblocks for our implementations. 

Before concluding this section, it is worth pointing out that instead of using single electron spins as our qubits, it is also possible to use singlet-triplet qubits~\cite{levyUniversalQuantumComputation2002} or exchange-only qubits~\cite{divincenzoUniversalQuantumComputation2000} in our pipelining architectures. An advantage is that faster single-qubit gates could then be available. Furthermore, if we are transporting all the constituent spins together, some of the shuttling noise we mentioned above may have trivial effects since these qubits live in decoherence-free subspaces~\cite{lidarReviewDecoherenceFreeSubspaces2014}. However, there will be additional challenges in shuttling and additional noise mechanisms involved like leakage from the qubit subspace. Hence, the performance and trade-offs for such qubit encodings in the pipelining architecture can be an interesting future direction to explore.

\section{Application to Quantum Error Mitigation}\label{sec:QEM}
Besides quantum error correction, pipelining can also find an application in recently proposed purification-based quantum error mitigation using multiple copies of the noisy state, which is called \emph{error suppression by derangement}~\cite{koczorExponentialErrorSuppression2021} or \emph{virtual distillation}~\cite{hugginsVirtualDistillationQuantum2021}. In these methods, the ideal state we want to prepare is some \emph{pure state} $\rho_0$, but due to the noise in the circuit, we have the noisy state $\rho$ instead. If the dominant eigenvector of $\rho$ is $\rho_0$, then we can construct a $M^\text{th}$-degree purified state
\begin{align*}
    \rho^{(M)}_{\mathrm{pur}} = \frac{\rho^M}{\Tr(\rho^M)}
\end{align*}
which has removed up to the $(M-1)^\text{th}$-order errors in the noisy state, i.e. the errors in the state is exponentially suppressed with the increase of the degree of purification $M$. If the dominant eigenvector of $\rho$ is not $\rho_0$, the error suppression will still work well as long as it is not too far from $\rho_0$, but now there is an upper-bound on the amount of noise it can remove in the name of \emph{noise floor}~\cite{hugginsVirtualDistillationQuantum2021} or \emph{coherent mismatch}~\cite{koczorExponentialErrorSuppression2021}. In some practical cases, such coherent mismatch is shown to be small~\cite{koczorDominantEigenvectorNoisy2021}.

Now in practice, instead of trying to obtain the ideal state, we are often interested in the expectation value of some observable $O$ on the ideal state $\rho_0$: $\Tr(O \rho_0)$. Then instead of trying to construct the purified state $\rho^{(M)}_{\mathrm{pur}}$, we can construct the ``purified'' expectation value $\Tr(O\rho^{(M)}_{\mathrm{pur}})$:
\begin{align}\label{eqn:pur_expect}
    \Tr(O\rho^{(M)}_{\mathrm{pur}}) = \frac{\Tr(O\rho^M)}{\Tr(\rho^M)}.
\end{align}
Hence, all we need do is to estimate $\Tr(O\rho^M)$ for some observable $O$, and obtain $\Tr(\rho^M)$ using the same method but with $O = I$. Then $\Tr(O\rho^{(M)}_{\mathrm{pur}})$ is obtain by dividing $\Tr(O\rho^M)$ by $\Tr(\rho^M)$. 

As shown in \cite{hugginsVirtualDistillationQuantum2021,koczorExponentialErrorSuppression2021}, $\Tr(O\rho^M)$ can be rewritten as:
\begin{align}\label{eqn:O_rho_M}
    \Tr(O\rho^M) = \Tr(O^{(1)}\overline{C}_M\rho^{\otimes M})
\end{align}
where $\rho^{\otimes M}$ denotes $M$ copies of the noisy state $\rho$, $\overline{C}_M$ is the cyclic permutation operators among these $M$ copies, and $O^{(1)}$ means the operator $O$ is \emph{only applied to the first copy}. Hence, the target expectation value $\Tr(O\rho^M)$ can be obtained by preparing $M$ copies of the same noisy state $\rho^{\otimes M}$ and measuring the product of the observable $O^{(1)}$ and the copy-cyclic-permutation operator $\overline{C}_M$. 

Without loss of generality, we can assume $O$ to be Pauli since any observable of interest can be decomposed into a linear sum of Pauli operators. Hence, $O$ can be written as a tensor product of \emph{single-qubit Pauli operators} $\{G_i\}$ acting on the different qubits in the first copy:
\begin{align*}
    O^{(1)} = \bigotimes_{i = 1}^N G_i^{(1)}.
\end{align*}
Here $N$ is \emph{the number of qubits in each copy} of $\rho$, and $i$ is the labelling of these qubits. We can measure $O^{(1)}$ simply by measuring $G_i$ on the $i^{\text{th}}$ qubit of the first copy. 

On the other hand, the $M$-copy cyclic permutation operator $\overline{C}_M$ is equivalent to applying the $M$-qubit cyclic-permutation operator $C_M$ transversally to the corresponding physical qubits in each of the copies:
\begin{align*}
    \overline{C}_M = C_M^{\otimes N}.
\end{align*}
Hence, the observable $O^{(1)}\overline{C}_M$ can be decomposed into the following tensor product:
\begin{align*}
    O^{(1)}\overline{C}_M = \bigotimes_{i = 1}^N G_i^{(1)}C_M
\end{align*}
which \emph{can be measured transversally}.

Consider using a conventional $2$D nearest-neighbour qubit architecture, in which different copies are stored in different 2D qubit arrays (for example, adjacent large regions of entire complete system's array). If we want to measure a multi-copy transversal operator like $O^{(1)}\overline{C}_M$, we would need to interlace the qubits arrays corresponding to different copies using a number of swaps gates that are proportional to the size of the qubit array. This would be extremely costly -- indeed it is for an analogous reason that in 2D topological code, it is conventinally more efficent to employ lattice surgery rather than transversal CNOTs. Instead we would ideally use the looped pipelining architecture proposed in the present paper, with all the $M$ qubit arrays in the same loop array (i.e. $M$ qubits in each loop structure). Then the transversal measurement of the $M$-copy operator $O^{(1)}\overline{C}_M$ can simply be carried out by measuring the $M$-qubit operator $G_i^{(1)}C_M$ in the $i^{\text{th}}$ loop.

\begin{figure}[htbp]
    \centering
    \includegraphics[width = 0.35\textwidth]{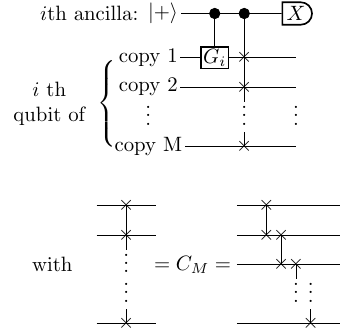}
    \caption{Measurement circuit in each loop for purification-based QEM.}
    \label{fig:individual_Hadarmard_test}
\end{figure}

To be more specific, the full process of implementing purification-based QEM using the pipelining scheme is now as follows: Using the qubit-array pipeline discussed in \cref{sec:array_pipeline}, we can have $M$ qubits in each loop and process them in the same way to prepare $M$ noisy copies of our target quantum states. However we require one further ancilla qubit in each loop, effectively adding another array of ancilla qubits on top of the $M$ noisy copies.The $i^\text{th}$ loop will contain the $i^\text{th}$ qubit of each copy of $\rho$ as well as the $i^\text{th}$ ancilla, for a total of $M+1$ qubits. We will implement the circuit in \cref{fig:individual_Hadarmard_test} in each loop. The $i^\text{th}$ ancilla will measure the observable $G_i^{(1)}C_M$, and the \emph{product of the measurement results of all ancilla} will measure the observable $O^{(1)}\overline{C}_M$ on the $M$ copies of state $\rho^{\otimes M}$, giving us the expectation value in \cref{eqn:O_rho_M}, which we can then use to obtain the error-mitigated expectation value in \cref{eqn:pur_expect}. 

As discussed before, in many hardware platforms, shuttling is an essential component for scaling up. In that case, if we scale up using a loop array as discussed in \cref{sec:array_pipeline}, we can pipeline extra qubits at zero spatial overhead. Moreover, the pipelining time cost is also negligible if the computation is deep as seen from \cref{eqn:steady_flow_pipeline_time}. Hence, purification-based QEM can be carried out \emph{at almost zero space-time overhead} in the pipelining architecture. In contrast, conventional architectures need multiple copies of the same machines and long-range interactions.

Of course, the above arguments assume that the number of qubit arrays that we pipeline for the error mitigation is not limited by the qubit collision problem mentioned in \cref{sec:qubit_pipeline}. This is not a serious limitations since in practice, due to the noise floor and the sampling overhead, it would be unlikely that one would wish to go beyond $4^\text{th}$ degree purification~\cite{hugginsVirtualDistillationQuantum2021,koczorExponentialErrorSuppression2021}, which can be implemented using only $3$ qubit arrays consisting of $2$ noisy copies and $1$ ancilla with the help of state verification~\cite{obrienErrorMitigationVerified2021,caiResourceefficientPurificationbasedQuantum2021}. For the main computation, we only need to compute on $2$ noisy copies which means $2$ qubits per loop, which should not pose any significant challenges as we have seen in the example of pipelining surface codes. At the error mitigation stage we need to add in the ancilla qubit to each loop, but any additional buffering time here can be expected to be insignificant since the main computation is presumably much deeper. Note that if we are interested in only $2^{\text{nd}}$ degree purification, it can be carried out by performing transversal measurements on two noisy copies without using any ancilla~\cite{hugginsVirtualDistillationQuantum2021}.

\section{Conclusion}\label{sec:concl}
Several proposed platforms for scalable quantum computing aim to employ physical shuttling of qubits in order to optimally space out the components of the technology, thus making room for classical control systems, permitting heat dissipation, avoiding crosstalk, and so on. Shuttling structures are typically implemented or envisaged as \emph{linear} `highways' that connect components in a sparse grid. 
In this article, we have examined an alternative realisation where 
one stores a sparse qubit array in an array of shuttling \emph{loops}. The key merit of this formulation is that in such a loop array structure, we can put additional qubits into each loop, to form a local stream of qubits where each follows same trajectory as the first qubit. Thus one can store and process multiple qubit arrays using the same architecture \emph{without adding any additional hardware}. 

With $k$ qubits in each loop, we have increased the qubit density (reduced the spatial overhead) by a factor of $k$ compared to the other shuttling-based schemes while still retaining all of their advantages for scaling up. This loop array can be viewed as a pipeline for processing a stack of $k$ qubit arrays layer by layer, i.e. a \emph{qubit-array pipeline}. Furthermore, by allowing interactions between qubits within the same loop, we can connect between different layers of the stack of qubit arrays. In this way, we can perform computations on an effectively 3D qubit lattice using a 2D loop array whose hardware requirements are similar to the those that are needed \emph{in any case} for 2D shuttling-based platform. The height of the resultant 3D qubit lattice is given by the number of qubits in the loop $k$ and thus cannot be increased indefinitely. We have shown that significant improvements in both NISQ and fault-tolerant applications can be achieved via our architecture despite this limited height.

Processing multiple topological logical qubits in one loop array using pipelining will reduce the space overhead of logical qubits. Furthermore, intra-loop qubit interactions enable transversal CNOT among logical qubits, which can significantly speed up most fault-tolerant applications since CNOT is usually one of the rate-limiting steps. We have outlined a possible implementation of the surface code pipeline for semiconductor spin qubits, and we estimate that one could achieve a reduction by a factor of $100$ in the space-time overhead for magic state distillation compared to a generic shuttling-based architecture. Moreover a reduction by a factor of $200$ in the space-time overhead can be achieved if we add two more measurement devices to each loop. Achieving such factors would of course require that certain bottlenecks are avoided, and we discuss these caveats. When considering the full process of fault-tolerant quantum computation, we have estimated that one to two orders of space-time saving can be achieved by our architecture depending on the implementation details.

For the case of silicon spin qubit devices, we have shown that one can easily fit $10$ qubits in each loop for implementing the surface code, which is enough for carrying out magic state distillation using only transversal CNOTs. Furthermore, we performed surface code threshold simulations using several models for the noise which may be introduced by shuttling. We concluded that the permissible noise levels are well within those that are expected for spin shuttling. 

In the final section of our analysis, we considered the utility of the looped pipeline architecture for an application that is more relevant to NISQ-era devices: certain powerful purification-based quantum error mitigation methods that have recently been proposed. These mitigation methods require two or more entire copies of the computer's output, and then interact these copies qubit-by-qubit; this is potentially costly and unwieldy for a canonical 2D grid system. We note that a natural solution is to arrange that the $i^\text{th}$ qubit of each copies of the state are stored within the $i^\text{th}$ loop of the architecture. In this way, we can perform the circuit for purification-based QEM at almost zero space-time overhead compared to the unmitigated circuit. 

There are many other possible applications using the pipelining architecture beyond what we have studied. For example, the pipelining architecture is a natural choice for concatenating other codes on top of a topological code (indeed the case of magic state distillation which we studied is essentially an instance of this). It would be interesting to see if such code concatenation would bring advantages in fault-tolerant memory or storage. It might also be possible to connect the boundaries of the code patches in the same pipeline to construct a long strip of folded code that might be robust against biased noise~\cite{tuckettTailoringSurfaceCodes2019}.

We focused on an application of the qubit-array pipeline where qubits form layers of a virtual 3D stack. Here the majority of gate operations are inter-loop and remain within each 2D qubit array. There is a high degree of regularity between different qubit arrays in the pipeline, e.g. the stabiliser checks for the 2D topological code pipeline and the noisy copy preparation for the QEM pipeline. Given this scenario we were able to analyse the qubit movement scheduling and time cost of the whole pipeline by studying only one of the arrays. However, relaxing these regularities may be perfectly possible in both electron spin qubit devices and ion traps; generalisations would include performing different operations on distinct qubit arrays in the same pipeline, having more frequent scheduling of transversal operations between different qubit arrays, or varying either the number of qubits in the loops or their cycling frequency. Given some of these capabilities, a particularly interesting application that can be explored in the future is the implementation of 3D codes. Note that one might need to use just-in-time decoding~\cite{bombin2DQuantumComputation2018,brownFaulttolerantNonCliffordGate2020} to overcome the limited height of the 3D lattice in the pipelining architecture. Going beyond 3D codes, there has been recent work on implementing more general LDPC codes using our pipelining architecture~\cite{strikisQuantumLDPCCodes2022}. When the end-to-end process of fault-tolerant computation for these more general LDPC codes is fully worked out, it will be interesting to perform more detailed optimisations and performance analysis for their pipelining implementations.

\section*{Acknowledgements}

The authors are grateful to Benjamin Brown, Hamza Jnane, B\'alint Koczor, Markus M\"uller and Armands Strikis for helpful comments. We thank Michael Fogarty for discussions about possible implementation methods for full 3D codes, which we intend to explore comprehensively in a subsequent study. The authors acknowledge support from Innovate UK project 133997: Multicore NISQ Processors on Silicon Chips, from the EPSRC QCS Hub EP/T001062/1, from EU H2020-FETFLAG-03-2018 under the grant agreement No 820495 (AQTION) and from the IARPA funded LogiQ project. Z.C. is supported by the Junior Research Fellowship from St John’s College, Oxford.

\appendix

\section{Synchronising an Array of Shuttled Qubits}\label{sec:shuttling_sync}
Let us again consider a large qubit array stored in a loop array similar to \cref{fig:array_shuttle}, which can also be viewed as a qubit-array pipeline processing just \emph{one} qubit array. The looped qubit pipelines in the array are engineered to have the \emph{same cycling period} at every round so that they can work in synchronisation, but they are \emph{at different phases} to allow the right qubit interactions to happen. The cycling period of the qubit-array pipeline is determined by the constituent looped qubit pipeline with the longest period, and other loops will synchronise with this rate-limiting looped qubit pipeline through buffering. Such a local buffering (idling) stage is no difference from those required for the gate scheduling in the other 2D architectures. 

We can define a \emph{global clock cycle} that takes the time of one cycling period, within which all the qubit streams flow around the loops exactly once, enabling us to implement single-qubit operations on any qubits and two-qubit operations between any neighbouring qubits for all the qubit arrays in the pipeline. Through multiple global clock cycles, we can implement any unitaries we want on the qubit arrays. Since the qubit streams in the loop array are flowing in synchronisation like an array of meshed cogwheels, we can study the time taken for the qubit-array pipeline by just looking at the time needed for any of its constituent looped qubit pipelines. The initialisation at the beginning and the measurement at the end for the whole qubit-array pipeline also have all of its constituent looped qubit pipelines working in perfect synchronisation, thus the same argument applies. Hence, when we want to perform any \emph{unitary} circuits on the qubit arrays using such a shuttling architecture (the qubit-array pipeline), the time required is directly given by the time needed for any of its constituent looped qubit pipelines adding the buffering time needed for synchronising the cycling period. 

Going beyond unitary circuits, we may want to perform mid-circuit non-destructive measurement on some qubits in the arrays. Suppose we want to perform a mid-circuit measurement on a given qubit pipeline within the array between the $n^{\text{th}}$ and the $n+m^{\text{th}}$ global clock cycle. The qubit stream in the corresponding qubit pipeline needs to break away from the global clock cycle after the $n^{\text{th}}$ round, flow to the measurement device for the operation and then flow back to the outer loop to rejoin the $n+m^{\text{th}}$ global clock cycle. If the measurement operation described here (including the additional shuttling) takes longer than $m-1$ global clock cycles, then the measured qubit pipeline would become the rate-limiting pipeline, and we need to add buffering on top of the $m-1$ global clock cycles for the other qubits in the array to synchronise with the measured qubit. Otherwise, the mid-circuit measurement of the given qubit would not affect the global schedule of the other qubits in the circuits, and thus it can be carried out without additional time cost. We can always add more measurement devices as discussed in \cref{sec:qubit_pipeline} to speed up the measurement process and avoid the additional time cost. All the discussion of mid-circuit non-destructive measurement above also applies to destructive measurement plus re-initialisation. 

\begin{figure*}[htbp]
    \centering
    \includegraphics[width = 0.9\textwidth]{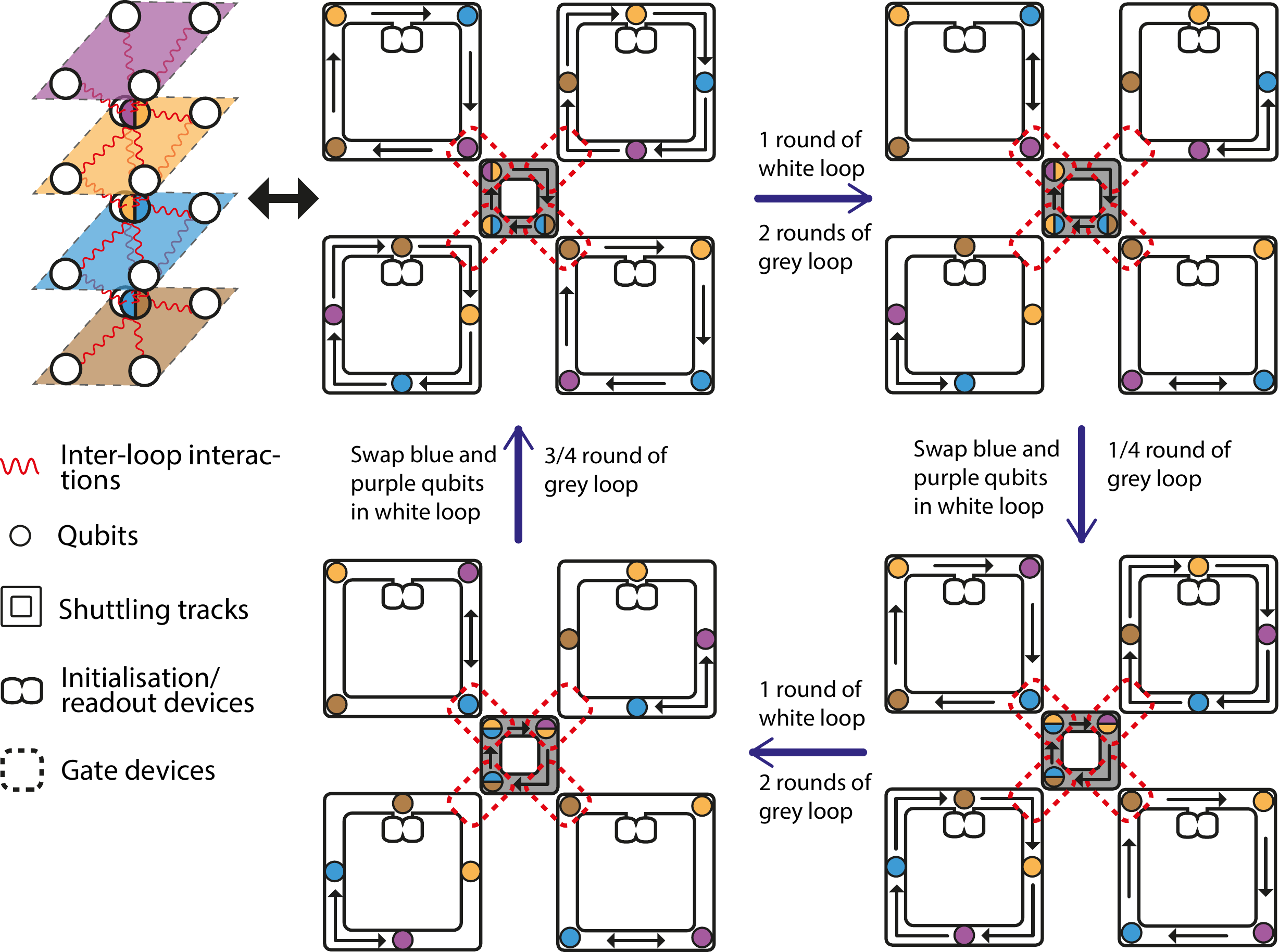}
    \caption{A qubit pipelining scheme to achieve the qubit connectivity shown on the left using loops of different cycling frequencies and different number of qubits. Here the only intra-loop operation we need is swapping the qubits in the same loop, which can be carried out using swap gates or via qubit shuttling if we allow one qubit to bypass another qubit in the loop. Qubit bypass in the loop can be achieved by adding structures to allow one qubit to move off the loop to give way to another qubit.}
    \label{fig:3D_qubit_connectivity}
\end{figure*}

\section{Colour Code Pipeline}\label{sec:colour_code_pipe}
\begin{figure*}[htbp]
    \centering
    \includegraphics[width = 0.9\textwidth]{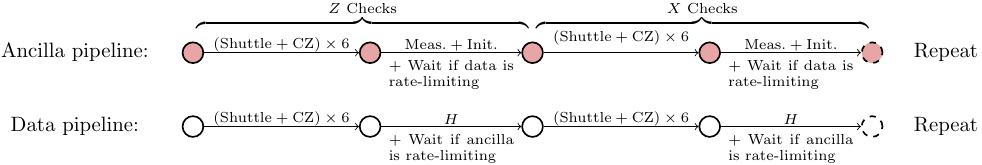}
    \caption{Pipeline workflow for a code cycle in colour code. As illustrated in the diagram, ``Wait'' needs to be added into the pipeline for synchronisation depending on which pipeline is the rate-limiting pipeline.  We have not include the initialisations of qubits before all code cycles and the measurement of all qubits after all code cycles.}
    \label{fig:colour_code_cycle_pipeline}
\end{figure*}

Similar to the surface code pipeline in \cref{sec:surface_code}, we will look at the data pipeline and the ancilla pipeline separately. A possible pipeline is shown in \cref{fig:colour_code_cycle_pipeline}, in which we carry out the $X$ and $Z$ checks in a strictly sequential manner. The time required for processing one qubit array is given by:
\begin{align}\label{eqn:cycle_1_color}
    \Tcirc^{\mathrm{data}} &= 2 \tsh + 12 \tcz + 2\tH\\
    \Tcirc^{\mathrm{anc}} &= 2 \tsh + 12 \tcz + 2\tinit + 2\tmeas.
\end{align}
They are both longer than the corresponding times in the surface code as expected. The overall code cycle time is again determined by the rate-limiting pipeline out of the data pipeline and the ancilla pipeline and is given by \cref{eqn:cycle_time_surface} with the new $\Tcirc^{\mathrm{data}}$ and $\Tcirc^{\mathrm{anc}}$ for the colour code.

The above process can be sped up by doubling the number of ancilla qubits in the ancilla pipeline, with the first half of them being the $Z$-check ancilla qubits and the second half of them being the $X$-check ancilla qubits. That is to say, when we have $k$ qubits in each data pipeline corresponding to $k$ different colour code patches, we will have $2k$ qubits in each ancilla pipeline. In this way, different operations of the $X$ and $Z$ checks might be applied at the same time, e.g. the CZs of the $Z$ checks and initialisation of the $X$ checks, or the measurements of the $Z$ checks and the CZs of the $X$ checks.

\section{Time Overhead of Pipelined Magic State Distillation}\label{sec:time_overhead}
\subsection{Colour Codes}

Let us first consider the case in which the base code is a colour code. For colour codes, the only operation requiring $\order{d}$ code cycles are CNOTs via lattice surgeries. Hence, the time required for a given circuit is largely determined by the number of rounds of CNOTs via lattice surgeries we need to implement, with each round costing $2d$ code cycles (\cref{tab:op_time_cost}). In this way, the number of code cycles $D$ we need for the distillation circuit for different pipelining schemes are given below. 

\begin{itemize}[leftmargin=*]
    \item \textbf{One layer (no pipelining).}  We need to perform $5$ rounds of multi-target CNOTs in \cref{fig:fowler_circ} and one additional round of CNOT for teleporting the $T$ gates. Hence, we have
    \begin{align}\label{eqn:time_cost_no_pipe}
        D = (5+1)2d = 12d
    \end{align}

    \item \textbf{Five layers.} We need to implement $3$ rounds of CNOTs between the $T$-state stack and the distillation stack via lattice surgeries for $T$ gate teleportation. All the other CNOTs are implemented transversally. Hence, we have 
    \begin{align}\label{eqn:time_cost_two_pipe}
        D = 3\times2 d = 6d,
    \end{align}
    i.e. half the time needed without pipelining in \cref{eqn:time_cost_no_pipe}. 
    
    \item \textbf{Ten layers.} All CNOTs can be implemented transversally and thus no CNOTs via lattice surgeries are required. However, as mentioned in \cref{tab:op_time_cost}, the initialisation process requires the whole circuit to last at least $d$ code cycles. Hence, we have
    \begin{align}\label{eqn:time_cost_one_pipe}
        D  =  d,
    \end{align}
    which is $12$ times smaller than the time needed without pipelining in \cref{eqn:time_cost_no_pipe}. 
\end{itemize}

\subsection{Surface Codes}
For the surface code, the only difference is that the $S$ gate for the correction in the $T$ gate teleportation will also require $\order{d}$ code cycles. Hence, for each round of $T$ gate teleportation, we need to add $d$ code cycles to the overall time cost. The number of code cycles $D$ we need for the distillation circuit for different pipelining schemes are given below. 
\begin{itemize}[leftmargin=*]
    \item \textbf{One layer (no pipelining).} One round of $T$ gate teleportation, which means adding $d$ to \cref{eqn:time_cost_no_pipe} and thus we have:  
    \begin{align}\label{eqn:time_cost_no_pipe_sur}
        D = 13d. 
    \end{align}
    
    \item \textbf{Five layers.} Three rounds of $T$ gate teleportation, which means adding $3d$ to \cref{eqn:time_cost_two_pipe}. However, note that in this case we can again view initialisation as instantaneous and thus we have: 
    \begin{align*}
        D = 9d,
    \end{align*}
    which is $1.4$ times smaller than the time needed without pipelining in \cref{eqn:time_cost_no_pipe_sur}. 
    
    \item \textbf{Ten layers.} Three rounds of $T$ gate teleportation, which means adding $3d$ to \cref{eqn:time_cost_one_pipe} and thus we have: 
    \begin{align*}
        D = 3d,
    \end{align*}
    which is $4.3$ times smaller than the time needed without pipelining in \cref{eqn:time_cost_no_pipe_sur}. 
\end{itemize}

We have not discussed the pipelining time cost in \cref{eqn:time_surface_code}. This is because magic state distillation is just a subroutine and not the full circuit while the pipelining time cost is for implementing the full circuit. As mentioned in \cref{sec:QEC_pipe_time}, when we look at the full circuit, for most of the interesting applications we will have relatively deep computations such that we can neglect the pipelining time cost. Even if we focus only on the magic state distillation circuit itself, the number of qubits we need in each pipeline is at most $k = 10$. This is only a fraction of the code distance in practice ($d \sim 30$), which is in turn smaller than the number of code cycles $D$. Hence, following \cref{eqn:time_surface_code}, the pipelining time cost is only a fraction of the time needed for one round of the magic state distillation subroutine and thus should be negligible compared to the time needed for the full circuit.

    \section{Space-time Overhead of Pipelined Surface Code Fault-tolerant Computation}\label{sec:full_FT}
    The full process of fault-tolerant quantum computation can be viewed as a pipeline in which magic states are distilled and then consumed. At each time step, we will input $N_0$ magic states into the pipeline, which are distilled into $N_1 \leq N_0$ magic states after the first round of magic state distillation (MSD) and then further distilled into $N_2 \leq N_1$ magic states after the second round. Practical implementations of surface codes rarely go beyond two rounds of MSD, thus these $N_2$ magic states will go straight into the main computation to be consumed for implementing $T$ gates. Magic states input at \emph{different} time steps can be processed concurrently in the first round of MSD, the second round of MSD and the main computation, and thus these three steps can form a pipeline for the full fault-tolerant computation. 
    
    We have analysed the space-time saving achievable by the first round of MSD and the main computation in \cref{sec:magic_state_distill} and \cref{sec:multi_stack}, respectively. Unlike the first round of MSD in which the input $T$ states are prepared in place in constant time (\cref{tab:op_time_cost}), for the second round of MSD, \emph{all} $15$ of its input $T$ states need to coexist with the qubits for the distillation code before we can carry out the distillation circuit. Hence, when using the pipelining architecture, unless we can fit $20$ logical qubits into one stack, we need to have multiple data stacks to carry out the second round of MSD. For the $5$-layer pipelining scheme, we can still achieve the same time saving as the first round of MSD, but now the number of code stacks needed is $A = 5$ (including one ancilla stack). For the $10$-layer pipelining scheme, additional time is needed for the second round of MSD to move the input $T$ states into the same stack as the distillation code \emph{or} we need to carry out CNOTs between different stacks for $T$ gate teleportation. Correspondingly, the number of code cycles needed becomes $D = 7d$ (slightly less than the $5$-layer scheme since the distillation code can be initialised in the same stack with $5$ of the $T$ states) and the number of code stacks needed is $A = 3$ (including one ancilla stack). 
    
    The space-time saving brought by different pipelining schemes at the different steps of the full fault-tolerant computation is summarised in \cref{tab:comp_saving}. There for simplicity we have assumed the code cycle times stay the same as we fit $5$ and $10$ qubits into each loop: $\Tcycle^{(1)} = \Tcycle^{(5)} = \Tcycle^{(10)}$, which is achievable in silicon architecture by adding measurement devices as shown in \cref{sec:silicon_code_cycle}. \emph{The speed of the full computation is determined by the rate-limiting steps in the whole computation pipeline}. In practice, one can only fit in a limited number of MSD factories due to the significant space overhead involved, and thus the rate-limiting step is usually one of the two MSD steps~\cite{babbushEncodingElectronicSpectra2018,fowlerLowOverheadQuantum2019,litinskiGameSurfaceCodes2019}. Hence, the time-saving factor achievable by pipelining for the full computation is given by the time-saving factor achievable by the rate-limiting MSD step, which is a $1.4$ times reduction in the time overhead for the $5$-layer pipelining scheme. For the $10$-layer pipelining scheme, the factor of achievable time overhead reduction is $4.3$ and $1.9$ when the rate-limiting step is the first and second rounds of MSD, respectively. On the other hand, the space overhead saving for the full computation lies between the saving achieved by the different steps, and is dependent on the fraction of the total spaces used for different steps. One can \emph{at least} achieve a $x$ time reduction in the spatial overhead by using the $x$-layer pipelining scheme, and this space saving will increase as we increase the fraction of spaces used for MSD. Hence, the $5$-layer pipelining scheme can achieve a $5$ to $16$ times reduction in the space overhead, while the $10$-layer pipelining scheme can achieve a $10$ to $50$ times reduction in the space overhead.
    
    \begin{table}[tbhp!]
        \centering
        \renewcommand{\arraystretch}{2}
        \setlength{\tabcolsep}{0.2ex}
        \subfloat[$5$ qubits per loop.]{
            \begin{tabular}{ccc}
                \toprule
                \heading{14ex}{Steps in the pipeline}& \heading{13ex}{Space saving factor}&\heading{12ex}{Time saving factor}\\
                \hline
                1st round MSD&$16.6$&$1.4$\\
                2nd round MSD&$10$&$1.4$\\
                Main computation&$\geq 5$&\raisebox{0pt}{\makecell{Intra-stack CNOTs: $\order{d}$ \\ Inter-stack CNOTs: $\leq 5$}}\\
                \botrule
            \end{tabular}
        }\\
        \subfloat[$10$ qubits per loop.]{
            \begin{tabular}{ccc}
                \toprule
                \heading{14ex}{Steps in the pipeline}& \heading{13ex}{Space saving factor}&\heading{12ex}{Time saving factor}\\
                \hline
                1st round MSD&$50$&$4.3$\\
                2nd round MSD&$16.6$&$1.9$\\
                Main computation&$\geq 10$&\raisebox{0pt}{\makecell{Intra-stack CNOTs: $\order{d}$ \\ Inter-stack CNOTs: $\leq 10$}}\\
                \botrule
            \end{tabular}
        }
        \caption{A summary of the space and time saving of the various pipelining schemes for different steps in surface code fault-tolerant computation. Here $d$ is the code distance. We have assumed the code cycle time does not change as we fit $5$ and $10$ qubits into the pipeline, which can be achieved by having one or two more measurement devices in each loop for silicon platforms as noted in \cref{sec:silicon_code_cycle}.
        Without the additional measurement devices, the savings indicated above will reduce by a factors of $1.2$ and $2.2$ for the $5$-layer scheme and the $10$-layer scheme respectively.
        }
        \label{tab:comp_saving}
    \end{table}
    
    At the early stage of fault-tolerant computation, we will only have enough resources for one round of magic state distillation, or a very limited number of first-round MSD factories in a two-round MSD process. In such cases, the rate-limiting step will be the first round of MSD, which means a $7$ to $20$ times reduction in the space-time overhead for the $5$-layer pipelining scheme, and a $40$ to $200$ times reduction in the space-time overhead for the $10$-layer pipelining scheme. In the later stage of fault-tolerant computation, since it is possible to fit in more MSD factories for the first round, sometimes the second round of MSD might become the rate-limiting step, which will lead to the \emph{same} reduction in the space-time overhead for the $5$-layer pipelining scheme, and a $20$ to $100$ times reduction in the space-time overhead for the $10$-layer pipelining scheme.
    
    It is worth noting that the analysis above is just a crude estimate of the overall saving for the full fault-tolerant computation. It is impossible to get an exact figure for the space-time saving without knowing the specific hardware constraints and the exact logical circuits that we try to implement. Another subtlety that we have not mentioned above is the stochastic nature of MSD, i.e. inputting $N_0$ magic states into the pipeline will not guarantee to output $N_1$ and  $N_2$ magic states after the first and second round of MSD, respectively. The output numbers only indicate the expected behaviour in each step. Nonetheless, the above analysis gives a good intuition about the expected saving achievable via pipelining, at least in the limit of large $N_0$.
    
    If we are allowed to shuttle qubits across different loops using e.g. the $Y$ junctions shown in \cref{fig:corner_structure} that connect different loops, then it is possible to move logical qubits from one stack to another using shuttling to enable intra-stack transversal CNOTs rather than performing inter-stack lattice surgery CNOTs. The time needed for such shuttling of logical qubits from one stack to another will still scale with the code distance $d$. However, since shuttling is usually much faster than the other operations, such movement of logical qubits may still be much faster than the other logical operations for practical code sizes. Therefore, this may be an interesting option to explore to effectively enable transversal CNOTs among all qubits even in a multi-stack picture.

\section{Surface Code Pipeline using Semiconductor Spin Qubits}\label{sec:steady_flow_around_loop}
Now let us look at surface code pipelines using semiconductor spin qubits following a similar argument to \cref{sec:silicon_code_cycle}, there the ranking of the processing time for different steps in the pipeline is $\tmeas > \tcz > \tH$ while the individual shuttling step and the initialisation step take negligible time. Note that $\tmeas = \tmeas^{m=1}/m$ where $m$ is the number of measurement devices in each loop. An assumption that measurement remains the rate-limiting step, implies that 
\begin{align}\label{eqn:restrict_m}
    \tmeas = \tmeas^{m=1}/m \geq \tcz\quad  \Rightarrow \quad m \leq \tmeas^{m=1}/\tcz.
\end{align}
In \cref{sec:silicon_code_cycle}, we have discussed the steady-flow scheme in which the time gap is $\tgap =\tmax = \tmeas$ and the minimum cycle period is inside the data pipeline with the value of \begin{align}\label{eqn:tloopmin}
    \Tloop^{\mathrm{min}} = \tsh + \tcz + \tH.
\end{align}
The maximum number of qubits we can fit into the pipeline is given by \cref{eqn:max_qubit_on_loop}:
\begin{align}\label{eqn:max_qb_on_loop}
    \Kloop = \frac{\Tloop^{\mathrm{min}}}{\tgap} + 1.
\end{align}
Hence, in order to fit in more qubits, we can either reduce the time gap such that $\tgap < \tmax = \tmeas$ and/or increase the minimum cycle period such that $\Tloop^{\mathrm{min}} > \tsh + \tcz + \tH $. 

In this section, we will only focus on time gaps that are larger than the rate-limiting component time on the data pipeline $\tgap \geq \tcz$, i.e.
\begin{align}\label{eqn:kloop_restrict}
    \frac{\Tloop^{\mathrm{min}}}{\Kloop - 1} \geq \tcz,
\end{align}
so that no time-gap buffering would be needed in the data pipeline.

There is one cycle around the loop in the ancilla pipeline while there are three cycles in the data pipeline. Out of these cycles, for the given operation times, the minimum cycle period  $\Tloop^{\mathrm{min}}$ is most likely inside the data pipeline. Hence, increasing $\Tloop^{\mathrm{min}}$ will mean adding buffering to the data pipeline, but have no effects on the ancilla pipeline. On the other hand, reducing the time gap $\tgap$ below $\tmax = \tmeas$ means we need to add time-gap buffering to the measuring step in the ancilla pipeline following \cref{eqn:constant_gap_buffer}:
\begin{align*}
    \Delta T_{\mathrm{gap},\mathrm{anc}}^{\mathrm{loop}} &= (\Kloop-1)(\tmeas - \tgap) \\
    &= (\Kloop - 1) \tmeas -\Tloop^{\mathrm{min}}
\end{align*}
where we have used \cref{eqn:max_qb_on_loop}. Combining with the ancilla circuit time given by \cref{eqn:silicon_anc_circ}, we can obtain the effective ancilla circuit time to be
\begin{align}
    T_{\mathrm{eff}}^{\mathrm{anc}}  &= \Tcirc^{\mathrm{anc}} + \Delta T_{\mathrm{gap},\mathrm{anc}}\nonumber\\
    &= \tsh + 4 \tcz + \tmeas + (\Kloop - 1) \tmeas -\Tloop^{\mathrm{min}}\nonumber\\
    &= \tsh + 5 \tcz + \Kloop \tmeas -\Tloop^{\mathrm{min}} \label{eqn:anc_pipe_time}
\end{align}
when we increase $\Tloop^{\mathrm{min}}$ and/or decrease $\tgap$. Note that $\tgap$ does not explicitly appear in the equation here since we can re-express $\Tloop^{\mathrm{min}}$ in terms of $\tgap$ for a given $\Kloop$ using \cref{eqn:max_qb_on_loop}.

For the time needed for the data pipeline, we need to consider several parameter regimes.

\subsection{No buffering in the data pipeline}
No buffering in the data pipeline means that $\Tloop^{\mathrm{min}}$ will not change, which is given by \cref{eqn:tloopmin}. The time gap used can be obtained using \cref{eqn:max_qb_on_loop,eqn:tloopmin} for different $\Kloop$:
\begin{align}\label{eqn:time_gap_no_buffer}
    \tgap = \frac{\tsh + \tcz + \tH}{\Kloop-1}
\end{align}
this gives $\tgap = 0.28\ \si{\us}$ for $\Kloop=5$ and  $\tgap = 0.125\ \si{\us}$ for $\Kloop=10$. 

No buffering in the data pipeline also means $\tgap \geq \tcz$, which simply means that 
\begin{align}\label{eqn:critical_qb_number}
    \Kloop &\leq \frac{\tsh + \tcz + \tH }{\tcz} + 1 = 12
\end{align}
i.e. we can at most achieve $\Kloop = 12$ without adding any buffering to the data pipeline. 

In this way, the effective circuit time for the data pipeline is the \emph{same} as before in \cref{eqn:silicon_data_circ}
\begin{align}\label{eqn:data_pipe_time}
    T_{\mathrm{eff}}^{\mathrm{data}}  = \Tcirc^{\mathrm{data}} = 3 \tsh + 8 \tcz + 2\tH
\end{align}
which is independent of the time gap. 

Substituting \cref{eqn:tloopmin} into \cref{eqn:anc_pipe_time} we get the effective ancilla circuit time in this case to be
\begin{align}
    T_{\mathrm{eff}}^{\mathrm{anc}} &=  4 \tcz - \tH + \Kloop \frac{\tmeas^{m=1}}{m} \label{eqn:anc_pipe_time_2}
\end{align}

\subsubsection{Data pipeline is rate-limiting}
For the data pipeline to be rate-limiting, the number of measurement devices we need per loop is:
\begin{align}
    T_{\mathrm{eff}}^{\mathrm{anc}} &\leq T_{\mathrm{eff}}^{\mathrm{data}} \nonumber\\
    m &\geq \left\lceil \frac{\tmeas^{m=1}\Kloop}{3 \tsh + 4 \tcz + 3\tH}\right\rceil = \left\lceil \frac{\Kloop}{3.5}\right\rceil
    \label{eqn:m_requirement}
\end{align}
In this way, the code cycle time is just the data circuit time which in turn is unchanged since no buffering is added to the data pipeline. Hence, the code cycle time $\Tcycle$ is just the \emph{same} as the steady-flow scheme in \cref{sec:silicon_code_cycle} with $\Tcycle = 3.85\ \si{\us}$. 

Hence, following \cref{eqn:m_requirement}, to maintain the same code cycle time, we can fit in $\Kloop = 5$ qubits to carry out the $5$-layer distillation scheme if we have at least $m = \lceil 5/3.5\rceil = 2$ measurement devices per loop. In order to fit in $\Kloop = 10$ qubits to carry out the $10$-layer distillation scheme, we need $m = \lceil 10/3.5\rceil = 3$ measurement devices per loop. 

\subsubsection{Ancilla pipeline is rate-limiting}
For the ancilla pipeline to be rate-limiting, we need the reverse of \cref{eqn:m_requirement} to be true, i.e. $m \leq \lfloor \Kloop/3.5\rfloor$. If we want to fit in $\Kloop = 5$ qubits, this implies $m =1$. Since the ancilla pipeline is rate-limiting, the code cycle time is simply the effective ancilla circuit time given by \cref{eqn:anc_pipe_time}. For $m = 1$ and $\Kloop = 5$ we have:
\begin{align*}
    \Tcycle = T_{\mathrm{eff}}^{\mathrm{anc}} = 4 \tcz - \tH + \Kloop \frac{\tmeas^{m=1}}{m} \approx 5.4\ \si{\us}
\end{align*}
If we want to fit in $\Kloop = 10$ qubits, this implies $m \leq 2$, which gives a code cycle time of:
\begin{align*}
    \Tcycle = 4 \tcz - \tH + \Kloop \frac{\tmeas^{m=1}}{m} \approx \begin{cases}
        10.4\ \si{\us}  \quad& m =1\\
        5.4\ \si{\us}  \quad& m =2
    \end{cases}
\end{align*}

\subsection{Adding buffering to the data pipeline}
Here we will consider the case in which we add buffering to the data pipeline to increase $\Tloop^{\mathrm{min}}$ in order to fit in more qubits. For simplicity, we will only consider the case in which $\Tloop^{\mathrm{min}}$ is increased until it is larger than the slowest cycle in the data pipeline
\begin{align}\label{eqn:tloop_strict}
    \Tloop^{\mathrm{min}} \geq \tsh + 4\tcz,
\end{align}
so that all three cycles in the data pipeline can take the same cycling period. (We also require $\tgap \geq \tcz$ which is true as long as $\Kloop \leq 12$). In such a case, the effective circuit time for the data pipeline is simply:
\begin{align}
    T_{\mathrm{eff}}^{\mathrm{data}}  &= 3\Tloop^{\mathrm{min}} \label{eqn:data_pipe_time_3}
\end{align}

For a given $\Kloop$, this effective circuit time for the data pipeline and that for the ancilla pipeline in \cref{eqn:anc_pipe_time}  will be equal when:
\begin{align}
    \Tloop^{\mathrm{min}*} & = \frac{\tsh + 4 \tcz + \tmeas^{m=1}\Kloop/m}{4} \label{eqn:critical_loop_time}
\end{align}
When $\Tloop^{\mathrm{min}}$ increase beyond $\Tloop^{\mathrm{min}*}$, we will have $T_{\mathrm{eff}}^{\mathrm{data}}> T_{\mathrm{eff}}^{\mathrm{anc}}$ while if we have $\Tloop^{\mathrm{min}} < \Tloop^{\mathrm{min}*}$, then $T_{\mathrm{eff}}^{\mathrm{data}} < T_{\mathrm{eff}}^{\mathrm{anc}}$. 

Hence, the surface code cycle time is given as:
\begin{align*}
    \Tcycle &= \max(T_{\mathrm{eff}}^{\mathrm{data}}, T_{\mathrm{eff}}^{\mathrm{anc}}) \\
    & = \begin{cases}
        3\Tloop^{\mathrm{min}} & \Tloop^{\mathrm{min}} \geq \Tloop^{\mathrm{min}*}\\
        \tsh + 4 \tcz + \Kloop\tmeas - \Tloop^{\mathrm{min}} & \Tloop^{\mathrm{min}} < \Tloop^{\mathrm{min}*}
    \end{cases}
\end{align*}
which has the minimum at $\Tloop^{\mathrm{min}} = \Tloop^{\mathrm{min}*}$, which gives:
\begin{align*}
    \Tcycle^* &= 3\Tloop^{\mathrm{min}*} = \frac{3\tsh + 12 \tcz + 3\Kloop\tmeas}{4}.
\end{align*}

Substituting the time needed for the different operations into \cref{eqn:critical_loop_time}, the minimum code cycle time is given by
\begin{align*}
    \Tloop^{\mathrm{min}} = \Tloop^{\mathrm{min}*} = (0.35 + \frac{\Kloop}{4m})\ \si{\us}
\end{align*}
The restriction on $\Tloop^{\mathrm{min}}$ in \cref{eqn:tloop_strict,eqn:kloop_restrict} translate into $\Tloop^{\mathrm{min}} \geq 1.4$ and $\Tloop^{\mathrm{min}} \geq 0.1\Kloop - 0.1$, which will always be true if we have $\Kloop/m \geq 5$ and $m \leq 2$. 

For $\Kloop = 10,\ m = 2$ and $\Kloop = 5,\ m = 1$, we have  $\Kloop/m = 5$ and 
\begin{align*}
    \Tloop^{\mathrm{min}*} &= 1.6\ \si{\us}\\
    \Tcycle^* &= 3\Tloop^{\mathrm{min}*} = 4.8\ \si{\us}. 
\end{align*}
The corresponding time gap is $\tgap = \Tloop^{\mathrm{min}^*}/9 = 0.18\ \si{\us}$ for $\Kloop = 10$ and $\tgap = \Tloop^{\mathrm{min}^*}/4 = 0.4\ \si{\us}$ for $\Kloop = 5$.

For $\Kloop = 10,\ m = 1$, we have  $\Kloop/m = 10$ and 
\begin{align*}
    \Tloop^{\mathrm{min}*} &= 2.85\ \si{\us}\\
    \Tcycle^* &= 3\Tloop^{\mathrm{min}*} = 8.55\ \si{\us}. 
\end{align*}
The corresponding time gap is $\tgap = \Tloop^{\mathrm{min}^*}/9 = 0.32\ \si{\us}$.

\subsection{Others}
Further pipelining schemes can be explored where $\tgap < \tcz$. However, in that case, we need to take into account the time-gap buffering needed in the data pipeline and the corresponding change in the minimum cycling period $\Tloop^{\mathrm{min}}$.

\section{Semi-conductor spin qubit surface code pipeline using ESR}\label{sec:ESR_spacetime_saving}
In \cref{sec:steady_flow_around_loop}, we have assumed the single-qubit gates are performed using EDSR with the time required being $\tH = 25 \si{\ns}$. However, EDSR has required the incorporation of micromagnets into the architecture which is not always possible in practice. In this section, we will perform the same analysis but with the single-qubit gates carried out using electron spin resonance (ESR) instead. High-fidelity Rabi oscillation using ESR with a period of $\sim 1\ \si{\us}$ has been demonstrated for both local field and global field control~\cite{veldhorstAddressableQuantumDot2014,vahapogluCoherentControlElectron2021}. Hence, the $\pi/2$ $Y$ rotation: $Y^{\frac{1}{2}}$, which corresponds to a quarter of the Rabi cycle and can be used in place of the Hadamard gates, should be able to be carried out in $\tH \sim 250\ \si{\ns}$. The $Z$ rotation used for constructing CZ can be carried out using stark shift~\cite{hwangImpactFactorsValleys2017} in a similar time scale. Hence, we will assume the CZ gate can be carried out in $\tcz \sim 300\ \si{\ns}$. To realise a many-qubit processor, targeting ESR effects to specific qubit(s) could be a fundamental challenge. In that respect it may be a favourable feature of homogeneous codes that multiple Hadamard gates should be performed simultaneously. Certainly it is interesting to explore the potential performance under the assumption that the architectural challenges can be met.

\subsection{No buffering in the data pipeline}
In this case, the time gap used can be obtained for different $\Kloop$ can be obtained using \cref{eqn:time_gap_no_buffer}, which gives $\tgap = 0.39\ \si{\us}$ for $\Kloop=5$ and  $\tgap = 0.17\ \si{\us}$ for $\Kloop=10$. 

The restriction $\tgap \geq \tcz$ translate into (see \cref{eqn:critical_qb_number})
\begin{align}
    \Kloop &\leq \frac{\tsh + \tcz + \tH }{\tcz} + 1 = 6
\end{align}
i.e. we can at most achieve $\Kloop = 6$ without adding any buffering to the data pipeline. 

The effective circuit time for the data pipeline is
\begin{align}\label{eqn:data_pipe_time_4}
    T_{\mathrm{eff}}^{\mathrm{data}}  = \Tcirc^{\mathrm{data}} = 3 \tsh + 8 \tcz + 2\tH = 5.9\ \si{\us}.
\end{align}

\subsubsection{Data pipeline is rate-limiting}\label{sec:two_pipe_time_new}
In this way, the code cycle time is just the data circuit time given by \cref{eqn:data_pipe_time_4}: $\Tcycle = 5.9\ \si{\us}$. This includes the steady-flow scheme ($\Tcycle^{(1)} = 5.9\ \si{\us}$). 

The requirement on the number of measurement devices is given by \cref{eqn:m_requirement}:
\begin{align*}
    m &\geq \left\lceil \frac{\tmeas^{m=1}\Kloop}{3 \tsh + 4 \tcz + 3\tH}\right\rceil = \left\lceil \frac{\Kloop}{5}\right\rceil
\end{align*}
Hence, with $m = \lceil 5/5\rceil \approx 1$ measurement devices per loop, we can fit in $\Kloop = 5$ qubits to carry out the $5$-layer distillation scheme and maintain the same code cycle time ($\Tcycle^{(5)} = 5.9\ \si{\us}$).

\subsubsection{Ancilla pipeline is rate-limiting}
We need $m  < \Kloop/5$. If we want to fit in $\Kloop = 10$ qubits, this implies $m = 1$. The corresponding code cycle time is just the ancilla circuit time given by \cref{eqn:anc_pipe_time_2}:
\begin{align*}
    \Tcycle = T_{\mathrm{eff}}^{\mathrm{anc}} = 4 \tcz - \tH + \Kloop \frac{\tmeas^{m=1}}{m} \approx 10.95\ \si{\us}.
\end{align*}

\subsection{Adding buffering to the data pipeline}\label{sec:one_pipe_time_new}

Substituting the time needed for the different operations into \cref{eqn:critical_loop_time} and focusing on the case of $\Kloop = 10$, the minimum code cycle time is given by
\begin{align*}
    \Tloop^{\mathrm{min}} = \Tloop^{\mathrm{min}*} = \frac{\tsh + 4 \tcz + \Kloop\tmeas}{4} = (0.55 + \frac{5}{2m})\ \si{\us}.
\end{align*}
This will satisfy both \cref{eqn:tloop_strict,eqn:kloop_restrict} as long as $m =1$.

For $m = 1$, the minimum code cycle time is simply:
\begin{align*}
    \Tcycle^* =  3\Tloop^{\mathrm{min}*} = 9.15\ \si{\us}. 
\end{align*}
which allow us to fit $10$ qubits into the pipeline to carry out the $10$-layer distillation scheme ($\Tcycle^{(10)} = 9.15\ \si{\us}$). The corresponding time gap is $\tgap = \Tloop^{\mathrm{min}^*}/9 = 0.34\ \si{\us}$. 

\subsection{Space-time overhead saving for magic state distillation}
From the arguments above, with $m=1$ measurement devices per loop, the time required for no 1-layer, 5-layer and 10-layer schemes for the magic state distillation circuit are:
\begin{align*}
    \Tcycle^{(1)} &= 5.9\ \si{\us} \quad\text{(\cref{sec:two_pipe_time_new})}\\
    \Tcycle^{(5)} &= 5.9\ \si{\us} \quad\text{(\cref{sec:two_pipe_time_new})}\\
    \Tcycle^{(10)} &= 9.15\ \si{\us} \quad\text{(\cref{sec:one_pipe_time_new})}
\end{align*}
Substituting into \cref{tab:msd_saving}, we get $24$ times space-time saving by using the $5$-layer scheme and $140$ times space-time saving by using the $10$-layer scheme.

Note that with these new operation times using ESR, we are not able to simply increase the number of measurement devices $m$ in each loop to improve the efficiency of the pipeline. This is because the measurement time is much closer to the second slowest step: CZ gates. Therefore the measurement step will not be the rate-limiting step anymore when there are more than two measurement devices in each loop and improving its efficiency will not the pipelining efficiency. However, it is possible to add more gate devices at the same time to further parallelise the other steps as well, which can then improve the efficiency of the whole pipeline.

\section{Time Costs for Qubit Pipelines}\label{sec:pipeline_time_cost}
Here we will consider different ways $k$ qubits can flow through a pipeline with $M$ steps. The time required for the rate-limiting step is given as $\tmax$. For every step in the pipeline, we can put buffering regions before and after the step, and we will call them \emph{the entry buffer} and \emph{the exit buffer} of the step, respectively. These buffers can temporarily hold the qubits if necessary and otherwise cost zero time. When we say we have a qubit stream with a qubit time gap $\tau_{\mathrm{gap},m}$ in between the $m^\text{th}$ and $(m+1)^\text{th}$ steps, it means that after any given qubit in the qubit stream exits the $m^\text{th}$ exit buffer and enters the $m+1^{\text{th}}$ exit buffer, the next qubit will do the same after time $\tau_{\mathrm{gap},m}$. The time gap for the input qubit stream is denoted as $\tau_{\mathrm{gap},0}$ and the time gap of the qubit stream at the output is simply $\tau_{\mathrm{gap},M}$. We always have to make sure the time gap between the qubits entering the $m^\text{th}$ step is larger than the processing time of the $m^\text{th}$ step $\tau_m$. Hence, sometime there is a need for changing the time gap from $\tau_{\mathrm{gap},m-1}$ to $\tau_{m}$ before entering the $m^\text{th}$ step if $\tau_{\mathrm{gap},m-1} < \tau_{m}$. 

There are two ways to change the time gap:
\begin{itemize}
    \item Increasing the time gap from $\tgap$ to $\tgap'$, we need to buffer the $n^{\text{th}}$ qubit by the amount of $(n-1)(\tgap' - \tgap)$. Note that all qubits are buffered other than the first qubit. 
    \item Decreasing the time gap from $\tgap$ to $\tgap'$, we need to buffer the $n^{\text{th}}$ qubit by the amount of $(k-n)(\tgap - \tgap')$. Note that all qubits are buffered other than the last qubit. 
\end{itemize}

\subsection{Constant time gap}
Let us suppose we want to maintain a time gap of $\tgap$ for the qubit stream throughout the pipeline: $\tau_{\mathrm{gap},m}  = \tgap \quad \forall m$. Such a qubit stream can pass through all the steps that have $\tau_m \leq \tgap$ without any buffering. However, whenever we need to pass through a step with $\tau_m > \tgap$, we need to temporary increase the time gap from $\tgap$ to $\tau_m$, passing through the step, and the reduce the time gap from $\tau_m$ to $\tgap$ again. This will buffer the \emph{first qubit} by the amount of $(k-1)(\tau_{m} - \tgap)$. Hence, after passing through all steps in the pipeline, the first qubit will be buffered by an amount of
\begin{align}\label{eqn:time_gap_buffer_gen}
    \Delta T_{\mathrm{gap}} = (k-1)(\sum_{m = 1}^M \max(\mathrm{\tau_{m}-\tgap},0))
\end{align}
Hence, the effective time needed to process the first qubit is now 
\begin{align}\label{eqn:eff_circuit_time}
    T_{\mathrm{eff}} = \Tcirc + \Delta T_{\mathrm{gap}}.
\end{align}
Time needed to process any additional qubit is simply given by the time gap $\tgap$ and thus the total time needed to process $k$ qubits using this pipeline is
\begin{equation}\label{eqn:gen_pipeline_time}
    \begin{aligned}
        T_{\mathrm{pipe}}(k) &= T_{\mathrm{eff}} + (k-1)\tgap.
    \end{aligned}
\end{equation} 

If we have $\tgap  = \tmax$, we then have $\max(\mathrm{\tau_{m}-\tgap},0) = 0$ for all $m$ and hence the pipelining time is simply given by \cref{eqn:steady_flow_pipeline_time} as expected, in which the first qubit will need $\Tcirc$ to complete the pipeline while and any additional qubit will need $\tmax$, regardless of the depth of the pipeline $M$. 

Using \cref{eqn:time_gap_buffer_gen}, we can rewrite \cref{eqn:gen_pipeline_time} as
\begin{equation}
    \begin{aligned}
        T_{\mathrm{pipe}}(k) &= \Tcirc \\
        &\quad + (k-1) (\tgap + \sum_{m = 1}^M \max(\tau_{m}-\tgap,0)) \label{eqn:equal_gap_time_cost}. 
    \end{aligned}
\end{equation}
which can also be viewed as \cref{eqn:gen_pipeline_time} with $\Delta T_{\mathrm{gap}} = 0$ and $\tgap' = \tgap + \sum_{m = 1}^M \max(\tau_{m}-\tgap,0)$. Hence, if we have $\tgap  < \tmax$, then the time needed for processing an additional qubit through the pipeline is now $\tgap + \sum_{m = 1}^M \max(\mathrm{\tau_{m}-\tgap},0)$, which is dependent on the depth of the pipeline $M$. 

\subsection{Steady flow around the loop}\label{sec:steady_loop_flow}
Due to the repeated use of the loop in the loop pipeline, an interesting pipelining scheme besides trying to achieve a steady flow throughout the whole pipeline will be achieving a steady flow only on the loop instead (excluding measurement and initialisation), which we will simply call the \emph{steady-loop-flow} scheme. This is essentially the same scheme as the steady-flow scheme in \cref{sec:classical_pipeline}, but only taking into account the steps on the loop. Hence, the smallest time gap we can have for the \emph{steady-loop-flow} scheme is simply:
\begin{align}\label{eqn:rate_limiting_time_loop}
    \tmax^{\mathrm{loop}} = \max_{\text{on loop}} \tau_i.
\end{align}
which is the time needed for the slowest step on the loop. 

The \emph{steady-loop-flow} scheme will be different from the the \emph{steady-flow} scheme \emph{only when the rate-limiting step is initialisation or measurement}, so that we have $\tmax = \tau_{\mathrm{init/meas}} > \tmax^{\mathrm{loop}}$. By definition, the steady-loop-flow scheme does not require any buffers on the loop. Hence, we have the same cycling period $\Tloop$ as the steady-flow scheme. Due to the smaller time gap $\tmax^{\mathrm{loop}} < \tmax$, we will be able to use the steady-loop-flow scheme to fit more qubits in the pipeline compared to the steady-flow scheme according to \cref{eqn:max_qubit_on_loop}. 

However, using the steady-loop-flow scheme comes with additional time costs. While the steady-loop-flow scheme does not need any buffers on the loop, it will require buffers before and after it passes through the rate-limiting step (initialisation or measurement) to maintain the qubit time gap $\tmax^{\mathrm{loop}}$. The amount of such buffering put on the first qubit is denoted as $\Delta T_{\mathrm{gap}}^{\mathrm{loop}}$. As shown in \cref{sec:pipeline_time_cost}, to process $k$ qubits using the steady-loop-flow scheme, the amount of buffering we need to apply is:
\begin{align}\label{eqn:constant_gap_buffer}
    \Delta T_{\mathrm{gap}}^{\mathrm{loop}} = (k-1)(\sum_{m} \max(\tau_m - \tmax^{\mathrm{loop}}, 0)).
\end{align}
Here we sum over all steps in the pipeline and only the steps are slower than the rate-limiting elements on the loop (i.e. measurements and/or initialisations) are effectively included.

Hence, the effective time needed to process the first qubit is given by \cref{eqn:eff_circuit_time} with $\Delta T_{\mathrm{gap}} = \Delta T_{\mathrm{gap}}^{\mathrm{loop}}$:
\begin{align}\label{eqn:eff_circuit_time_loop_flow}
    T_{\mathrm{eff}}^{\mathrm{loop}} = \Tcirc + \Delta T_{\mathrm{gap}}^{\mathrm{loop}}.
\end{align}
The total time needed to process $k$ qubits using this pipeline is given by \cref{eqn:gen_pipeline_time} with $T_{\mathrm{eff}} = T_{\mathrm{eff}}^{\mathrm{loop}}$ and $\tgap = \tmax^{\mathrm{loop}}$:
\begin{equation}\label{eqn:pipeline_time_loop_flow}
    \begin{aligned}
        T_{\mathrm{pipe}}(k) &= T_{\mathrm{eff}}^{\mathrm{loop}} + (k-1)\tmax^{\mathrm{loop}}.
    \end{aligned}
\end{equation}

\subsection{Varying time gap}
If we allow the time gap to vary, we will have a series of peaks (local maxima) and troughs (local minima). Each transition from a peak to a trough will be called a \emph{fall}. Suppose there are $L$ falls throughout with the $\ell^{\text{th}}$ fall being the transition from $t_{\mathrm{pk}, \ell}$ to $t_{\mathrm{tr}, \ell}$, then the buffering we need to apply to the first qubit to carry out these $L$ falls is simply 
\begin{align*}
    \Delta T_{\mathrm{gap}} = (k-1) (\sum_{\ell = 1}^L \tau_{\mathrm{pk}, \ell} - \tau_{\mathrm{tr}, \ell}). 
\end{align*}
Hence, using \cref{eqn:gen_pipeline_time}, the total time needed for pipelining all $k$ qubits (i.e. the time taken for the last qubit to exit the pipeline) is:
\begin{align*}
    T_{\mathrm{pipe}}(k) = \Tcirc + (k-1) (\tau_{\mathrm{gap},M} + \sum_{\ell = 1}^L \tau_{\mathrm{pk}, \ell} - \tau_{\mathrm{tr}, \ell}). 
\end{align*}

\begin{figure*}[htbp]
    \centering
    \includegraphics[width = 0.85\textwidth]{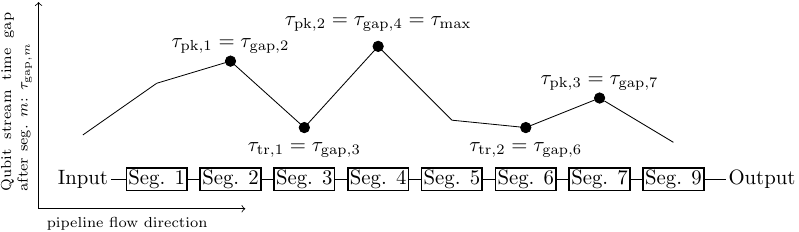}
    \caption{A diagram showing the change of the time gap of the qubit stream as it pass through different steps in the pipeline.}
    \label{fig:time_gap_variation}
\end{figure*}

If $\tau_{\mathrm{gap},M}$ is a trough, then $\tau_{\mathrm{tr}, L} = \tau_{\mathrm{gap},M}$ and we have
\begin{align}\label{eqn:time_cost_tr}
    T_{\mathrm{pipe}}(k) =  \Tcirc + (k-1) (\tau_{\mathrm{pk}, L} + \sum_{\ell = 1}^{L -1} \tau_{\mathrm{pk}, \ell} - \tau_{\mathrm{tr}, \ell}).
\end{align}
Otherwise, if $\tau_{\mathrm{gap},M}$ is a peak, then $\tau_{\mathrm{tr}, L+1} = \tau_{\mathrm{gap},M}$ and we have:
\begin{align}\label{eqn:time_cost_pk}
    T_{\mathrm{pipe}}(k) =  \Tcirc + (k-1) (\tau_{\mathrm{pk}, L + 1} + \sum_{\ell = 1}^{L} \tau_{\mathrm{pk}, \ell} - \tau_{\mathrm{tr}, \ell}).
\end{align}
In either case, if there are $Q$ peaks throughout the whole process with their time gaps denoting as $\{\tau_{\mathrm{pk}, q}\ |\ 1\leq q\leq Q\}$. In between these $Q$ peaks, there are $Q-1$ troughs (i.e. excluding any troughs at the beginning and end of the pipeline) with their corresponding time gaps denoting as $\{\tau_{\mathrm{tr}, q}\ |\ 1\leq q\leq Q-1\}$. A example is shown in \cref{fig:time_gap_variation}. Then by defining $\tau_{\mathrm{tr}, 0} = 0$, the additional time cost in \cref{eqn:time_cost_tr,eqn:time_cost_pk} can be combined into:
\begin{align}\label{eqn:pipeline_time_cost_general}
    T_{\mathrm{pipe}}(k) =  \Tcirc +  (k-1) (\sum_{q = 1}^Q \tau_{\mathrm{pk}, q} - \tau_{\mathrm{tr}, q-1}) .
\end{align}

We will restrict our scheme such that at all peaks, we have $\tau_{\mathrm{pk}, q} = \tau_{\mathrm{gap}, m} = \tau_m$ in which we have the $q^{\text{th}}$ peak occurs at the $m^\text{th}$ step. Hence, $\sum_{q = 1}^Q \tau_{\mathrm{pk}, q}$ is simply the sum of time required for some subset of steps where the time peak happens, and thus it will be smaller than the overall time require for the all steps $\sum_{q = 1}^Q \tau_{\mathrm{pk}, q} < \Tcirc$, which implies that 
\begin{align*}
    T_{\mathrm{pipe}}(k) &=  \Tcirc +  (k-1) (\sum_{q = 1}^Q \tau_{\mathrm{pk}, q} - \tau_{\mathrm{tr}, q-1})\\
    & < k \Tcirc.
\end{align*}
i.e. the pipelining scheme will always have time saving over the sequential scheme.

For the sum $\sum_{q = 1}^Q \tau_{\mathrm{pk}, q} - \tau_{\mathrm{tr}, q-1}$, we know that one of the peak would be $\tmax$. Hence, we can extract out $\tmax$ and pair up the peaks and troughs before the $\tmax$ peak using ``falls'' and pair up the peaks and troughs after the $\tmax$ peak using ``rises''. In this way, we can show that $\sum_{q = 1}^Q \tau_{\mathrm{pk}, q} - \tau_{\mathrm{tr}, q-1} \geq \tmax$. i.e. any time gap variation will never outperform the scheme with a \emph{constant} time gap of $\tmax$. 

When there is only $1$ peak, i.e. $Q=1$, which would occur at the rate-limiting element with $\tau_{\mathrm{pk}, 1} = \tmax$, we would have:
\begin{align*}
    T_{\mathrm{pipe}}(k) =  \Tcirc +  (k-1) \tmax
\end{align*}
which is simply \cref{eqn:steady_flow_pipeline_time,eqn:rate_limiting_time}. i.e. there is no additional time cost if there is just one peak in the time gap variation.

%

\end{document}